\documentclass{aa}  

\usepackage{graphicx}
\usepackage{txfonts}
\usepackage[]{hyperref}
\hypersetup{
   colorlinks=true,
   linkcolor= blue,
   filecolor= magenta,
   urlcolor= blue,
   citecolor= blue}

\usepackage{float,amsmath}
\usepackage{color}
\usepackage[version=4]{mhchem}
\usepackage{booktabs}
\usepackage{soul} 
\usepackage{orcidlink} 
\usepackage{multirow}
\usepackage{subcaption}
\usepackage{comment}
\usepackage{placeins}

\begin{document}

   \title{CoCCoA: Complex Chemistry in hot Cores with ALMA}

   \subtitle{The chemical evolution of acetone from ice to gas}

   \author{Y. Chen\inst{1}\orcidlink{0000-0002-3395-5634}
    \and R. T. Garrod\inst{2}\orcidlink{0000-0001-7723-8955}
    \and M. Rachid\inst{3}\orcidlink{0000-0001-5874-1838}
    \and E. F. van Dishoeck\inst{1,4}\orcidlink{0000-0001-7591-1907}
    \and C. L. Brogan\inst{5}\orcidlink{0000-0002-6558-7653}
    \and R. Loomis\inst{5}\orcidlink{0000-0002-8932-1219}
    \and A. Lipnicky\inst{5}\orcidlink{0000-0002-6667-7773}
    \and B. A. McGuire\inst{6,5}\orcidlink{0000-0003-1254-4817}
    }

   \institute{Leiden Observatory, Leiden University, P.O. Box 9513, 2300RA Leiden, The Netherlands \\
              \email{ychen@strw.leidenuniv.nl}
         \and Departments of Chemistry \& Astronomy, University of Virginia, Charlottesville, VA 22904, USA
         \and Netherlands Organization for Applied Scientific Research (TNO), 2628 CK Delft, The Netherlands
         \and Max Planck Institut für Extraterrestrische Physik (MPE), Giessenbachstrasse 1, 85748 Garching, Germany
         \and National Radio Astronomy Observatory, Charlottesville, VA 22903, USA
         \and Department of Chemistry, Massachussetts Institute of Technology, Cambridge, MA 02139,USA
        }

   \date{Received 11 Dec 2025 / Accepted 7 Mar 2025}

 
  \abstract
   {Acetone (\ce{CH3COCH3}) is one of the most abundant three-carbon oxygen-bearing complex organic molecules (O-COMs) that have been detected in space. The previous detections were made in the gas phase toward star-forming regions that are chemically rich, mostly in the hot cores of protostellar systems. Recently, acetone ice has also been reported as (tentatively) detected toward two low-mass protostellar systems, which enables the comparison of acetone abundances between gas and ice. The detection of acetone ice warrants a more systematic study of its gaseous abundances which is currently lacking.} 
   {We aim to measure the gas-phase abundances of acetone in a larger sample obtained from the CoCCoA survey and investigate its chemical evolution from ice to gas in protostellar systems.}
   {We fit the ALMA spectra to determined the column density, excitation temperature, and line width of acetone in 12 high-mass protostars as part of the CoCCoA survey. We also constrained the physical properties of propanal (\ce{C2H5CHO}), ketene (\ce{CH2CO}), and propyne (\ce{CH3CCH}), which might be chemically linked with acetone. We discuss the possible formation pathways of acetone by making comparisons in its abundances between gas and ice and between observations and simulations.}
   {We firmly detect acetone, ketene, and propyne in the 12 high-mass protostars. The observed gas-phase abundances of acetone are surprisingly high compared to those of two-carbon O-COMs (especially aldehydes). Propanal is considered as tentatively detected due to lack of unblended lines covered in our data. The derived physical properties suggest that acetone, propanal, and ketene have the same origin from hot cores as other O-COMs, while propyne tends to trace the more extended outflows. The acetone-to-methanol ratios are higher in the solid phase than in the gas phase by one order of magnitude, which suggests gas-phase reprocessing after sublimation. There are several suggested formation pathways of acetone (in both ice and gas) from acetaldehyde, ketene, and propylene. The observed ratios between acetone and these three species are rather constant across the sample, and can be well reproduced by astrochemical simulations.} 
   {The observed high gas-phase abundances of acetone along with dimethyl ether (\ce{CH3OCH3}) and methyl formate (\ce{CH3OCHO}) may hint at specific chemical mechanisms that favor the production of ethers, esters, and ketones over alcohols and aldehydes. On the other hand, the overall low gas-phase abundances of aldehydes may result from destruction pathways that are overlooked or underestimated in previous studies. The discussed formation pathways of acetone from acetaldehyde, ketene, and propylene seem plausible from observations and simulations, but more investigations are needed to draw solid conclusions. We emphasize the importance of studying acetone, which is an abundant COM that deserves more attention in the future.} 


   \maketitle
%

\section{Introduction}
The formation of complex organic molecules (COMs), usually considered as carbon-bearing molecules with at least six atoms \citep{2009ARAA}, has been a hot topic in astrochemistry since their first detections in the gas phase toward hot molecular cores (i.e., the compact chemically rich regions around protostars) in the last century \citep[e.g.,][]{Cummins1986, Blake1987}. In the past decade or so, powerful radio telescopes, including both single dishes and interferometer arrays, have detected more than 100 gas-phase COMs in space \citep{2020ARAA, McGuire2022}, and both the number and complexity keep increasing\footnote{\url{https://cdms.astro.uni-koeln.de/classic/molecules}}. In particular, the Atacama Large Millimeter/submillimeter Array (ALMA), known for its unprecedented sensitivity and resolution, has significantly increased the number of the identified COM-rich sources. Thanks to ALMA's high efficiency of detecting weak lines and resolving the compact hot core regions, observations on gas-phase COMs are moving from case studies \citep[e.g., toward Sgr~B2, IRAS~16293-2243, and G31.41+0.31;][]{EMoCA_2016, Jorgensen2016, Mininni2023} to large-sample surveys \citep[e.g.,][]{vanGelder2020, PEACHES_2021, ALMASOP_2022, Nazari2022_NCOM, Chen2023}, where the COM abundances can be accurately measured and systematically analyzed.

Although the gas-phase COMs have been intensively studied, the exploration of their icy counterparts has just been enabled by the \textit{James Webb} Space Telescope, or specifically, its Mid-Infrared Instrument (MIRI) and the Medium Resolution Spectroscopy (MRS) mode, which cover the fingerprint range of COM ices around 7--8 $\mu$m. Before the era of JWST, only methanol (\ce{CH3OH}), the simplest and most abundant COM, was firmly detected in interstellar ice \citep{Boogert2015}, with some tentative attribution of the absorption bands at 7.24 and 7.41~$\mu$m to ethanol (\ce{C2H5OH}) and acetaldehyde ices \citep[\ce{CH3CHO};][]{Schutte1999, Oberg2011}. Recently, more quantitative detections of COM ices, including ethanol, acetaldehyde, dimethyl ether (\ce{CH3OCH3}), methyl formate (\ce{CH3OCHO}), and acetone (\ce{CH3COCH3}), have been claimed toward three protostars using the JWST/MIRI-MRS spectra \citep{Rocha2024, Chen2024}. Methyl cyanide (\ce{CH3CN}), the most abundant nitrogen-bearing COM, has also been searched in ices and considered as tentatively detected in three protostars \citep{Nazari2024_CH3CNice}.
With the new detections of COM ices, it is now feasible to make gas-to-ice comparisons in their abundances, which can help us probe into the chemical evolution of COMs from ice (on dust grains) to gas (in hot cores). These comparisons have been conducted for four O-COMs (\ce{CH3CHO}, \ce{C2H5OH}, \ce{CH3OCH3}, and \ce{CH3OCHO}) toward two low-mass protostars \citep{Chen2024}, and two O-COMs (\ce{CH3CHO} and \ce{C2H5OH}) show signs of undergoing reprocessing in the gas phase. However, the conclusions remain to be strengthened by investigating more species and sources.


Besides the aforementioned four O-COMs, acetone is also considered as detected in ice in \cite{Chen2024} but is not included in the gas-to-ice comparisons, largely because of the lack of statistics on its gas-phase abundances. In previous observational studies on gas-phase COMs, acetone was either overlooked or served as a `byproduct' of those more abundant two-carbon COMs. However, previous measurements show that the relative abundances of acetone with respect to methanol are 10$^{-3}$--10$^{-2}$ \citep{Isokoski2013, Fuente2014, Lykke2017, vanGelder2020, Chahine2022, Baek2022, Mininni2023}, which is well above the detection limit of ALMA when covering the strong transitions. In Sgr~B2 and IRAS~16293-2243 (hereafter IRAS~16293), acetone was found to be much more abundant than its isomer, propanal (\ce{C2H5CHO}), by around one order of magnitude \citep{Belloche2013, Lykke2017}. This raises an intriguing question why acetone is such abundant as a three-carbon and ten-atom molecule.
 
To answer this question, we conduct a follow-up study of \cite{Chen2023} (hereafter C23), where the physical properties of six selected O-COMs (i.e., \ce{CH3CHO}, \ce{C2H5OH}, \ce{CH3OCH3}, \ce{CH3OCHO}, \ce{CH2OHCHO}, and \ce{(CH2OH)2}) were measured toward 14 high-mass protostellar sources in the Complex Chemistry in hot Cores with ALMA (CoCCoA) survey. In this work, we extend our attention from two-carbon O-COMs to acetone and propanal, which will increase the sample size by more than ten sources for the detection of three-carbon O-COMs. We also check ketene (\ce{CH2CO}) and propyne (\ce{CH3CCH}) which may be relevant to the formation of acetone \citep[see discussion in Sect.~\ref{sect:discussion} for details;][]{Sung1988, Hudson2017}. By making comparisons of acetone abundances between gas and ice and between observations and simulations, we can shed light on the chemical evolution of acetone from ice to gas in protostellar systems. 
This work is also helpful for future studies on the formation mechanisms of O-bearing COMs in general.


\section{Observations}\label{sect:observations}
The Complex Chemistry in hot Cores with ALMA survey (CoCCoA; PI: B. McGuire) includes data from four ALMA projects: 2019.1.00246.S, 2019.2.00112.S, 2022.1.00499.S, and 2023.1.00466.S. This paper makes use of the original 2019.1.00246.S project, which observed 24 high-mass star-forming regions that are known to be chemically rich. The observations aimed to cover two spectral tunings per source in Band 6, which are 238.0--241.7~GHz in the lower tuning, and 258.0--261.7~GHz in the upper one. In the 2019.1.00246.S project, both tunings were observed for 16 out of 24 sources at a uniform angular resolution of $\sim$0.3$\arcsec$, except one source (Orion KL) was observed at a lower resolution of $\sim$0.7$\arcsec$. The remaining seven sources only have the lower tuning observed but at a higher resolution of $\sim$0.17$\arcsec$. In the follow-up ALMA project 2022.1.00499.S, the upper-tuning observations of these seven sources were executed, and all the targeted sources except for Orion KL were observed again in both tunings at a lower resolution of $\sim$1.0$\arcsec$ to cover the molecular emission on a larger scale.

So far, only the upper-tuning data (258.0--261.7~GHz) of the 16 sources observed in project 2019.1.00246.S have been manually reduced, which already yield fruitful results on six selected O-COMs in C23. In this work, we switch our focus to a larger O-COM that was not studied in C23, acetone, of which the emission lines are mostly blended in the upper tuning. Therefore, we want to make use of the lower tuning as well, in order to cover as wide a frequency range as possible when fitting the spectra. 

Although the lower-tuning data have not been manually reduced, we found that the pipeline data provided in the ALMA Science Archive are already of very good quality. In C23, the upper-tuning spectra were extracted from the peak pixels of the integrated intensity (moment 0) maps of a representative \ce{CH3OH} line (19$_{3,17}$--19$_{2,18}$ at 258.78~GHz). This line is the strongest unblended \ce{CH3OH} line in the upper tuning. Since \ce{CH3OH} is the simplest and most abundant COM, its emission is likely to share a similar morphology with other O-COMs. We found that the pipeline-reduced datacubes can produce similar spectra if extracted from the same positions (i.e., the \ce{CH3OH} emission peaks). The only difference is that the manually-reduced data have a larger beam (0.33$\arcsec$) than the pipeline data (0.24$\arcsec$--0.27$\arcsec$), which introduces 10--20\% difference in the line intensities. To unify the beam size, we implemented the \texttt{imsmooth} task in CASA and adjusted the beam size of the pipeline data to 0.33$\arcsec$. In this way, the difference is very small ($\lesssim5$\%) between the manually-reduced and the pipeline data, which is smaller than the calibration uncertainties in ALMA Band 6 ($\sim$10\%) according to our quality assurance stage 2 (QA2) report.

By concatenating the pipeline spectra in the lower tuning and the manually-reduced spectra in the upper tuning extracted from the same coordinates, we now obtain `complete' spectra covering both tunings (7.5~GHz in total) for 12 sources: G19.01-0.01, G19.88-0.53, G22.04+0.22, G23.21-0.37, G34.30+0.20, G34.41+0.24, G345.5+1.5, G35.03+0.35, G35.20-0.74N, IRAS~16547-4247 (hereafter IRAS~16547), IRAS~18151-1208 (hereafter IRAS~18151), and NGC~6334-38. The source luminosities are listed in Table~\ref{tab:fit_results_acetone}, and we refer to C23 for other information on coordinates and distances. Reusing the manually-reduced data in the upper tuning also means that the fitting results of the six selected O-COMs in C23 are still valid.

\begin{figure*}[!h]
    \centering
    \includegraphics[width=\textwidth]{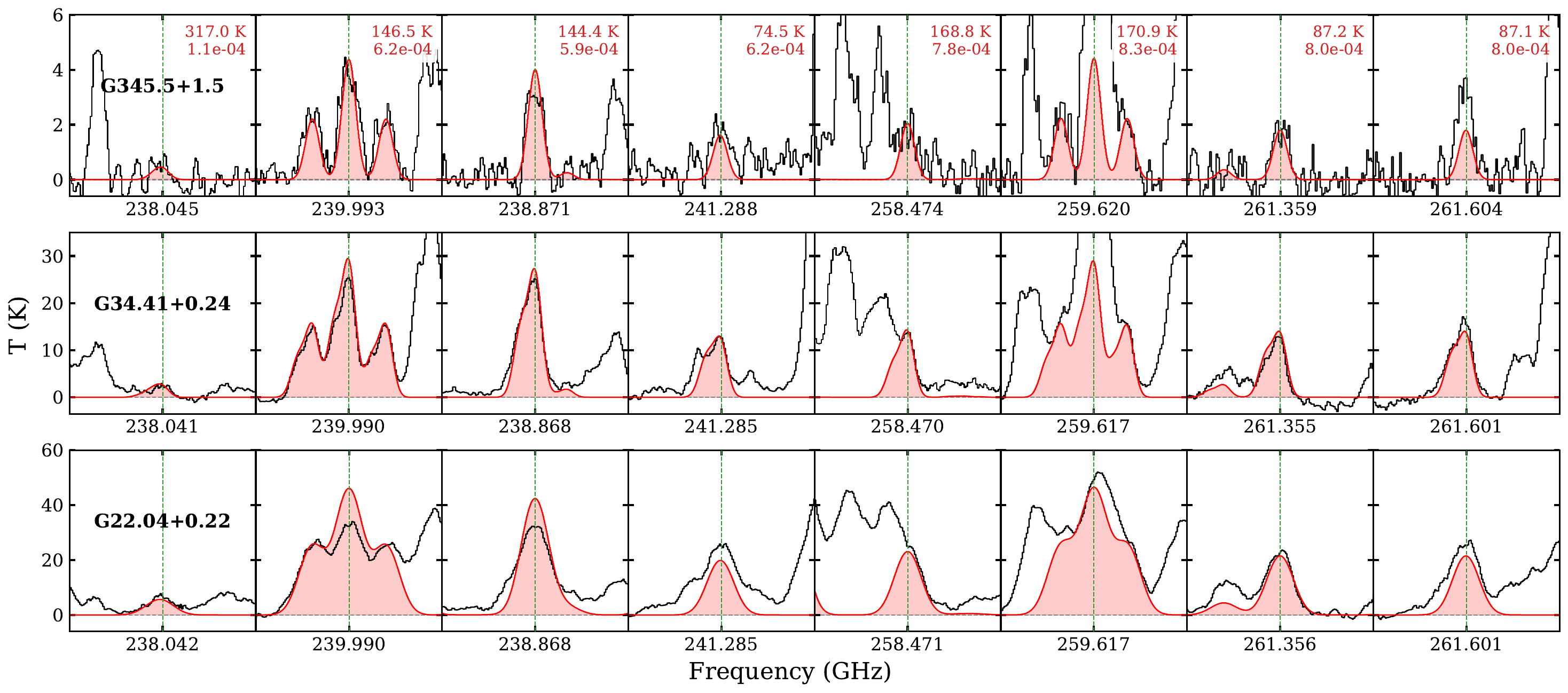}
    \caption{Best-fit LTE models of acetone spectra (red) at eight strong transitions overlaid on the ALMA spectrum of three example sources in the CoCCoA survey (black). The upper energy levels and Einstein A coefficients are annotated in red text in the upper right. On the x axis, only the central frequency of each line is labeled, but the velocity span of each panel is fixed at [--15, +15] km~s$^{-1}$.}
    \label{fig:AT_fit}
\end{figure*}

\section{Methods}\label{sect:methods}
We performed local thermodynamic equilibrium (LTE) fitting to the ALMA spectra using the spectral analysis software CASSIS\footnote{\url{http://cassis.irap.omp.eu/}} \citep{Vastel2015CASSIS}. We derived the column densities ($N$), excitation temperatures ($T_\mathrm{ex}$), line widths (FWHM), and velocities ($v_\mathrm{lsr}$) of acetone and several relevant species in the gas phase. Here LTE means assuming the populations of all levels can be characterized by a single temperature, that is, assuming one temperature component in the fitting, whereas this temperature does not need to be the same as the kinetic temperature. This assumption has been shown to work well in C23 and other observational studies \citep[e.g.,][]{Jorgensen2016}.

In addition to acetone, we also checked ketene, propyne, and propanal, 
which may be chemically relevant to acetone according to astrochemical simulations and experiments. For this section, we only introduce how we handled the spectral fitting, and we reserve the discussion about their relations to acetone in Sect.~\ref{sect:discussion}.

\subsection{Acetone (\ce{CH3COCH3})}\label{sect:method_AT}
The concatenated spectra cover tens of acetone lines above the detection limit, but only about a small portion is unblended or marginally blended in the fitting (i.e., unblended within the FWHM) due to the large amount of emission lines emerging from the hot cores. The line blending issue is more severe for bright sources in which the lines are stronger and wider. Figure~\ref{fig:AT_fit} shows the LTE-modeled spectra of acetone in comparison with the observed spectra, taking three sources with different line widths as examples. The key transitions are the triplets around 239.99~GHz and the one at 261.36~GHz (i.e., the second and the last second panels in Fig.~\ref{fig:AT_fit}). 
The multiple lines in 238.03--238.35 GHz are useful in constraining the $T_\mathrm{ex}$, since they have high upper energy levels ($E_\mathrm{up}$) of $\sim$300~K, whereas the $E_\mathrm{up}$ of other strong transitions fall in 77--171 K. These high-$E_\mathrm{up}$ lines are weak (optically thin) and unblended, thus are also good for monitoring overestimation in the fitting. More information on the transitions and the LTE-modeled spectra can be found in Appendices~\ref{appendix:transitions}--\ref{appendix:full_fit}.

\begin{figure}[!h]
    \centering
    \includegraphics[width=\linewidth]{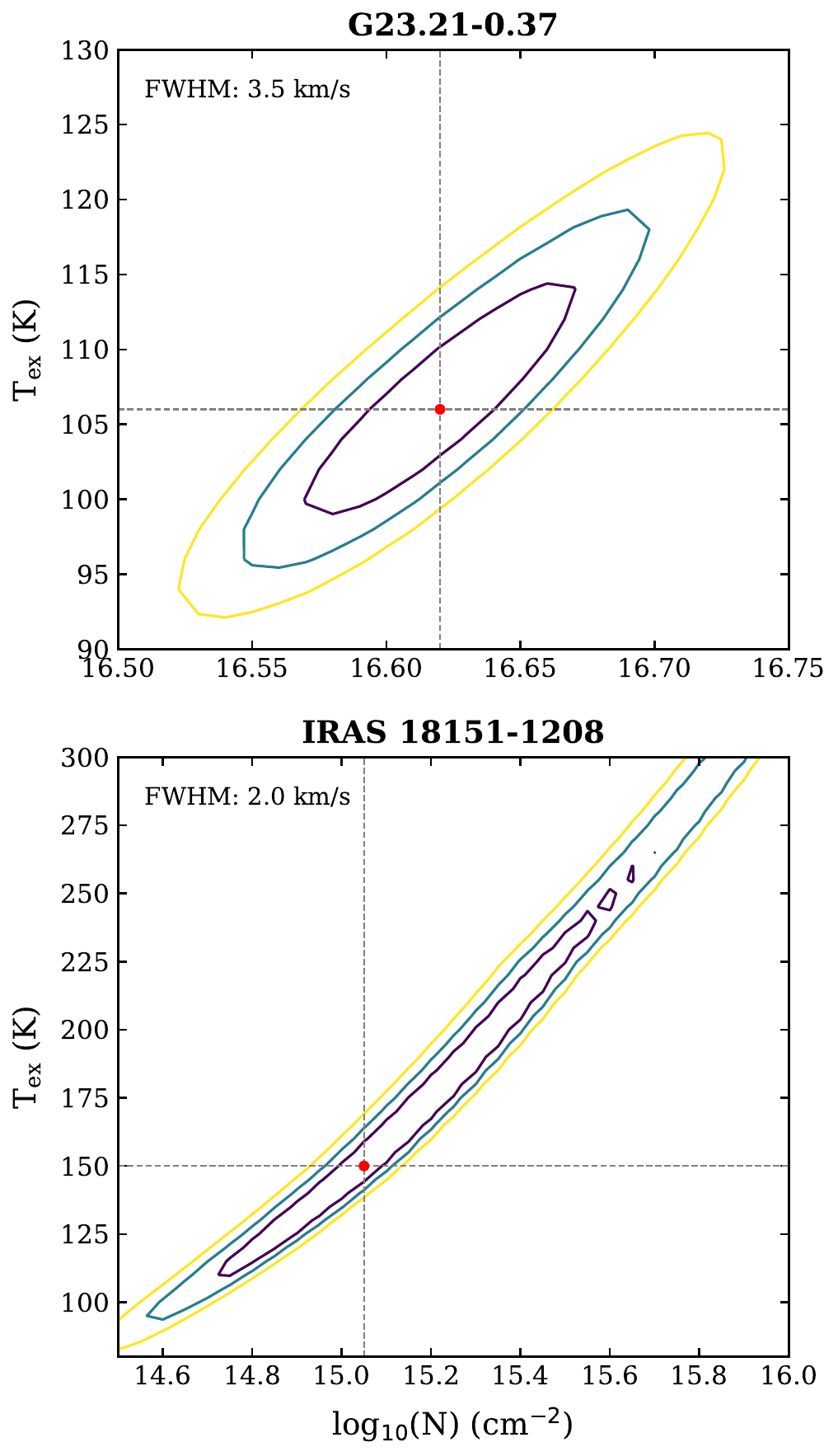}
    \caption{The $\chi^2$ contour plots in the $\log_{10}(N)-T_\mathrm{ex}$ plane for two example sources, G23.21-0.37 (top) and IRAS~18515-1208 (bottom).  The contours in yellow, blue, and purple correspond to the 1, 2, and 3 $\sigma$ around the minimum of $\chi^2$. The best-fit $\log_{10}(N)$ and $T_\mathrm{ex}$ are labeled with a red dot, and the best-fit FWHM is annotated in the top left corner.}
    \label{fig:AT_chi2_contour}
\end{figure}

Similar to C23, we tried both manual fitting (by visual inspection) and grid fitting (by running script of $\chi^2$ minimization). The manual fitting was used for sources that have more than one velocity component (e.g., G19.01-0.03) or bright sources (e.g., IRAS~16547) that suffer too much from line blending, and the grid fitting was more suitable for weak and intermediate sources. In the latter case, a contour plot on the $N$--$T_\mathrm{ex}$ plane was plotted to determine the best-fit $N$ and $T_\mathrm{ex}$ as shown in Fig.~\ref{fig:AT_chi2_contour}, and in the meantime, estimate the uncertainties on a 2$\sigma$ level.

For some sources such as IRAS~18151 (see the bottom panel of Fig.~\ref{fig:AT_chi2_contour}), there is a strong degeneracy between $N$ and $T_\mathrm{ex}$ of acetone (i.e., similar fitting results can be produced by increasing $N$ and $T_\mathrm{ex}$ simultaneously). As a consequence, the possible $T_\mathrm{ex}$ ranges from 100~K to 300~K, and the corresponding best-fit $N$ varies across one order of magnitude , which exceeds the normally estimated uncertainty of 30\%. Interestingly, this problem was also encountered by \cite{Lykke2017}, whose data are in a different band and cover a larger range of frequency (329.15--362.90~GHz, literally half of the ALMA Band 7). They chose to assign a fixed $T_\mathrm{ex}$ and determine the best-fit $N$. In our sample, most sources have a contour plot between the cases of G23.21-0.37 and IRAS~18151. Since we have derived $T_\mathrm{ex}$ for other O-COMs in C23, the range of $T_\mathrm{ex}$ set by all the other O-COMs can be used as a constraint on the $T_\mathrm{ex}$ of acetone.


\begin{figure*}[!h]
    \centering
    \includegraphics[width=\textwidth]{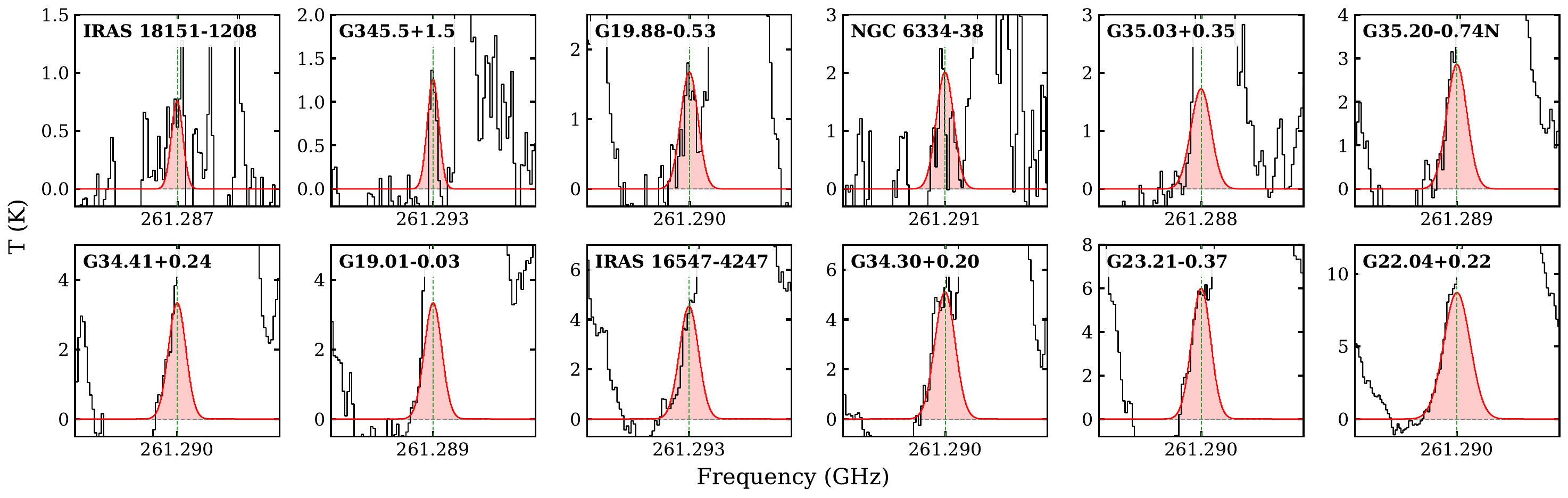}
    \caption{Best-fit LTE models of $s$-propanal spectra (red) overlaid on the ALMA spectrum of the 12 CoCCoA sources (black). All panels are centered on the strongest $s$-propanal line at 261.290~GHz. The velocity span of each panel is [--15, +15] km~s$^{-1}$.}
    \label{fig:spropanal_fit}
\end{figure*}

\subsection{Ketene (\ce{CH2CO})}\label{sect:method_CH2CO}
For \ce{CH2CO}, two strong lines, one in each tuning, are covered in the CoCCoA data. Unfortunately, these two lines are partially blended with strong lines of \ce{CH2DOH} and \ce{NH2CHO}, respectively. In our sample, the \ce{NH2CHO} line is optically thick and overfits the emission feature in most sources, and the \ce{CH2DOH} is optically thick in about half of the sources. In the cases that both \ce{CH2DOH} and \ce{NH2CHO} are optically thick, we could only estimate an upper limit of $N$(\ce{CH2CO}) at a fixed $T_\mathrm{ex}$ of 100 K. If only \ce{NH2CHO} is optically thick, a column density was fit also at $T_\mathrm{ex}$=100 K, given that the $E_\mathrm{up}$ of the two \ce{CH2CO} lines are around 88~K and 100~K, respectively. In the cases that both \ce{CH2DOH} and \ce{NH2CHO} are optically thin, the $T_\mathrm{ex}$ was constrained to be 50--120~K. Although $T_\mathrm{ex}$ could not be constrained very well by only two lines with similar $E_\mathrm{up}$, changing $T_\mathrm{ex}$ between 50~K and 150~K would only make a small difference (<20\%) in the best-fit $N$(\ce{CH2CO}). A higher $T_\mathrm{ex}$ of 200--300~K would increase the best-fit $N$(\ce{CH2CO}) by 50--100\%, but is also less plausible given that the $T_\mathrm{ex}$ of O-COMs are generally lower than 200~K (see the fifth column in Table~\ref{tab:fit_results_acetone} for the $T_\mathrm{ex}$ range of the two-carbon O-COMs studied in C23).

The Einstein A coefficients ($A_\mathrm{ij}$) of the two \ce{CH2CO} lines are (1.5--2.0)$\times10^{-4}$~s$^{-1}$, which corresponds to an modeled optical depth of $\tau\sim$0.1--1.1 under the best-fit conditions (specifically, six sources have $\tau\sim$0.1--0.25, two $\tau\sim$0.5, two $\tau\sim$0.7, and two $\tau\sim$1.1), which does not fully meet the criterion of optically thin lines ($\tau\ll1$). To avoid underestimation of $N$(\ce{CH2CO}), we checked the minor isotopologs of \ce{CH2CO}, and only CH$_2^{13}$CO has two lines covered in our data. Since the CH$_2^{13}$CO lines are next to the strong lines of the main isotopolog, only upper limits were estimated. We compared these upper limits ($N_\mathrm{uplim}$(CH$_2^{13}$CO)) with the supposed column densities of CH$_2^{13}$CO, which is the measured $N$(\ce{CH2CO}) divided by the $^{12}$C/$^{13}$C ratio. The $^{12}$C/$^{13}$C ratio empirically scales with the distance to the Galactic Center \citep[$D_\mathrm{GC}$;][]{Milam2005}:
\begin{equation}
    R(^{13}\mathrm{C}) \equiv {^{12}\mathrm{C}}/^{13}\mathrm{C} = (6.21 \pm 1.00)D_\mathrm{GC} + (18.71 \pm 7.37).
\end{equation}
According to the $D_\mathrm{GC}$ provided in C23, the $^{12}$C/$^{13}$C ratios of the considered sources range from 45 to 60, which are smaller than the value of local ISM ($\sim$70). If 
\begin{equation}\label{eq:R_13C}
    N_\mathrm{uplim}(\ce{CH2}^{13}\ce{CO}) > \frac{N(\ce{CH2CO})}{R(^{13}\mathrm{C})},
\end{equation}
then \ce{CH2CO} is likely to be optically thick and its column densities were underestimated; otherwise, \ce{CH2CO} is likely to be optically thin, or optically thick to a small degree (as the column densities of CH$_2^{13}$CO were not well constrained). For all of our sources, eq.~\ref{eq:R_13C} is not satisfied, so we consider the measured $N$(\ce{CH2CO}) to be reliable.

\subsection{Propyne (\ce{CH3CCH})}\label{sect:method_CH3CCH}
Propyne lines are only present in the lower tuning, and most of them are strong and unblended, enabling good constraints on the physical properties ($N$, $T_\mathrm{ex}$, $v_\mathrm{lsr}$, and FWHM) of \ce{CH3CCH}. The line at 239.234~GHz is blended with a strong line of $a$-\ce{(CH2OH)2}; to crosscheck, we plugged in the fitting results of $a$-\ce{(CH2OH)2} reported by C23, and the overall fitting looks decent when \ce{(CH2OH)2} is not optically thick. We also checked the $^{13}$C isotopolog of \ce{CH3CCH} as we did for \ce{CH2CO}: only CH$_3^{13}$CCH is covered in our data, and its upper limits suggest that \ce{CH3CCH} is likely to be optically thin.

We noticed that some sources (e.g., G19.01-0.03, see Table~\ref{tab:fit_results_other_sp} for details) show clearly more than one velocity components in the propyne lines. In some sources such as G345.5+1.5 (Fig.~\ref{fig:fit_all_G345.5}) and G35.20-0.74N (Fig.~\ref{fig:fit_all_G35.20}), the velocity separation between different components is larger than the FWHM of propyne lines, so that each velocity component can be fit individually by running the $\chi^2$ minimization script. For G19.01-0.03, G35.03+0.35, and IRAS~16547, the two velocity components are not separated enough, and we fit them by visual inspection with the relative uncertainty on $N$ estimated to be 25\%.

\subsection{Propanal (\ce{C2H5CHO}) and other three-carbon O-COMs}\label{sect:method_C3H6O}
Propanal drew attention as an isomer of acetone. The \textit{syn} conformer ($s$-\ce{C2H5CHO}) is the lowest energy conformer and has many transitions with $E_\mathrm{up}$ of 140--200~K and $A_\mathrm{ij}$~>~10$^{-4}$~s$^{-1}$ covered in our data, while the \textit{gauche} conformer only has weak transitions (i.e., $E_\mathrm{up}>600$~K and $A_\mathrm{ij}<10^{-5}$~s$^{-1}$). Although the $s$-\ce{C2H5CHO} lines look `strong' in terms of $E_\mathrm{up}$ and $A_\mathrm{ij}$, they are mostly clusters of hyperfine transitions and correspond to weak or (partially) blended emission features in the spectra. Therefore, we only estimated upper limits for $N$($s$-\ce{C2H5CHO}), mainly based on the strongest line at 261.29~GHz along with other unblended weaker lines (see Fig.~\ref{fig:spropanal_fit} for the fitting and Table~\ref{tab:key_trans} for the information of transitions and references). The $v_\mathrm{lsr}$ of $s$-\ce{C2H5CHO} was also determined by the strongest line. However, we noticed that the modeled spectra have slight offset from the observed spectra in the central frequencies of several lines, which was also spotted in the fitting of \cite{Lykke2017} (see Fig.~A.2 therein). This may result from the inaccuracy of laboratory measurements, and may lead to some difference in the $v_\mathrm{lsr}$ between $s$-\ce{C2H5CHO} and other O-COMs in our case where too few strong and unblended lines are detected.

\begin{table*}[!h]
    \setlength{\tabcolsep}{0.1cm}
    \renewcommand{\arraystretch}{1.3}
    \centering
    \caption{Luminosities of the 12 CoCCoA sources and the best-fit parameters of acetone.}
    \begin{tabular}{lcccccccc}
    \toprule
         \multirow{2}{*}{Source} & \multirow{2}{*}{\shortstack{$L$\\(10$^4~L_\odot$)}} & \multirow{2}{*}{\shortstack{$N_\mathrm{CH_3COCH_3}$\\(cm$^{-2}$)}} & \multicolumn{2}{c}{$T_\mathrm{ex}$ (K)} & \multicolumn{2}{c}{FWHM (km~s$^{-1}$)} & \multicolumn{2}{c}{$v_\mathrm{lsr}$ (km~s$^{-1}$)} \\
         \cmidrule(lr{0.1em}){4-5}\cmidrule(lr{0.1em}){6-7}\cmidrule(lr{0.1em}){8-9}
          & & & \ce{CH3COCH3} & O-COMs & \ce{CH3COCH3} & O-COMs & \ce{CH3COCH3} & O-COMs \\
         \midrule
         \multirow{2}{*}{G19.01-0.03} & 1 & 2.6$^{+2.0}_{-1.0}\times10^{16}$ & 160$^{+45}_{-30}$ & \multirow{2}{*}{120$\sim$180} & 3.5 & 3.0$\sim$4.0 & 62.3 & 61.8$\sim$63.0 \\
          & & 4.4$^{+1.3}_{-1.2}\times10^{16}$ & 200$\pm$25 & & 3.5 & 3.2$\sim$3.5 & 57.5 & 57.8$\sim$58.5 \\
         \hline
         \multirow{2}{*}{G19.88-0.53} & \multirow{2}{*}{0.47} & 7.2$^{+2.8}_{-1.9}\times10^{15}$ & 110$\pm$15 & 100$\sim$220 & 3.0 & 2.5$\sim$4.0 & 46.7 & 46.0$\sim$46.7 \\
          & & 4.5$^{+1.6}_{-1.3}\times10^{15}$ & 95$\pm$15 & -- & 3.0 & -- & 43.7 & -- \\
         \hline
         G22.04+0.22 & 0.497 & (1.5$\pm$0.2)\,$\times10^{17}$ & 145$\pm$10 & 130$\sim$170 & 6.5 & 4.5$\sim$5.5 & 52.5 & 52.0$\sim$53.0 \\
         \hline
         G23.21-0.37 & 1.3 & 4.2$^{+0.8}_{-0.7}\times10^{16}$ & 105$^{+15}_{-10}$ & 120$\sim$200 & 3.5 & 3.0$\sim$3.8 & 77.2 & 76.8$\sim$77.2 \\
         \hline
         G34.30+0.20 & 4.6 & 2.8$^{+0.4}_{-0.4}\times10^{16}$ & 120$\pm$10 & 130$\sim$180 & 3.0 & 3.2$\sim$3.5 & 56.2 & 55.5$\sim$56.5 \\
         \hline
         \multirow{2}{*}{G34.41+0.24} & \multirow{2}{*}{0.48} & (3.7$\pm$0.5)\,$\times10^{16}$ & 138$\pm$8 &  \multirow{2}{*}{130$\sim$180} & 3.3 & 2.5$\sim$3.0 & 59.8 & 60.0 \\
          & & 9.5$^{+3.0}_{-2.5}\times10^{15}$ & 93$\pm$15 & & 3.5 & 3.0$\sim$3.3 & 56.8 & 56.3$\sim$57.0 \\
         \hline
         G345.5+1.5 & 4.8 & 7.9$^{+3.8}_{-2.8}\times10^{16}$ & 160$\pm$30 & 120$\sim$140 & 3.5 & 2.0$\sim$2.2 & -14.5 & -16.3$\sim$-15.0\\
         \hline
         G35.03+0.35 & 0.63 & 5.5$^{+2.5}_{-1.8}\times10^{15}$ & 110$\pm$20 & 110$\sim$220 & 4.4 & 2.5$\sim$4.0 & 44.5 & 43.5$\sim$45.3 \\
         \hline
         G35.20-0.74N & 3 & 3.8$^{+0.5}_{-0.3}\times10^{16}$ & 118$^{+8}_{-5}$ & 100$\sim$180 & 4.2& 3.0$\sim$4.0 & 31.5 & 30.6$\sim$31.2 \\
         \hline
         \multirow{2}{*}{IRAS 16547-4247} & \multirow{2}{*}{6.3} & (6.5$\pm$\textit{2.1})\,$\times10^{16}$ & 180$\pm$20 & 140$\sim$200 & 3.5 & 2.2$\sim$3.3 & -35.5 & -35.7$\sim$-35.2 \\
          & & (8.0$\pm$\textit{2.0})\,$\times10^{16}$ & 110$\pm$20 & -- & 3.0 & -- & -38.7 & -- \\
         \hline
         IRAS 18151-1208 & 2.2 & 1.5$^{+0.5}_{-0.6}\times10^{15}$ & 110$\sim$180 & 110$\sim$180 & 2.0 & 1.7$\sim$2.0 & 35.0 & 35.0$\sim$35.3 \\
         \hline
         NGC 6334-38 & $<$20 & 2.7$^{+2.0}_{-0.9}\times10^{16}$ & 170$^{+50}_{-30}$ & 100$\sim$190 & 3.5 & 2.6$\sim$3.2 & -5.4 & -5.2$\sim$-5.0 \\         
         \bottomrule
    \end{tabular}
    \label{tab:fit_results_acetone}
\end{table*}

We used a fixed $T_\mathrm{ex}$ of 150~K for all sources, considering that this is a typical temperature of hot cores and also agrees well with the $T_\mathrm{ex}$ range set by other O-COMs (see Table~\ref{tab:fit_results_acetone}). Using a different $T_\mathrm{ex}$ between 80 and 220~K (i.e., 150$\pm$70~K) would only change the best-fit $N$ by $\sim20$\%, since the strongest lines used in the fitting have similar $E_\mathrm{up}$ of 140--200~K (mostly around 180~K). 

Other \ce{C3H6O} isomers were also searched for, but no valid fitting could be performed. Propylene oxide ($c$-\ce{C3H6O}) has no transition covered in our data. Propenol (\ce{CH3CHCHOH}) and isopropenol (\ce{CH3COHCH2}) are not included in the databases. Oxetane (\ce{CH2CH2CH2O}) and methyl vinyl ether (\ce{CH3OCHCH2}) are too rare in the astrochemical context and therefore not considered in this work.

We also tried fitting other three-carbon O-COMs such as propanol (\ce{C3H7OH}) that do have transitions covered in our data, but their lines are either too weak or blended with others, making the constraints on column densities highly uncertain. We therefore did not report the results of these fitting attempts in this paper.

\subsection{Methanol (\ce{CH3OH})}\label{sect:method_CH3OH}
In C23, the column densities of \ce{CH3OH} were estimated from the minor isotopolog, CH$_3^{18}$OH, by multiplying its column densities with the $^{16}$O/$^{18}$O ratios, which also scale with $D_\mathrm{GC}$ as follows:
\begin{equation}\label{eq:R_18O}
    R(^{18}\mathrm{O}) \equiv {^{16}\mathrm{O}}/^{18}\mathrm{O} = (58.8 \pm 11.8)D_\mathrm{GC} + (37.1 \pm 82.6).
\end{equation}
This indirect determination of $N$(\ce{CH3OH}) was applied because almost all (except one) detected \ce{CH3OH} lines in the upper tuning are optically thick and cannot be used in the fitting. However, in some weak sources such as IRAS~18501-1208, the CH$_3^{18}$OH lines are not clearly detected, which may lead to inaccurate estimation of $N$(\ce{CH3OH}). Fortunately, there are many more \ce{CH3OH} lines covered in the lower tuning, and several of them are likely to be optically thin given their high $E_\mathrm{up}$ ($\gtrsim$~800--1000~K) and small $A_\mathrm{ij}$ ($\lesssim3\times10^{-5}$~s$^{-1}$). We therefore revisited the fitting of \ce{CH3OH} by only taking those optically thin lines into account. The newly derived $N$(\ce{CH3OH}) do not differ too much from the values given in C23; the difference is within a factor of two for all sources but G34.41+0.24. The O-COM ratios with respect to \ce{CH3OH} involved in this paper were recalculated using the new values of $N$(\ce{CH3OH}).

\section{Results}\label{sect:results}
We claim firm detections of acetone, ketene, and propyne in all the 12 CoCCoA sources. The transitions with $E_\mathrm{up}<800$~K and $A_\mathrm{ij}>5\times10^{-5}$~s$^{-1}$ covered in the dataset are listed in Table~\ref{tab:key_trans}. 
Figure~\ref{fig:AT_fit} shows the best-fit spectra of acetone at key lines in three example sources; the fitting of more lines and all sources are provided in Appendix~\ref{appendix:full_fit} (Figs.~\ref{fig:fit_all_G19.01}--\ref{fig:fit_all_NGC6334-38}). Overall, the LTE models fit the observations well, except that in some bright sources, some strong lines (mostly \ce{CH3OH} lines, and one to two acetone lines composed of hyperfine transitions with high $A_\mathrm{ij}$) are optically thick and overestimated by the models.



\subsection{Emission morphology}\label{sect:result_map}
\begin{figure*}[!h]
    \centering
    \includegraphics[width=\textwidth]{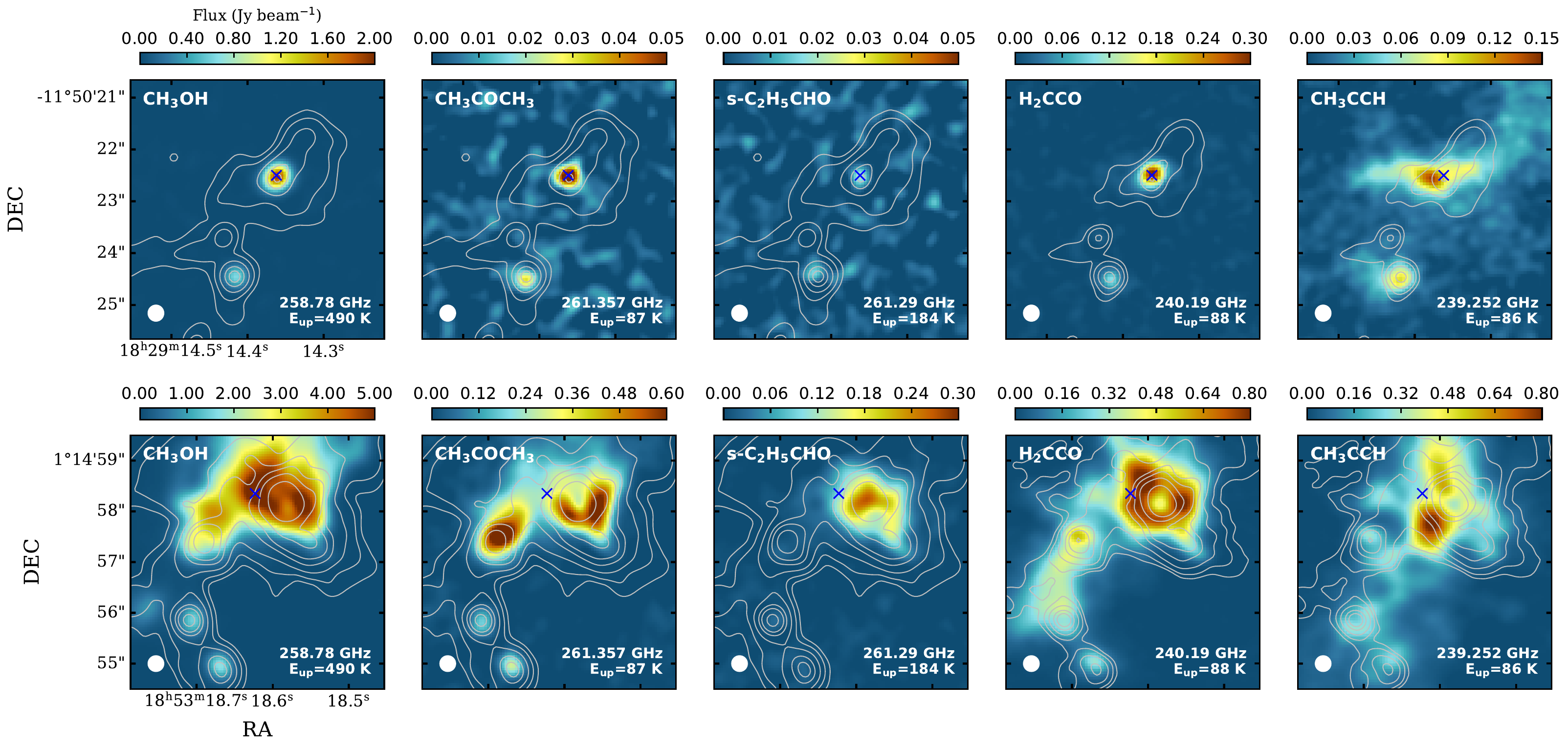}
    \caption{Moment 0 maps of the methanol, acetone, propanal, ketene, and propyne emission in G19.88-0.53 and G34.30+0.20. In each panel, the rest frequency and the upper energy level of the corresponding line is annotated in the bottom right corner, and the beam size is indicated in the bottom left corner. Contours in gray denote the 3, 5, 10, and 20$\sigma$ levels of the continuum emission at 258.465~GHz ($\sim$1.16~mm), where $\sigma=0.03$~Jy~beam$^{-1}$. The continuum maps are displayed in Fig.~1 of C23. The location where the ALMA spectrum was extracted is indicated by a blue cross in each panel.}
    \label{fig:mom0_maps}
\end{figure*}

Figure~\ref{fig:mom0_maps} shows the integrated intensity maps of representative emission lines of methanol, acetone, propanal, ketene, and propyne. The selection criteria of these emission lines are: they should be unblended within their FWHM, and also bright enough to depict the emitting regions of the molecules. For methanol, we chose the same line as used in C23, which has $E_\mathrm{up}=490.59$~K and $A_\mathrm{ij}=9.01\times10^{-5}$~s$^{-1}$. Even though this line has an $E_\mathrm{up}$ that is considerably higher than the typical $T_\mathrm{ex}$ of hot cores (100--200 K), and many more methanol lines were included after adding the lower tuning, it is still the brightest among all the unblended methanol lines (our data only cover methanol lines with $E_\mathrm{up}<90$~K and $>320$~K).

Two sources are taken as examples: G19.88-0.53 with compact (i.e., not spatially resolved) methanol emission, and G34.30+0.20 with the most extended (i.e., spatially resolved) methanol emission in the CoCCoA sample. Most sources in the CoCCoA sample are more similar to G19.88-0.53 in that the methanol emission is not or only partially resolved by the beam of 0.33$\arcsec$.

In G19.88-0.53, the emission maps of acetone, propanal, and ketene show a similar compact morphology to methanol and other O-COMs around the continuum peak, indicating that they all come from the hot core region (see Fig.~A.1 in C23 for the moment 0 maps of other O-COMs). As for propyne, the emission peak is slightly offset from the continuum peak, and the morphology shows an east-west elongation in G19.88-0.53. It seems that propyne is not tracing the hot core but some outflow-like structures. This is verified by previous observational studies that this source has a bipolar outflow in the east–west direction revealed by \ce{H2} and CO (isotopolog) emission \citep{Varricatt2010, Issac2020}. The emission morphology of propyne is also similar to that of \ce{C2H} shown in \cite{Saha2022}, suggesting that they are probably tracing the outflow cavity walls where the UV intensity is higher \citep{Tychoniec2021} and the abundances of hydrocarbons are enhanced. Propyne is also possibly present in the warm envelope, which is colder ($<$100~K) and more extended than the hot core. The supposed origins of propyne emission are also supported by its physical properties ($T_\mathrm{ex}$, FWHM, and $v_\mathrm{lsr}$; see discussion in Sect.~\ref{sect:results_N_T}).

In G34.30+0.20, there are four continuum peaks and the ALMA spectrum was extracted toward the brightest one on the upper right (the other three are much weaker on the left side). The hot core region is also around the continuum peak, but much more extended than that of G19.88-0.53. The molecular emission is attenuated by dust at the brightest continuum peak and shows an annular feature, which is particularly evident in the emission map of ketene. The emission of acetone, however, is asymmetric and brighter on the south side of the annulus. We noticed that the emission peak of methanol (also where the ALMA spectrum was extracted in C23) is actually different from that of acetone emission. This may cause underestimation of the column density of acetone and its ratios with respect to other species, (e.g., methanol), which will be discussed in Sect.~\ref{sect:discussion}. However, this is not the case in other sources where the hot regions are not well spatially resolved and the emission of O-COMs all comes from a similar compact region. We think G34.30+0.20 is an interesting source for investigating the spatial distribution of COM emission, which we decided to leave for future studies.  

Similar to the case of G19.88-0.53, the propyne emission in G34.30+0.20 again does not follow the morphology of the other species and shows a bipolar feature instead of an annulus shape.
According to the emission maps of these two sources, we consider ketene to be co-distributed with O-COMs in the hot core regions, while propyne tends to trace different physical components in the protostellar systems.

\subsection{Physical properties}\label{sect:results_N_T}
Table~\ref{tab:fit_results_acetone} lists the derived $N$, $T_\mathrm{ex}$, line width (FWHM), and $v_\mathrm{lsr}$ of acetone along with the ranges of these parameters obtained from other O-COMs in C23. The same results for propanal, ketene, and propyne are provided in Table~\ref{tab:fit_results_other_sp}.
In most sources, acetone has comparable $T_\mathrm{ex}$, FWHM, and $v_\mathrm{lsr}$ to other six O-COMs, suggesting similar origins from hot cores. 

For ketene, as described in Sect.~\ref{sect:method_CH2CO}, the only two covered lines are blended with strong \ce{CH2DOH} and \ce{NH2CHO} lines. As a result, we could only fit $N$ at a fixed $T_\mathrm{ex}$. The FWHM and $v_\mathrm{lsr}$ were estimated normally, but less accurately than for other species (the determination of FWHM also affects the best-fit $N$). The $v_\mathrm{lsr}$ of ketene agrees well with O-COMs, and the FWHM is comparable to or slightly larger, suggesting an origin from hot cores.

For $s$-propanal, we constrained the upper limits of its column densities at a fixed $T_\mathrm{ex}$ of 150~K by fitting the strongest line at 261.29~GHz, as shown in Fig.~\ref{fig:spropanal_fit} ($T_\mathrm{ex}$ could not be constrained due to very limited detected lines). For some sources such as G345.5+1.5, G19.88-0.53, NGC~6334-38, G34.30+0.20, and G23.21-0.37, the profile of this line can still be recognized, but for others it cannot, and the constraints had to depend on other weaker lines (see Fig.~\ref{fig:fit_all_G19.01}--\ref{fig:fit_all_NGC6334-38}). Although a full constraint of $N$ and $T_\mathrm{ex}$ cannot be well achieved with the frequency coverage of our data, the upper limits are reliable at the order-of-magnitude level, or even better that a factor of two for those sources with a distinguishable line profile at 261.29~GHz. 

For propyne, the physical properties were generally well constrained. The $T_\mathrm{ex}$ of propyne is overall consistent with those of O-COMs (usually 100--200~K), except in some sources (e.g., IRAS~18151 and NGC~6334-38) lower than 100~K. The FWHM of propyne is generally comparable to that of O-COMs (except for propanal), with some exceptions where the propyne lines are wider or narrower, but this may due to the merge or split of different velocity components. For instance, in G345.5+1.5, G35.05+0.35, and G35.20-0.74N, only propyne shows more than one velocity component, and each of them has smaller FWHM. The match and mismatch in FWHM can correspond to origins from warm envelopes and outflow cavity walls, respectively. The $v_\mathrm{lsr}$ of propyne lines in the main velocity component tend to slightly deviate from the O-COM range; in most cases the difference is small ($\sim$0.5~km~s$^{-1}$), but sometimes can be as large as 1.5--2.0~km~s$^{-1}$, and even larger in the secondary or third velocity components if there is one. The offset in $v_\mathrm{lsr}$ may suggest a different emitting region propyne from O-COMs, which can be displayed more straightforwardly in the moment 0 maps.

\begin{figure*}[!h]
  \sidecaption
  \includegraphics[width=12cm]{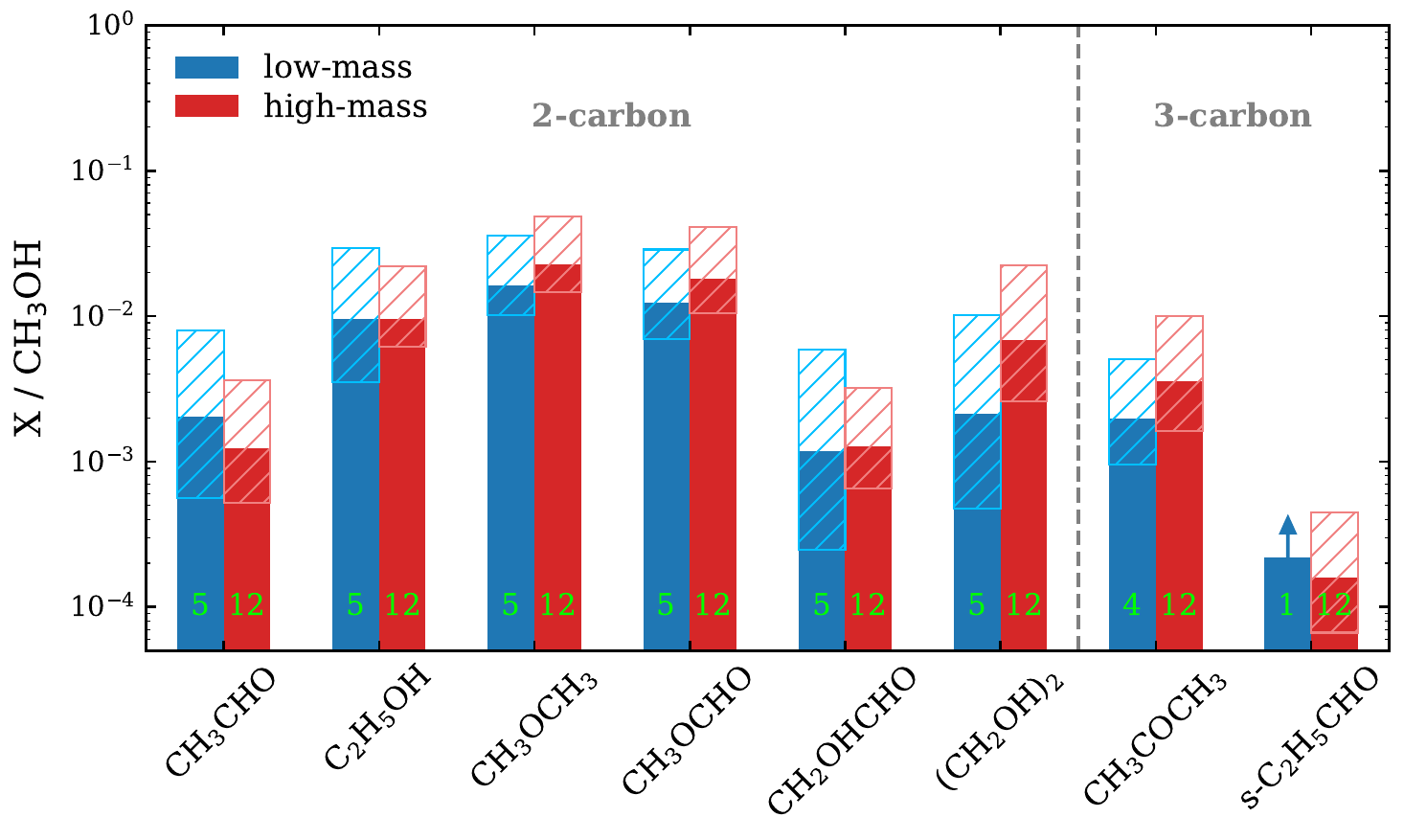}
     \caption{Column density ratios of six two-carbon and two three-carbon O-COMs with respect to methanol in low-mass (blue) and high-mass (red) protostars. The hatched regions indicate the uncertainties. The sample size (i.e., the number of sources) is labeled at the bottom of each bar in green. The column density of $s$-propanal was only measured in one low-mass source (IRAS~16293-2243 B), and its methanol column density was not measured at the same location as propanal but closer to the continuum peak, which generally has higher column densities; therefore the calculated ratio between propanal and methanol of this source is considered as lower limit.}
     \label{fig:OCOM_ratios_bar}
\end{figure*}

\section{Discussion}\label{sect:discussion}
\subsection{Acetone and other O-COMs}\label{sect:discuss_acetone_vs_OCOM}
With the column densities of two three-carbon O-COMs (i.e., acetone and propanal) measured for the CoCCoA sample, we can now compare their relative abundances with other two-carbon O-COMs that have been studied in C23. Figure~\ref{fig:OCOM_ratios_bar} summarizes the column density ratios of these eight O-COMs with respect to methanol. The column densities of the six two-carbon O-COMs in the CoCCoA sample were originally taken from C23, but their ratios with respect to methanol were recalculated using the updated methanol column densities (see Sect.~\ref{sect:method_CH3OH}). Here the high-mass sample size is reduced from 14 in C23 to 12, since the new measurements of methanol were only applied to the 12 sources with concatenated spectra (Sect.~\ref{sect:observations}). We also do not add other literature values in Fig.~\ref{fig:OCOM_ratios_bar} to ensure that the column densities are uniformly measured from the same high-quality dataset.
The low-mass sample remains the same as in C23, including B1-c, B1-bS, S68N, IRAS~16293~A and B \citep{vanGelder2020, Manigand2020, Jorgensen2018}. For acetone, there is no measurements toward IRAS~16293~A, and propanal was only quantified in IRAS~16293~B, so the low-mass sample size is smaller for these two species.

For the first time we show the column density ratios of eight O-COMs (or seven if excluding the less constrained propanal) with respect to methanol in a uniform sample of a dozen sources. On one hand, as a three-carbon O-COM, acetone is not much less abundant than the two-carbon ones (within one order of magnitude), and even tends to be more abundant than \ce{CH3CHO} and \ce{CH2OHCHO}.
On the other hand, propanal, as an isomer of acetone, is significantly less abundant than other O-COMs. Actually, species with aldehyde groups (i.e., \ce{CH3CHO}, \ce{CH2OHCHO}, and \ce{C2H5CHO}) are the least abundant among the COMs with the same amount of carbon atoms. 
There are two possible explanations: aldehydes are easily destroyed or converted into other O-COMs in the gas phase, or aldehydes are less produced than other-type O-COMs. However, it is not practical to prove these two cases only by observations; insights from theoretical and experimental studies are needed. 

Besides the low abundances of aldehydes, it is also noticeable that alcohols are not the most abundant species among two-carbon O-COMs. Instead, those with \ce{CH3O} radicals (\ce{CH3OCH3} and \ce{CH3OCHO}) are even more abundant than alcohols (\ce{C2H5OH} and \ce{(CH2OH)2}). A similar trend may also hold for three-carbon O-COMs, where acetone can be more abundant than propanol (\ce{C3H7OH}), or at least comparably abundant. Propanol has only been detected toward two high-mass star-forming regions G+0.693-0.027 and Sgr~B2(N) \citep{Jimenez-Serra2022, Belloche2022}, and its column densities can not be reliably constrained using our data. In Sgr~B2(N), the column density ratio of propanol with respect to methanol is 10$^{-3}$--10$^{-2}$, which is comparable to the acetone ratios in our sample. 
While acetone (\ce{CH3COCH3}) is not a \ce{CH3O}-bearing species, it has a similar structure to \ce{CH3OCH3} in that there are two \ce{CH3} on the sides and an oxygen atom in between. According to the available measurements, it seems that O-COMs with \ce{CH3O} or similar components are systematically more abundant in the gas phase.

However, the high abundances of \ce{CH3OCH3} and \ce{CH3OCHO} are difficult to reproduce in simulations and experiments. Instead, \ce{CH3OCH3} and \ce{CH3OCHO} were often underpredicted in simulations before the introduction of the barrierless reaction 
\begin{equation}\label{eq:CH2}
    \ce{C}\,+\,\ce{H2}\,\rightarrow\,\ce{CH2}
\end{equation}
and the nondiffusive grain-surface and ice-mantle chemistry in \cite{Garrod2022}, although there are different opinions on whether reaction \ref{eq:CH2} is barrierless or not \citep{Krasnokutski2016, Henning2019, Lamberts2022, Potapov2024}. In many experimental studies on solid-phase formation of O-COMs under non-energetic \citep[e.g.,][]{Fedoseev2015, Fedoseev2022, Chuang2020} or energetic \citep{Chuang2017, Chuang2021} conditions, \ce{CH3OCH3} is the only two-carbon O-COM that was not reported in the products, and \ce{CH3OCHO} is often less produced than its isomer \ce{CH2OHCHO}. Although \ce{CH3OCH3} has a viable gas-phase production route through the reaction of methanol with protonated methanol, its constant ratio of $\sim$1 with respect to \ce{CH3OCHO} across a large number of star-forming regions suggest non-negligible formation in ices \citep{Coletta2020}. The production of \ce{CH3OCH3} was only seen in the experiment of UV-irradiated \ce{CH3OH}:\ce{CH4}/CO ice mixtures by \cite{Oberg2009}. In comparison, all the experiments with non-detection of \ce{CH3OCH3} did not include \ce{CH4} in their initial ingredients; instead, they used C atoms, CO, \ce{C2H2}, and \ce{CH3OH}. 
It seems that irradiating \ce{CH4} is necessary to produce \ce{CH3} radicals and boost the formation of \ce{CH3OCH3} in experiments. A question then comes up why \ce{CH3} radicals are hard to generate experimentally, but the relevant molecules, represented by \ce{CH3OCH3}, are observed to be considerably abundant in space. 
This question is well-known among experimentalists, and becomes more evident after the revelation of the systematic high-abundance of acetone in this work.




\subsection{Acetone: ice versus gas}\label{sect:discuss_gas_vs_ice}
The recent detection of COM ices with JWST observations in two low-mass protostars (IRAS~2A and B1-c) enables the gas-to-ice comparisons for O-COMs including \ce{CH3CHO}, \ce{C2H5OH}, \ce{CH3OCH3}, and \ce{CH3OCHO} \citep{Rocha2024, Chen2024}. By conducting gas-to-ice comparisons in their abundances we can probe into their chemical evolution during the phase transition from ice in cold envelopes to gas in hot cores. 

In addition to these two-carbon O-COMs, acetone ice is also considered as detected in B1-c and tentatively detected in IRAS~2A as the only known candidate for the observed absorption feature at 7.3~$\mu$m \citep[detailed discussion on this point can be found at Sect.~4.2.3 of][]{Chen2024}. This conclusion may be modified in the future if new candidates are found to fit the observations better, but the currently measured ice abundances of acetone can at least serve as upper limits.

\begin{figure}[!h]
    \centering
    \includegraphics[width=0.5\textwidth]{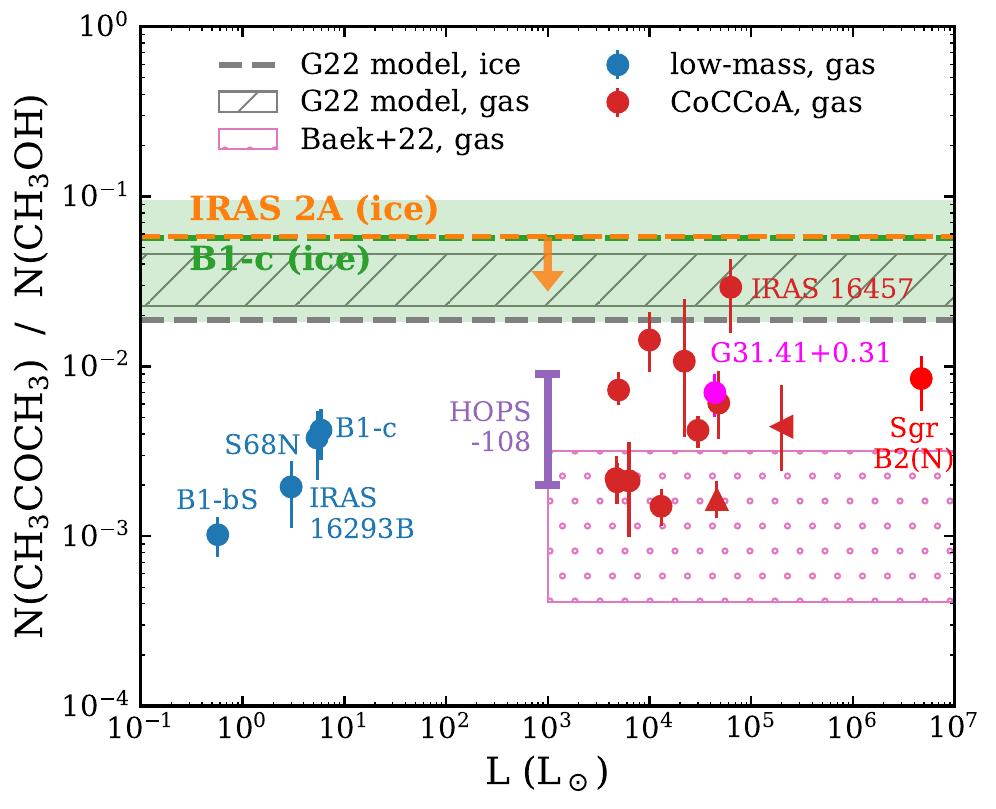}
    \caption{Column density ratios between acetone and methanol. The gas-phase ratios of low-, intermediate-, and high-mass sources are shown in blue, purple, and red points, respectively. The upward- and leftward-pointing triangles correspond to G34.30+0.20 (lower limit of the acetone-to-methanol ratio, see the last paragraph of Sect.~\ref{sect:discuss_gas_vs_ice}) and NGC~6334-38 (upper limit of luminosity, see Table~\ref{tab:fit_results_acetone}), respectively. The ice ratios two low-mass sources, B1-c and IRAS~2A, are indicated by dashed lines in green and orange, respectively. Since acetone ice is only considered as tentatively detected in IRAS 2A, the uncertainty is indicated by a green-shaded region only for B1-c. For the high-mass sample of \cite{Baek2022}, only a range is displayed since no luminosity information is provided. The dashed line and hatched region in gray indicate the modeling results of the \texttt{MAGICKAL} astrochemical simulations in \cite{Belloche2022}; the model is originally introduced in \cite{Garrod2022}, hence abbreviated as G22 in the legend.}
    \label{fig:AT_gas_vs_ice}
\end{figure}

Since the quantification of acetone ice has only been available in two sources, direct gas-to-ice comparisons in the same sources are still limited. However, with gas-phase acetone quantified in a larger sample of hot cores, we now have better constraints on its gas-phase abundances, making the gas-to-ice comparisons more reliable. The observed column density ratios between acetone and methanol in gas and ice are summarized in Fig.~\ref{fig:AT_gas_vs_ice}. For the gas-phase ratios, in addition to the 12 high-mass CoCCoA sources, we added literature data of four low-mass \citep[B1-c, B1-bS, S68N, and IRAS~16293B;][]{vanGelder2020, Jorgensen2018, Nazari2024_16293}, one intermediate \citep[HOPS-108;][]{Chahine2022}, and two additional high-mass sources \citep[G31.41+0.31 and Sgr~B2(N);][]{Mininni2023, Belloche2013}. We also considered the work of \cite{Baek2022} (hereafter B22) who report acetone column densities in 13 cores from eight high-mass protostellar sources. However, their methanol column densities $N$(\ce{CH3OH}) were likely measured from optically thick lines, and the reported $N$(\ce{CH3OH}) values are obviously underestimated (the ratios between the main and the minor isotopologs are significantly off from the isotope ratios). Since they also provide $N$(CH$_3^{18}$OH) and the source distance, more reliable values of $N$(\ce{CH3OH}) can be calculated by multiplying $N$(CH$_3^{18}$OH) and $R$($^{18}$O) given in eq.~\ref{eq:R_18O}. Nevertheless, in B22 the reported $N$(CH$_3^{18}$OH) are higher than $N$(\ce{^{13}CH3OH}), which is not normal since $R$(\ce{^{13}C}) is higher than $R$(\ce{^{18}O}), and $N$(\ce{^{13}CH3OH}) should be higher than $N$(CH$_3^{18}$OH). A possible explanation is that their $N$(CH$_3^{18}$OH) were overestimated for some reasons, for instance, the line blending issue was correctly handled. As a result, our re-calculated $N$(\ce{CH3OH}) from $N$(CH$_3^{18}$OH) may be overestimated, and the acetone-to-methanol ratios plotted in Fig.~\ref{fig:AT_gas_vs_ice} may be underestimated for the B22 sample.

Figure~\ref{fig:AT_gas_vs_ice} shows that the gas-phase abundances of acetone with respect to methanol are scattered between 10$^{-3}$ and 10$^{-2}$. In the high-mass sample, the CoCCoA sources are consistent with the three additional sources HOPS-108, G31.41+0.31, and Sgr~B2(N). The one lower limit in the CoCCoA sample is G34.30+0.20, where the spectrum was not extracted from an acetone-rich position, so that the acetone-to-methanol ratio could be underestimated. The B22 sources show generally lower ratios, but is likely because the methanol column densities are overestimated. The low-mass samples show slightly lower acetone abundance than the high-mass one, which is different from other O-COMs that show no trend with luminosity (see Fig.~4 in C23).
The ice abundance is about one order of magnitude higher than the mean gas abundance, following the same trend as \ce{CH3CHO} and \ce{C2H5OH} \citep[see Fig.~10 in ][]{Chen2024}. However, it should be taken into account that the sample size of ice ratios is very small (only two), and the quantification of ice abundances is much more uncertain than that of gaseous ones, as the current fitting results of COM ices may by changed in the future by possibly new identifications in the fingerprint range of 6.8--8.8 $\mu$m.
Assuming the ice abundances of acetone were reliably constrained, the higher ice ratios hint at gas-phase reprocessing after sublimation, which reduces the amount of acetone relative to methanol, but the detailed mechanisms still need to be investigated.

\subsection{Chemical evolution of acetone}\label{sect:discuss_AT}
To gain more insights into the formation and/or destruction mechanisms of acetone, we compared our observational results with the simulation results of the state-of-the-art astrochemical model \texttt{MAGICKAL} \citep{Garrod2013, Garrod2022}. We also searched for other theoretical and experimental studies that are relevant to acetone and a summary chemical network is shown in Fig.~\ref{fig:schematic_AT}.

\subsubsection{Insights from astrochemical simulations}\label{sect:discuss_simulation}
The \texttt{MAGICKAL} models simulate the chemistry of interstellar molecules in three phases (gas phase, ice surface, and bulk ice mantle) during two stages (cold collapse stage and warm-up stage) of the formation of protostellar systems. Three timescales are considered for the warm-up stage: fast ($5\times10^4$~yr, medium ($2\times10^5$~yr), and slow ($1\times10^6$~yr).

\begin{figure}[!h]
    \centering
    \includegraphics[width=\linewidth]{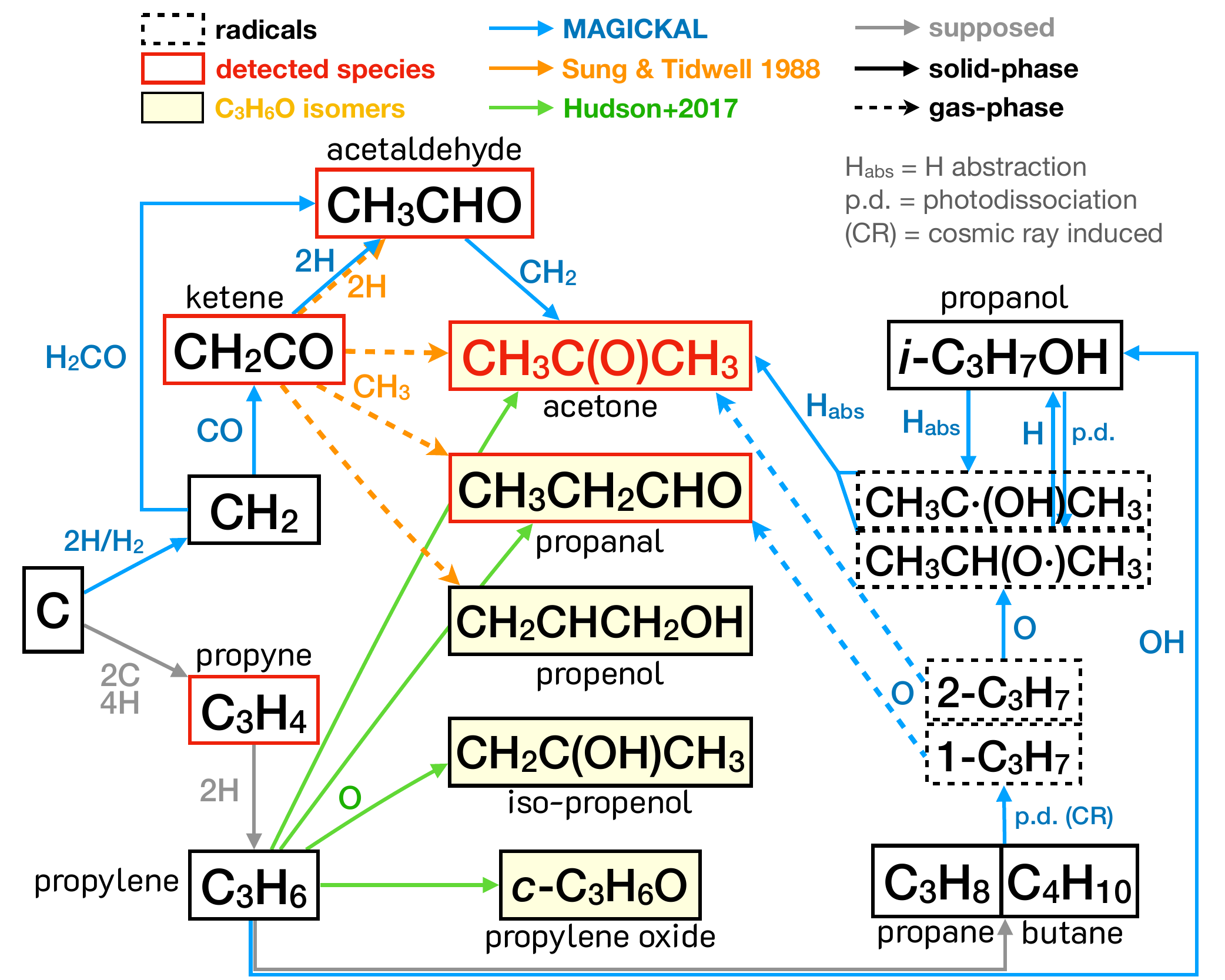}
    \caption{Schematics of the formation pathways of acetone, proposed by previous theoretical calculations \citep{Sung1988}, astrochemical simulations \citep[\texttt{MAGICKAL};][]{Garrod2022}, and experiments \citep{Hudson2017}. Except the first study that assumes normal temperature and pressure, the remaining two are intended for astrochemical conditions (i.e., low temperature and pressure). Legends are provided in the figure.}
    \label{fig:schematic_AT}
\end{figure}

Here we used an updated version of \texttt{MAGICKAL} which was applied in \cite{Belloche2022} for observational studies of propanol in Sgr~B2(N). The chemical network of this model involves new gas-phase and grain-surface chemistry for propanol and related species. Although \cite{Belloche2022} only detail the changes related specifically to propanol, additional reactions are also included for acetone and propanal. As illustrated by the blue routes in Fig.~\ref{fig:schematic_AT}, acetone is formed via two solid-phase and one gas-phase pathways. One solid-phase pathway is bottom-up from simpler species:
\begin{equation}\label{eq:AT_ice_CH2CO_CH3CHO}
    \ce{C} \xrightarrow{\ce{H,H2}} \ce{CH2} \xrightarrow{\ce{CO}} \ce{CH2CO} \xrightarrow{2\ce{H}} \ce{CH3CHO} \xrightarrow{\ce{CH2}} \ce{CH3COCH3},
\end{equation}
and the other is via H abstraction of two iso-\ce{C3H7O} radicals, \ce{CH3CH(\dot{\rm O})CH3} and \ce{CH3\dot{\rm C}(OH)CH3}:
\begin{align}
    \ce{H} + \ce{CH3CH(\dot{\rm O})CH3} &\rightarrow \ce{CH3COCH3} + \ce{H2} \ \ \text{and}\label{eq:AT_ice_Habs_1}\\
    \ce{H} + \ce{CH3\dot{\rm C}(OH)CH3} &\rightarrow \ce{CH3COCH3} + \ce{H2}\label{eq:AT_ice_Habs_2},
\end{align}
where the iso-\ce{C3H7O} radicals are produced from iso-propanol ($i$-\ce{C3H7OH}) and \ce{2-C3H7} radicals:
\begin{align}
    \ce{2-C3H7}+\ce{O} &\rightarrow \ce{CH3CH(\dot{\rm O})CH3} \quad{\rm or}\quad\label{eq:2-C3H7}\\
    \ce{i-C3H7OH} + h\nu &\rightarrow \ce{CH3CH(\dot{\rm O})CH3} + \ce{H};\label{eq:iC3H7O_1}\\
    \ce{i-C3H7OH}+\ce{H} &\rightarrow \ce{CH3\dot{\rm C}(OH)CH3} + \ce{H2}\label{eq:iC3H7O_2}.
\end{align}
The gas-phase pathway is:
\begin{equation}\label{eq:AT_gas}
    \ce{2-C3H7} + \ce{O} \rightarrow \ce{CH3COCH3} + \ce{H},
\end{equation}
with rates based on \cite{Tsang1988}.
The same mechanism is available to form propanal via the 1-\ce{C3H7} radical. Similar grain-surface reactions as for acetone are also involved in the production of propanal in the ice, which has origins in break-down of $n$-propanol.

Figure~\ref{fig:G22model_AT} shows the rate of change of acetone abundances (summed over all phases) in the \texttt{MAGICKAL} simulations. Both the cold collapse and the warm-up stages have considerable amount of acetone formed, and the contribution from the gas-phase production via reaction~\ref{eq:AT_gas} increases with the warm-up timescale (i.e., the longer the warm-up stage, the more acetone is formed in the gas phase). In the version of \texttt{MAGICKAL} models introduced in \cite{Garrod2022}, only one solid-phase pathway (reaction \ref{eq:AT_ice_CH2CO_CH3CHO}) is included, and more than 90\% of acetone is formed in the solid phase during the cold collapse stage. Without the other pathways (reactions \ref{eq:AT_ice_Habs_1}--\ref{eq:AT_gas}), the peak acetone abundance would decrease by more than one order of magnitude. 
This difference result from the mediated energy barriers of H abstraction reactions (e.g., \ref{eq:AT_ice_Habs_1}, \ref{eq:AT_ice_Habs_2}, and \ref{eq:iC3H7O_2}), of which the energy barriers are poorly constrained by theoretical calculations and experiments, and many of them are set as zero. 
 
The acetone-to-methanol ratios predicted by \texttt{MAGICKAL} simulations are plotted along with the observational results in Fig.~\ref{fig:AT_gas_vs_ice}. The gas-phase ratios are higher than the ice ones due to the gas-phase production in the warm-up stage (reaction \ref{eq:AT_gas}). The model-predicted ratios are comparable to the observed ice ratios, but are higher than the observed gas ratios by about half order of magnitude. The overestimation by simulations hint at gas-phase destruction of acetone that is not fully considered in the current chemical models, which is also suggested by the gas-to-ice comparisons discussed in Sect.~\ref{sect:discuss_gas_vs_ice}.

\begin{figure*}[h!]
    \centering
    \begin{subfigure}{0.49\textwidth}
        \centering
        \includegraphics[width=
        \linewidth]{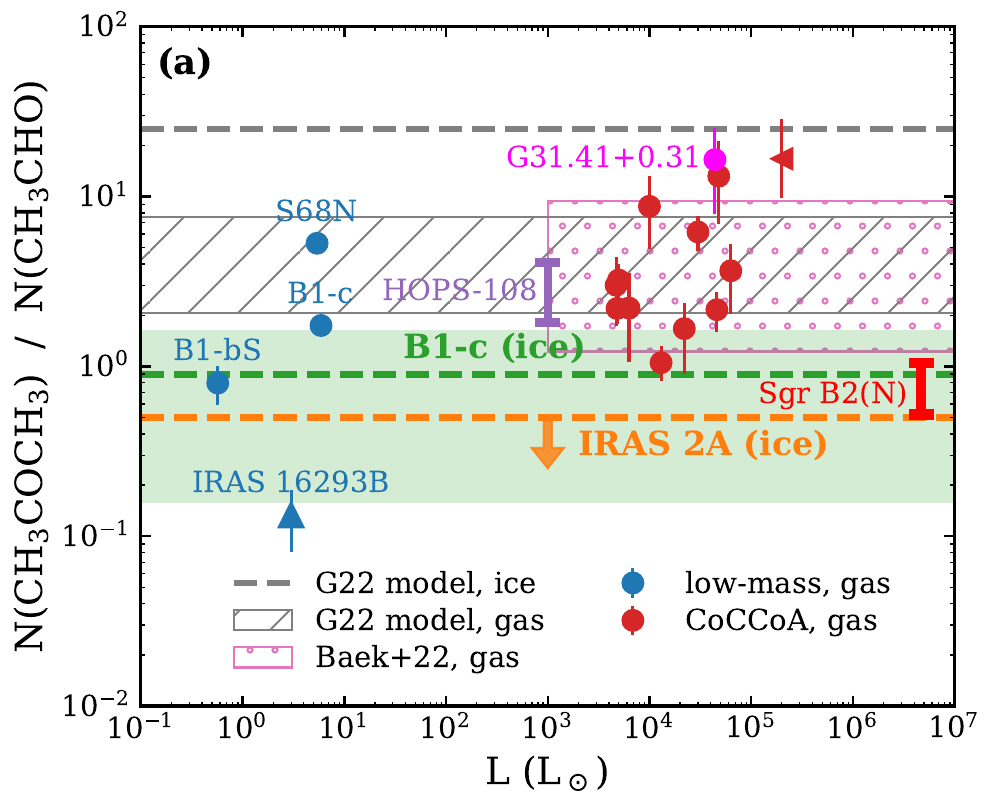}
    \end{subfigure}
    \hfill
    \begin{subfigure}{0.49\textwidth}
        \centering
        \includegraphics[width=\linewidth]{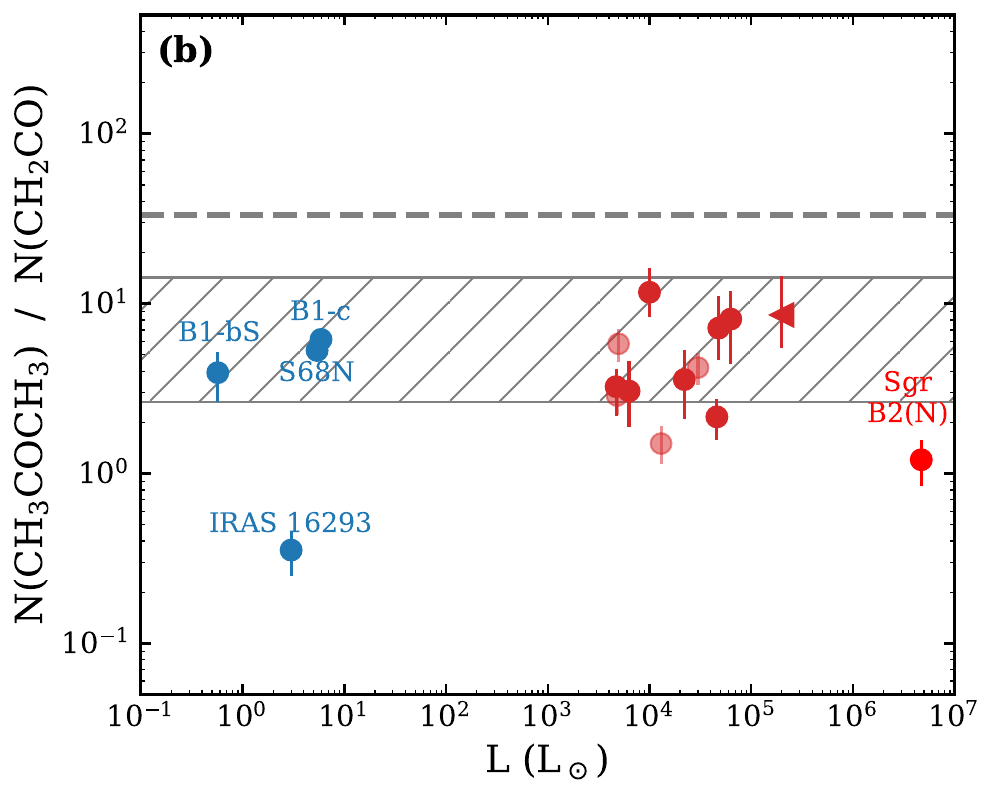}
    \end{subfigure}
    \vspace{0.2cm}
    \begin{subfigure}{0.49\textwidth}
        \centering
        \includegraphics[width=\linewidth]{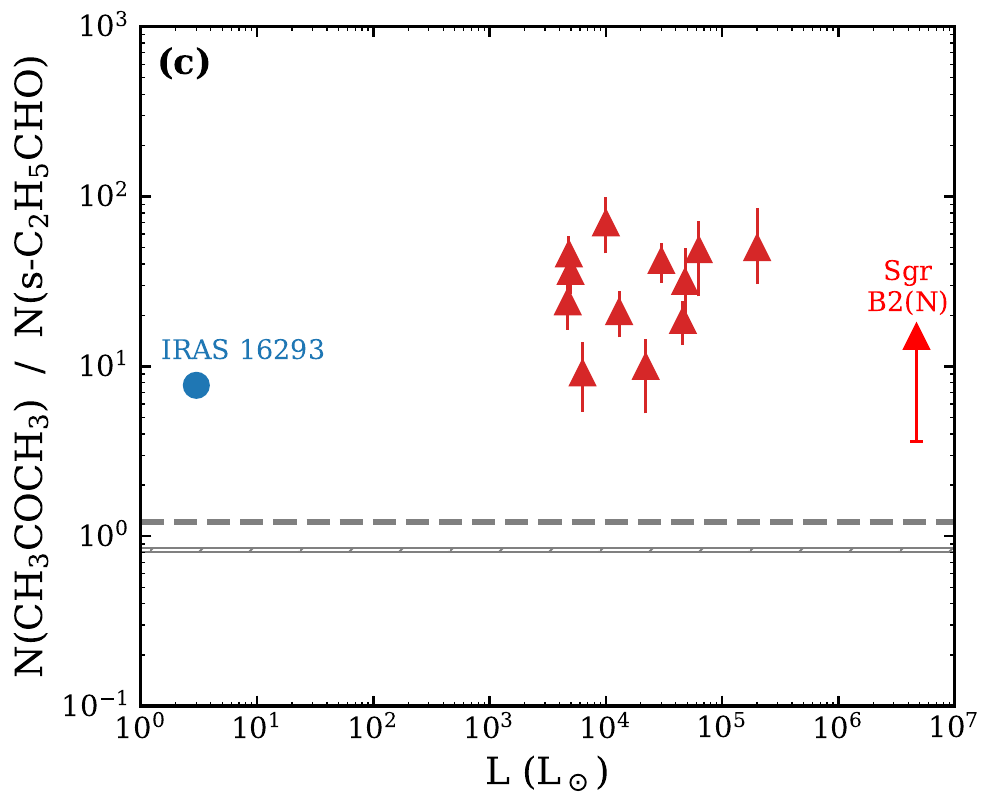}
    \end{subfigure}
    \hfill
    \begin{subfigure}{0.49\textwidth}
        \centering
        \includegraphics[width=\linewidth]{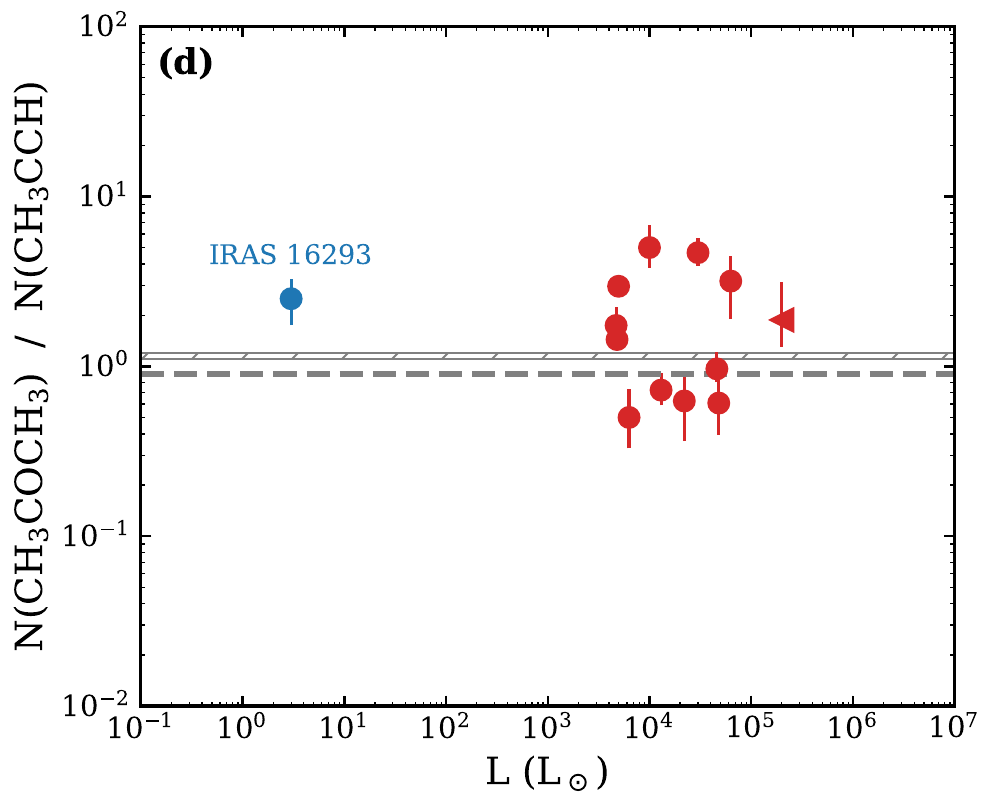}
    \end{subfigure}
    \caption{Column density ratios of acetone with respect to acetaldehyde (a), ketene (b), propanal (c), and propyne (d). 
    In panel (b), the data points in lighter red  refer to sources with less constrained column densities of ketene, of which the lines are blended with optically thick lines of \ce{CH2DOH} and \ce{NH2CHO} (see Sect.~\ref{sect:method_CH2CO}).}
    \label{fig:AT_ratios}
\end{figure*}

In the bottom-up formation pathway of acetone (\ref{eq:AT_ice_CH2CO_CH3CHO}), acetaldehyde is the direct precursor of acetone. We therefore checked the column density ratios between acetone and acetaldehyde, as shown in Fig.~\ref{fig:AT_ratios}a. The observed gas-phase ratios, especially in high-mass sources, are in general higher than the ice ratios, which means that acetaldehyde is relatively less abundant than acetone in the gas phase. This echos the possibility of aldehydes being destroyed or converted into other COMs in the gas phase (Sect.~\ref{sect:discuss_acetone_vs_OCOM}). Interestingly, the B22 sample shows high consistency with our CoCCoA sample in the acetone-to-acetaldehyde ratios, which implies that the lower acetone-to-methanol ratios are likely due to the overestimated methanol column densities (Sect.~\ref{sect:discuss_gas_vs_ice}).


The \texttt{MAGICKAL} simulations predict the gas-phase acetone-to-acetaldehyde ratios very well, while the ice ratios are overestimated significantly. This may due to the conflicting evolution histories of acetaldehyde suggested by simulations and observations. In the \texttt{MAGICKAL} simulations, acetaldehyde is abundantly formed in the gas phase, which can be seen in Fig.~\ref{fig:G22_n_t} that its abundance keeps increasing after sublimation. The main formation mechanism for gas-phase acetaldehyde is given by
\begin{equation}
    \ce{O} + \ce{C2H5} \rightarrow \ce{CH3CHO} + \ce{H}
\end{equation}
\citep{Tsang1986}. The production of \ce{C2H5}, and of the \ce{C3H7} radicals that form acetone in the gas phase, is dependent on the destruction of stable hydrocarbons such as propane (\ce{C3H8}), via cosmic-ray induced photodissociation. The rates of this process are poorly defined so far.
The predicted abundances of gas-phase acetaldehyde  in the \texttt{MAGICKAL} simulations are generally higher than in observations, in which the gas-phase ratios of acetaldehyde (with respect to methanol) are relatively lower than those of other O-COMs (Fig.~\ref{fig:OCOM_ratios_bar}) and also the ice ratios of itself \citep{Chen2024}; that is, there might be the opposite case that acetaldehyde is mainly formed in ices and then depleted in the gas phase. Taking the two discrepant evolution histories into account, it is possible that the overestimated ice ratios between acetone and acetaldehyde are due to the underproduction of acetaldehyde ice in the simulations, and the good match in the gas ratios could be just a coincidence. Although there is direct chemical link between acetone and acetaldehyde (i.e., the last step of pathway \ref{eq:AT_ice_CH2CO_CH3CHO}), this link is only considered in the solid phase, and both species are likely to undergo gas-phase reprocessing that seems not well simulated so far. However, we note that in high-mass protostellar systems, there might not be enough time for the gas-phase chemistry to significantly alter the abundances of these COMs if the reactions are not very effective.

The other formation pathways (i.e., reactions \ref{eq:AT_ice_Habs_1}--\ref{eq:AT_gas}) are difficult to examine from observations, since the relevant species are either radicals or large molecules that can hardly be quantified. We mentioned in Sect.~\ref{sect:method_C3H6O} that the column densities of iso-propanol, even in the form of upper limits, cannot be well constrained using the CoCCoA dataset. The efficiency of reactions \ref{eq:AT_ice_Habs_1}--\ref{eq:AT_gas}, especially the energy barriers, are also lack of constraints. However, the comparison between observations and simulations conveys an important message: the acetone abundances, along with its ratios with respect to other species, will be substantially underestimated if these reactions \ref{eq:AT_ice_Habs_1}--\ref{eq:AT_gas} are not included. This suggests that there must be more routes to form acetone than \ref{eq:AT_ice_CH2CO_CH3CHO}; we just have not fully understood them yet.

\subsubsection{Insights from computational chemistry}
Besides the aforementioned pathways considered in the \texttt{MAGICKAL} simulations, an early theoretical study on ketene \citep{Sung1988} suggested that acetone may form directly via ketene adding \ce{CH3} radicals (the orange routes in Fig.~\ref{fig:schematic_AT}). Since their calculations were performed assuming normal temperature and pressure (i.e., in the gas phase), their conclusions may not fully applicable to astrochemical situations. However, some parameters in the results are irrelevant to physical conditions, and the suggested pathways can still be inspiring to future theoretical and experimental studies.

According to their calculations, reactions between ketene and \ce{CH3} radicals can produce acetone and propanal with similar branching ratios; Propenol is also one of the possible products but is less favored:
\begin{equation}\label{eq:CH2CO+CH3}
    \ce{CH2CO} + \ce{CH3} \xrightarrow{+\ce{H}} \ce{CH3COCH3} / \ce{C2H5CHO} / \ce{CH2CHCH2OH}.
\end{equation}
We therefore checked the column density ratios between acetone and ketene in the gas phase, which are shown in Fig.~\ref{fig:AT_ratios}b. The observed ratios are moderately scattered within 1--10, and are comparable between the low-mass and the high-mass sources, except the outlier IRAS~16293 where acetone is less abundant than ketene. One possibility would be that the emission of acetone and ketene is inhomogenous in IRAS~16293 as what we see in G34.30+0.20, and the ALMA spectrum of IRAS~16293 was extracted from a biased location where ketene emission is very bright. Otherwise, there should be something interesting happening to the chemistry in this source. 

The observed gas-phase ratios between acetone and ketene are well reproduced by \texttt{MAGICKAL}, even though the only link between acetone and ketene in the model (i.e., pathway \ref{eq:AT_ice_CH2CO_CH3CHO}) is in the solid phase. According to the calculation results of \cite{Sung1988}, the energy barrier of reaction \ref{eq:CH2CO+CH3} is not low ($\sim$4000~K). If this is the case, the contribution of this formation pathway would be very limited even if it is included in the chemical network of \texttt{MAGICKAL}. Nevertheless, more reliable constraints on the efficiency as well as the branching ratios of reaction \ref{eq:CH2CO+CH3} should be set by experiments.

The simulated ice ratios between acetone and ketene are higher than the observed gas ratios by a few factors. Unfortunately, we are unable to measure the ice abundances of ketene by observations since no IR spectrum of ketene ice is available. Ketene is a very active species under normal conditions and therefore not commercially available for experiments. This also explains why ketene has not been used as an initial ingredient to form O-COMs in the laboratories. 


Since propanal is suggested to have a similar branching ratio to acetone in reaction~\ref{eq:CH2CO+CH3}, we also checked the column density ratios between acetone and propanal, which are shown in Fig.~\ref{fig:AT_ratios}c. These ratios are considered as lower limits, as only upper limits of $N$($s$-\ce{C2H5CHO}) were constrained. Propanal was only quantified in one low-mass source IRAS~16293~B. The observations show that acetone is more abundant than propanal in the gas phase by at least one order of magnitude. If acetone is mainly formed via reaction~\ref{eq:CH2CO+CH3}, the ratio should be around one. 
The observed high ratios between acetone and propanal are also significantly underestimated by \texttt{MAGICKAL}. The \texttt{MAGICKAL} simulations tend to predict comparable or slightly higher amount of propanal than acetone, since many aforementioned formation pathways of acetone also work for propanal (e.g., if replacing 2-\ce{C3H7} with 1-\ce{C3H7} in reactions \ref{eq:2-C3H7} and \ref{eq:AT_gas}). The 1- and 2-\ce{C3H7} radicals are mainly produced from the break-down of normal- and iso-propanol, and the efficiency of the relevant reactions is likely to affect the relative abundances between propanal and acetone. 

So far, all versions of \texttt{MAGICKAL} simulations \citep{Garrod2013, Garrod2022, Belloche2022} have not been able to reproduce the high acetone-to-propanal ratios in observations. This discrepancy implies that there may be some gas-phase destruction mechanisms for propanal (or more broadly speaking, aldehydes) that we are either unaware of or not paying adequate attention to. The destruction of aldehydes should be much more efficient than that of other O-COMs. However, it is hard to pinpoint the exact mechanisms only from observations; more investigation are needed by theorists and experimentalists.



\subsubsection{Insights from experiments}\label{sect:discuss_experiment}
Similar to astrochemical simulations, there is a lack of experimental studies that directly focus on acetone formation, though there are ones on large COMs such as propanol and glycine \citep{Qasim2019, Ioppolo2021}. In this case, we resort to an experimental study on propylene oxide, an isomer of acetone, in which the discussion involves possible formation pathways of acetone \citep{Hudson2017, Bergantini2018}. They propose that propylene oxide ice can be formed through epoxidation of propylene (\ce{C3H6}) by oxygen atoms yielded from photodissociation of \ce{CO2}:
\begin{equation}\label{eq:C3H6+O}
    \ce{C3H6} + \ce{O} \rightarrow \ce{C3H6O}.
\end{equation}
When the oxygen atom breaks the double bond in propylene molecules, it is possible to form four \ce{C3H6O} isomers: propylene oxide, acetone, propanal, and iso-propenol (green routes in Fig.~\ref{fig:schematic_AT}). The branching ratios of this reaction were not measured in \cite{Hudson2017}, but were provided by \cite{Bergantini2018}, which conducted similar experiments (i.e., irradiating ice mixtures of \ce{C3H6} and \ce{CO2} under low temperature, but with different irradiating conditions), and also performed quantum chemical calculations. In \cite{Bergantini2018}, iso-propenol was not observed in the products, and the branching ratios between propylene oxide, acetone, and propanal derived from experiments are (92$\pm$4):(23$\pm$7):1.

The relation between acetone and propanal is shown in Fig.~\ref{fig:AT_ratios}c and has been discussed in Sect.~\ref{sect:discuss_simulation}. It is worth noting that the branching ratio between acetone and propanal derived in \cite{Bergantini2018} is in good agreement with the observed ratios of the high-mass CoCCoA sample. Unfortunately, propylene oxide is not covered in our data, preventing further validation of reaction \ref{eq:C3H6+O} by observations.
Also, our data do not cover propylene lines, but do include propyne lines. We therefore checked the relation between acetone and propyne, a hydrogenation precursor of propylene. Figure~\ref{fig:AT_ratios}d shows the column density ratios between acetone and propyne. There is only one low-mass source (IRAS~16293~B) that has column density measurements of both acetone and propyne \citep{Calcutt2019}, and no ice detection has been reported. 
The observed gas-phase ratios are moderately scattered by about one order of magnitude, and seem to have no trend with luminosity. The acetone-to-propyne ratios predicted by \texttt{MAGICKAL} simulations in both gas and ice match the observations very well.


\begin{figure*}[!h]
    \centering
    \begin{subfigure}[c]{0.54\textwidth}
        \centering
        \includegraphics[width=\textwidth]{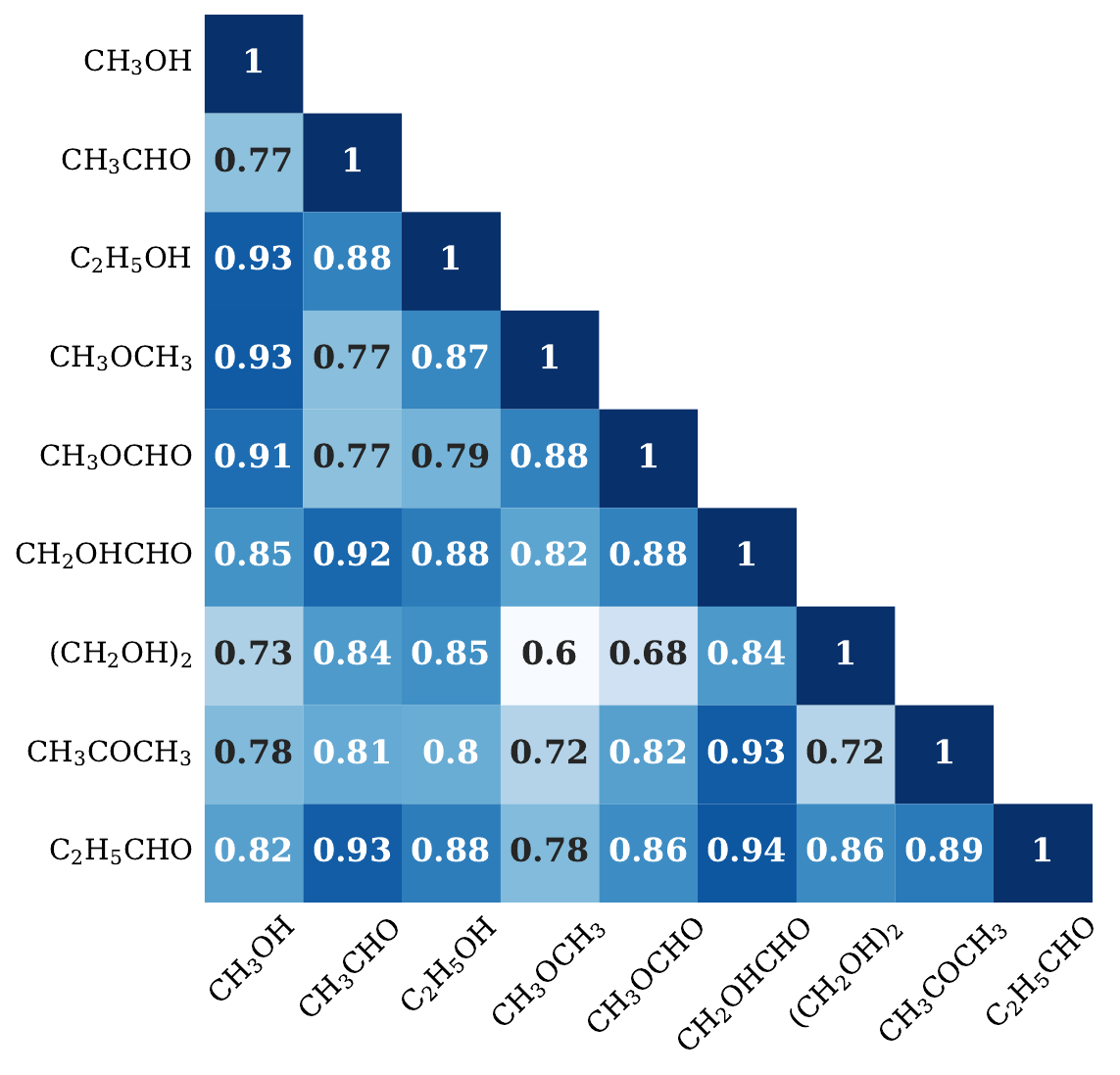}
    \end{subfigure}
    \hfill
    \begin{subfigure}[c]{0.45\textwidth}
        \centering
        \includegraphics[width=\textwidth]{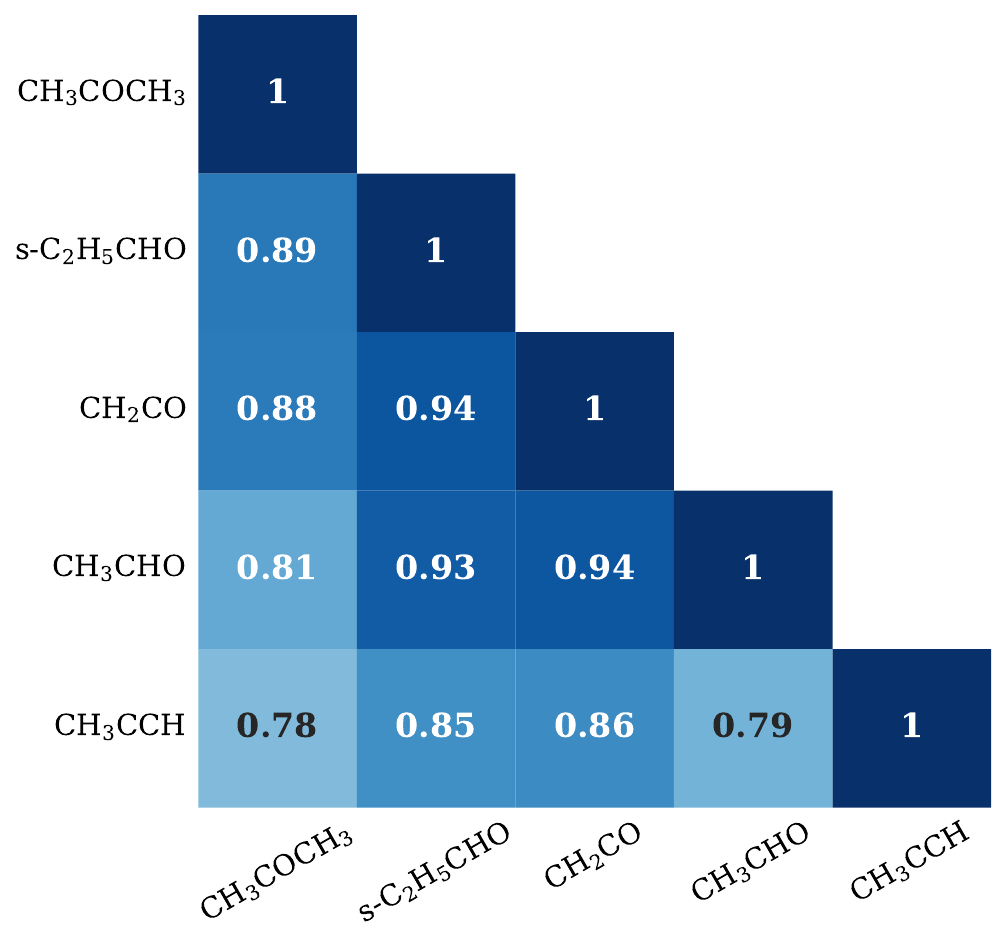}
    \end{subfigure}
    \caption{Correlation matrices of nine O-COMs (left) and five acetone-related species (right) in the 12 CoCCoA sources.}
    \label{fig:corr_matrices}
\end{figure*}

Although we cannot rule out the possible chemical links between acetone and propyne, the evidence is very limited. In the chemical network of \texttt{MAGICKAL}, acetone and propyne are distantly related, and the possible closer relation, reaction~\ref{eq:C3H6+O}, is not included. We still need more observations on the key species, propylene and propylene oxide, as well as modeling results from an updated chemical network. 



\subsection{Correlations among O-COMs and acetone relevant species}
To summarize the discussion on the abundance statistics of O-COMs and the possible formation pathways of acetone in Sects.~\ref{sect:discuss_acetone_vs_OCOM}--\ref{sect:discuss_experiment}, we calculated the Pearson correlation coefficients ($r$) of the column densities of two species groups: (1) all the nine O-COMs that have been quantified in the 12 CoCCoA sources in C23 and this work, including \ce{CH3OH}, \ce{CH3CHO}, \ce{C2H5OH}, \ce{CH3OCH3}, \ce{CH3OCHO}, \ce{CH2OHCHO}, \ce{(CH2OH)2}, \ce{CH3COCH3}, and \ce{C2H5CHO}; (2) the five species discussed in Sect.~\ref{sect:discuss_AT}, which are \ce{CH3COCH3}, \ce{C2H5CHO}, \ce{CH3CHO}, \ce{CH2CO}, and \ce{CH3CCH}. The correlation matrices for the two groups are displayed in Fig.~\ref{fig:corr_matrices}.

Among the nine O-COMs, the highest correlations belong to aldehyde pairs (i.e., between \ce{CH3CHO}, \ce{CH2OHCHO}, \ce{C2H5CHO}). Methanol (\ce{CH3OH}) also have several highly correlated species, which are \ce{C2H5OH}, \ce{CH3OCH3}, and \ce{CH3CHO}. Ethanol (\ce{C2H5OH}) belongs to the same group as methanol (alcohol), and the other two species are observed to have very constant ratios over methanol in both gas and ice \citep{Chen2023, Chen2024}. The mean correlation coefficient is 0.85.

The five acetone-related species are in general highly correlated ($r$ > 0.8). The highest correlated species with acetone are ketene and propanal. Ketene is also strongly correlated with acetaldehyde and propanal, which supports the chemical network suggested by \cite{Sung1988}. Propyne is relatively less correlated with other species, but still more correlated than about half of the O-COM pairs shown in the right panel of Fig.~\ref{fig:corr_matrices}. 
From an observational point of view, 
the discussed formation pathways of acetone in this section are possible to exist. However, more evidence is needed, 
in particular from the theoretical or experimental studies with the following focuses: 
\begin{itemize}
    \item destruction mechanisms and rates of O-COMs in the gas-phase, especially for aldehydes;
    \item theories or chemical networks that can properly reproduce the observed high gas-phase abundances of \ce{CH3OCH3}, \ce{CH3OCHO}, and \ce{CH3COCH3};
    \item the verification or exploration of the formation pathways of acetone and other three-carbon (or even larger) O-COMs that have been detected in space.
\end{itemize}

\section{Conclusions}
In this work, we continue the observational studies by C23 on six two-carbon O-COMs in the CoCCoA sample. We derived the physical parameters ($N$, $T_\mathrm{ex}$, FWHM, and $v_\mathrm{lsr}$) of four additional species in the gas phase, which are acetone, propanal, ketene, and propyne. Methanol was revisited by directly fitting its optically thin lines instead of inferring from its minor isotopolog. We compared our gas-phase results with those of ice observations by JWST and astrochemical simulations, which triggered a series of discussion about the formation mechanisms of O-COMs (with a special focus on acetone) and some lingering questions in this field. The conclusions are summarized below:

\begin{itemize}
    \item Acetone, ketene, and propyne are firmly detected in a dozen high-mass protostars as part of the CoCCoA survey.  Upper limits of propanal abundances were constrained.
    \item There are similarities in $T_\mathrm{ex}$, $v_\mathrm{lsr}$, and the emission morphology of acetone, ketene, and propanal to other O-COMs (represented by methanol), suggesting the same origin from hot cores. Unlike O-COMs, propyne has a slightly different $v_\mathrm{lsr}$ and tends to trace the more extended outflow cavity walls as inferred from the emission maps.
    \item The gas-phase column density ratios of O-COMs with respect to methanol in a combined sample of 12 high-mass CoCCoA sources and five low-mass sources from literature show systematically lower abundances of aldehydes (\ce{CH3CHO}, \ce{CH2OHCHO}, and \ce{C2H5CHO}) than other O-COMs with the same amount of carbon atoms. In contrast, \ce{CH3O}-bearing molecules (\ce{CH3OCH3} and \ce{CH3OCHO}) show higher abundances than alcohols (\ce{C2H5OH} and \ce{(CH2OH)2}). These observational trends hint at more efficient destruction of aldehydes in the gas phase and more efficient formation of \ce{CH3O}-bearing O-COMs (preferably in the solid phase).
    \item The column density ratios between acetone and methanol are higher in the solid phase than in the gas phase by about one order of magnitude, which is in the same case as \ce{CH3CHO} and \ce{C2H5OH} \citep{Chen2024}. This suggests that acetone may go through some destruction that is more severe than methanol in the gas phase. However, more detection and quantification of acetone ice need to be made in a larger sample.
    \item The state-of-the-art astrochemical model \texttt{MAGICKAL} includes two solid-phase and one gas-phase formation pathways of acetone. The contribution of gas-phase production increases with the warm-up timescale. The \texttt{MAGICKAL} simulations slightly overproduce the ratios between acetone and methanol, which may due to the mediated energy barriers of some relevant reactions.
    \item Previous theoretical and experimental studies suggest that acetone along with other \ce{C3H6O} isomers can be formed via ketene (\ce{CH2CO}\,+\,\ce{CH3}) and propylene (\ce{C3H6}\,+\,O). These pathways are plausible given that the observed gas-phase ratios between acetone and ketene or propyne (a precursor of propylene) are not very scattered and can be well reproduced by simulations. However, these studies did not specifically focus on acetone or did not assume astrochemical conditions, and more direct evidence is still needed to draw solid conclusions.
    \item It is urged for future COM studies, whether they are observations, simulations, or experiments, that acetone should receive adequate attention as an abundant three-carbon COM. Studying large COMs, especially those that have been detected in space, are very helpful for us to understand the broader picture of COM formation.
\end{itemize}

Observational results obtained for a variety of molecules and from a large sample of sources can bring us not only verification of existing theories, but also inspirations and guidance for future research. This observational study on acetone reveals the lack of consideration on this large but abundant molecule, and emphasizes the importance of resolving why simulations and experiments have so for not succeeded in reproducing the high abundance of \ce{CH3O}-bearing COMs. We look forward to more relevant investigation in the community, and hopefully those results will deepen our understanding of the chemical evolution of COMs in general.

\begin{acknowledgements}
This paper makes use of the following ALMA data: ADS/JAO.ALMA\#2019.1.00246.S. ALMA is a partnership of ESO (representing its member states), NSF (USA) and NINS (Japan), together with NRC (Canada), MOST and ASIAA (Taiwan), and KASI (Republic of Korea), in cooperation with the Republic of Chile. The Joint ALMA Observatory is operated by ESO, AUI/NRAO, and NAOJ. Astrochemistry in Leiden is supported by the Nether- lands Research School for Astronomy (NOVA), by funding from the European Research Council (ERC) under the European Union’s Horizon 2020 research and innovation programme (grant agreement No. 101019751 MOLDISK), by the Dutch Research Council (NWO) grants TOP-1 614.001.751 and 618.000.001, and by the Danish National Research Foundation through the Center of Excellence “InterCat” (Grant agreement no.: DNRF150). The National Radio Astronomy Observatory is a facility of the National Science Foundation operated under cooperative agreement by Associated Universities, Inc. Y.C. acknowledges Dijia Zou for the assistance in understanding chemical theories and linking them with astrochemical studies, and Julia Santos for the discussion on astrochemical experiments. R.T.G. thanks the National Science Foundation for funding through the Astronomy \& Astrophysics program (grant number 2206516).
\end{acknowledgements}

\bibliographystyle{aa} 
\bibliography{references} 

\begin{appendix}
\onecolumn
\section{Supplementary figures for Section \ref{sect:discussion}}\label{appendix:additional_fig}
\begin{figure*}[!h]
    \centering
    \begin{subfigure}[b]{\textwidth}
         \centering
         \includegraphics[width=\textwidth]{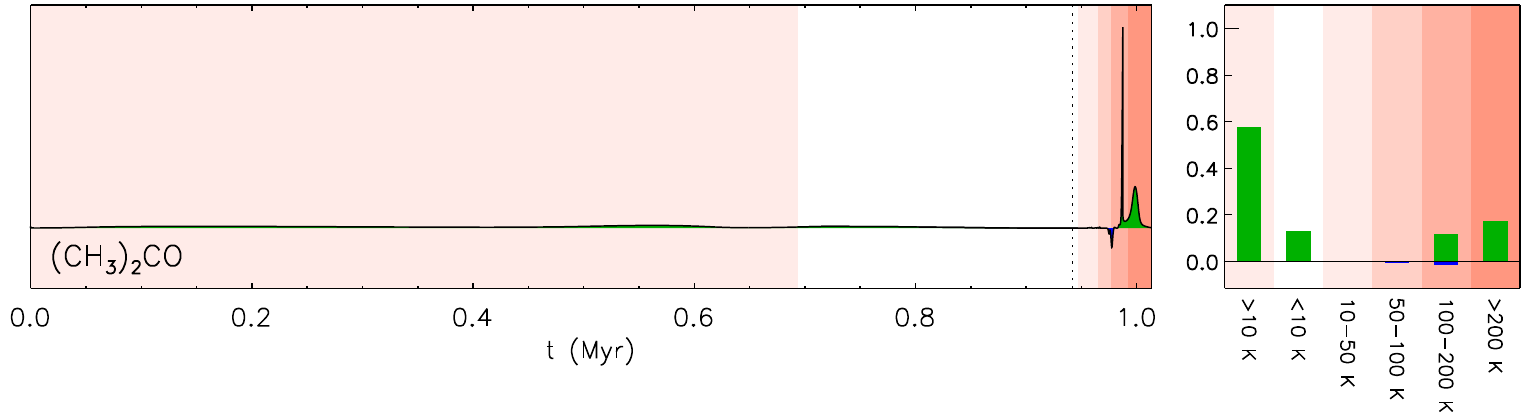}
     \end{subfigure}
     \hfill
     \begin{subfigure}[b]{\textwidth}
         \centering
         \includegraphics[width=\textwidth]{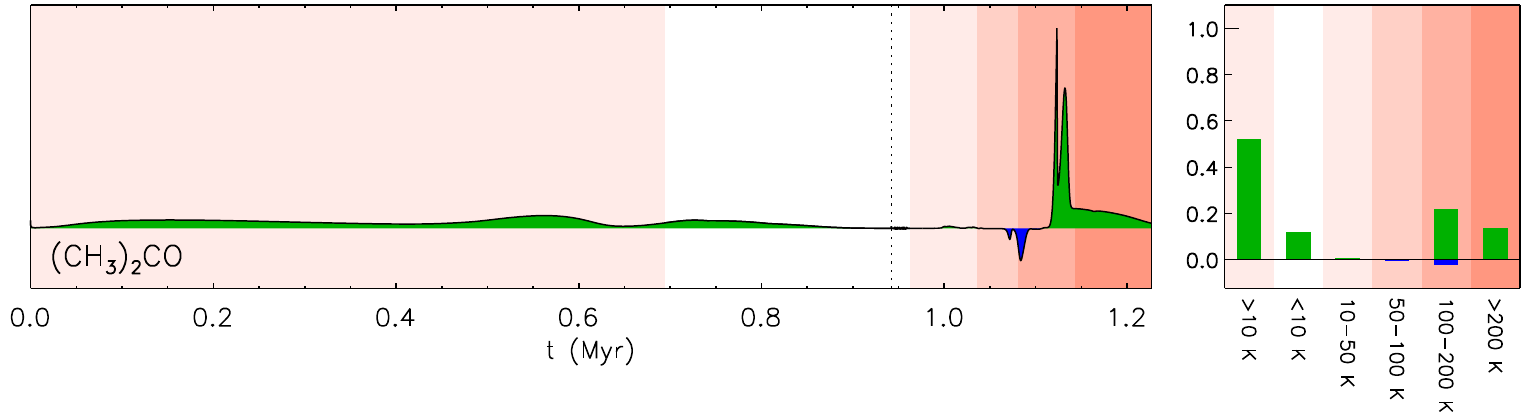}
     \end{subfigure}
     \hfill
     \begin{subfigure}[b]{\textwidth}
         \centering
         \includegraphics[width=\textwidth]{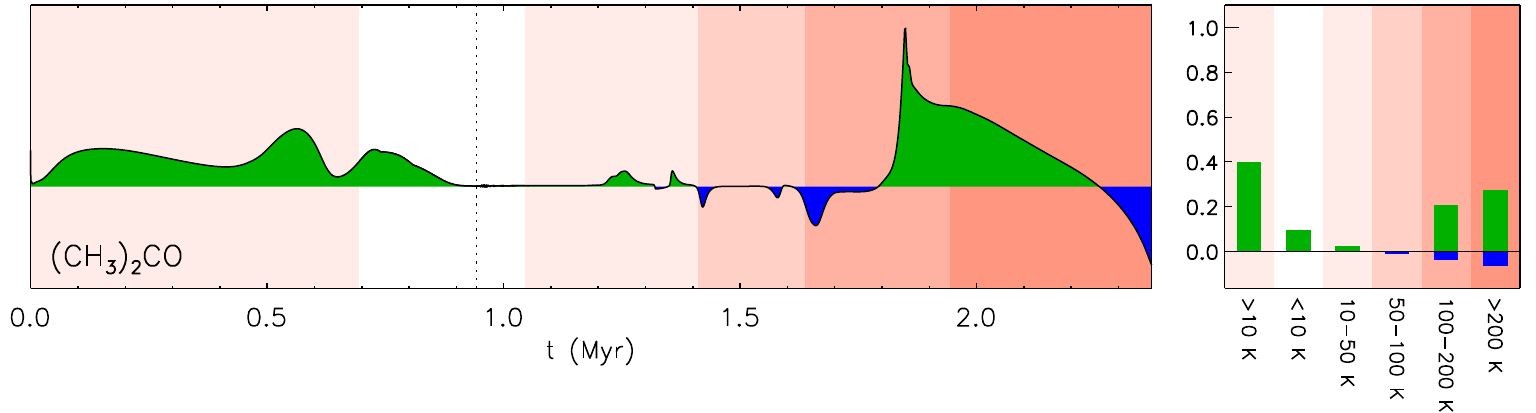}
     \end{subfigure}
    \caption{Net rate of change (arbitrary units) of acetone abundance as a function of time simulated by \texttt{MAGICKAL}. Here an updated version of \texttt{MAGICKAL} used in \citep{Belloche2022} is applied . The three rows from top to bottom correspond to the simulations with a fast ($5\times10^4$~yr), medium ($2\times10^5$~yr), and slow ($1\times10^6$~yr) warm-up stage, respectively. In each of the left panels, the vertical dotted line indicate the start of the warm-up stage. 
    The panels on the right side show the net rates of change that are normalized and integrated over each temperature range.}
    \label{fig:G22model_AT}
\end{figure*}

\begin{figure*}[h!]
    \centering
    \begin{subfigure}{0.49\textwidth}
        \centering
        \includegraphics[width=0.95\linewidth]{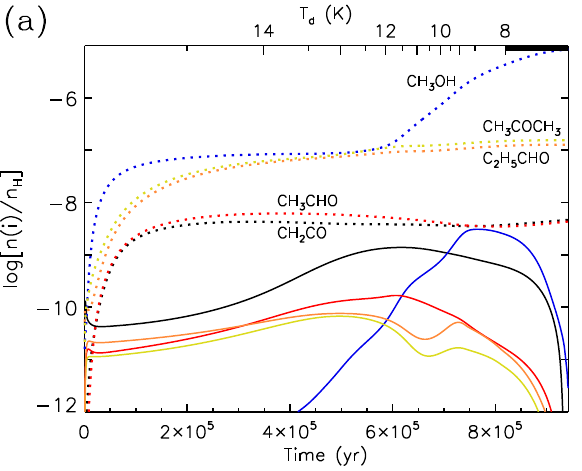}
    \end{subfigure}
    \hfill
    \begin{subfigure}{0.49\textwidth}
        \centering
        \includegraphics[width=\linewidth]{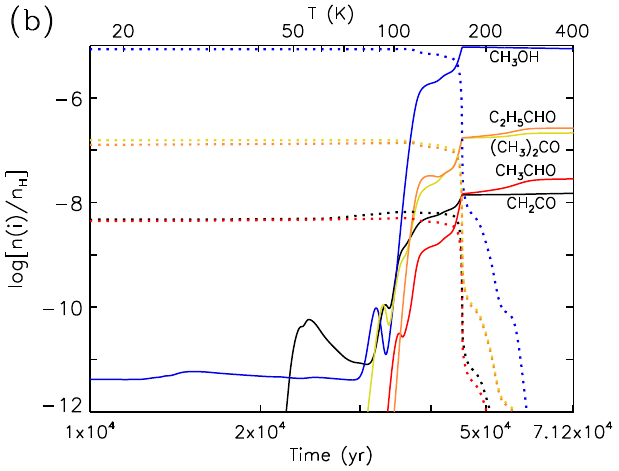}
    \end{subfigure}
    \begin{subfigure}{0.49\textwidth}
        \centering
        \includegraphics[width=\linewidth]{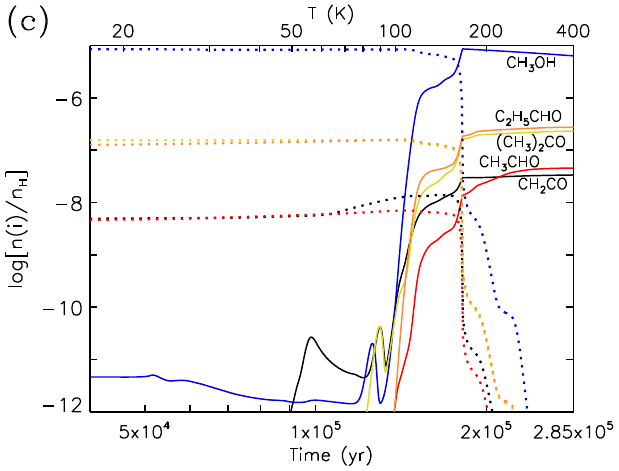}
    \end{subfigure}
    \hfill
    \begin{subfigure}{0.49\textwidth}
        \centering
        \includegraphics[width=\linewidth]{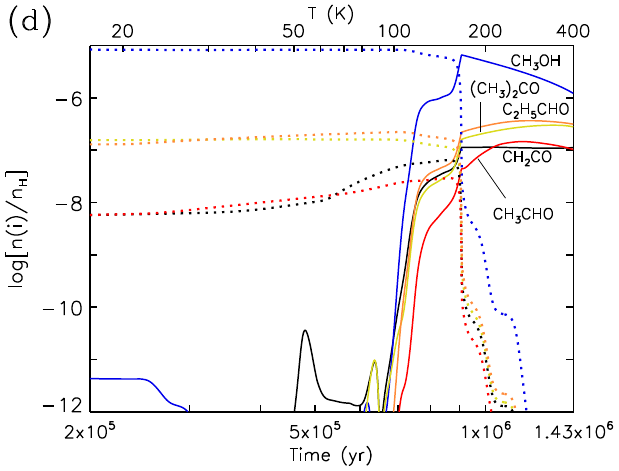}
    \end{subfigure}
    \caption{Abundances of \ce{CH3OH} (blue), \ce{CH2CO} (black), \ce{CH3CHO} (red), \ce{CH3COCH3} (yellow), and \ce{C2H5CHO} (orange) with respect to total H abundance during the collapse stage (a) and subsequent warm-up stage with fast (b), medium (c), and slow (d) timescales. The corresponding temperatures are indicated by the upper x-axis in each panel. The abundances in the gas phase and the solid phase are plotted in solid and dashed lines, respectively. The intersection of the solid and dashed lines in the same color (i.e., ice and gas abundances of the same species) corresponds to the sublimation of this species from ice to gas.}
    \label{fig:G22_n_t}
\end{figure*}

Figures~\ref{fig:G22model_AT}--\ref{fig:G22_n_t} provide more details about the \texttt{MAGICKAL} simulation results of acetone and other relevant species discussed in this paper. Figure~\ref{fig:G22model_AT} shows the net rate of change of acetone abundance as a function of time, from which we can learn at which stage acetone is efficiently produced or destroyed. Figure~\ref{fig:G22_n_t} shows the abundances of five molecules involved in the discussion (methanol, ketene, acetaldehyde, acetone, and propanal) in both ice and gas as a function of time. This can be seen as an integration results of Fig.~\ref{fig:G22model_AT}.

\FloatBarrier
\section{Best-fit physical parameters of four molecules additional to acetone}
Table~\ref{tab:fit_results_other_sp} is an extension of Table~\ref{tab:fit_results_acetone}, listing the best-fit physical parameters of three acetone-related species ($s$-propanal, ketene, and propyne) and the revisited methanol.
\begin{sidewaystable*}[!h]
    \setlength{\tabcolsep}{0.06cm}
    \renewcommand{\arraystretch}{1.3}
    \centering
    \caption{Best-fit parameters of methanol, propanal, and ketene in 12 CoCCoA sources.}
    \begin{tabular}{lcccccccccccccccc}
    \toprule
         Source & \multicolumn{4}{c}{\ce{CH3OH} (CDMS)} & \multicolumn{4}{c}{$s$-\ce{C2H5CHO} (CDMS)} & \multicolumn{4}{c}{\ce{CH2CO} (CDMS)} & \multicolumn{4}{c}{\ce{CH3CCH} (JPL)} \\
         \cmidrule(lr{0.2em}){2-5}\cmidrule(lr{0.2em}){6-9}\cmidrule(lr{0.2em}){10-13}\cmidrule(lr{0.2em}){14-17}
          & $N$ & $T_\mathrm{ex}$ & FWHM & $v_\mathrm{lsr}$ & $N$ & $T_\mathrm{ex}$ & FWHM & $v_\mathrm{lsr}$ & $N$ (cm$^{-2}$) & $T_\mathrm{ex}$ & FWHM & $v_\mathrm{lsr}$ & $N$ & $T_\mathrm{ex}$ & FWHM & $v_\mathrm{lsr}$ \\
          & (cm$^{-2}$) & (K) & (km s$^{-1}$) & (km s$^{-1}$) & (cm$^{-2}$) &  (K) & (km s$^{-1}$) & (km s$^{-1}$) & (cm$^{-2}$) & (K) & (km s$^{-1}$) & (km s$^{-1}$) & (cm$^{-2}$) & (K) & (km s$^{-1}$) & (km s$^{-1}$) \\
         \midrule
         \multirow{2}{*}{G19.01-0.03} & 1.7$^{+0.6}_{-0.4}$\,(18) & 162$\pm$10 & 3.5 & 63.0 & <1.0\,(15) & [150] & 3.0 & 61.0 & 1.5\textit{$\pm$0.3}\,(15) & [100] & 3.5 & 62.5 & 6.0\textit{$\pm$1.5}\,(15) & [200] & 3.0 & 62.3 \\
         & 3.2$^{+1.4}_{-1.2}$\,(18) & 165$\pm$15 & 5.5 & 58.0 & & & & & 4.5\textit{$\pm$1.0}\,(15) & [100] & 4.0 & 58.5 & 7.6$^{+0.6}_{-0.5}$\,(15) & 80$\pm$10 & 3.6 & 58.5 \\
         \hline
         G19.88-0.53 & 5.5$^{+1.3}_{-0.9}$\,(18) & 158$\pm$7 & 4.0 & 46.0 & <5.0\,(14) & [150] & 3.0 & 44.0 & 3.7\textit{$\pm$1.0}\,(15) & [100] & 3.2 & 46.2 & 6.9$^{+0.7}_{-0.6}$\,(15) & 145$\pm$20 & 3.2 & 45.5 \\
         \hline
         G22.04+0.22 & 2.0$^{+0.5}_{-0.3}$\,(19) & 135$\pm$5 & 6.0 & 52.2 & <4.0\,(15) & [150] & 4.5 & 51.5 & 2.5$\pm$0.5\,(16) & [100] & 7.0 & 52.2 & 4.9$\pm$0.4\,(16) & 135$\pm$15 & 7.0 & 51.7 \\
         \hline
         \multirow{2}{*}{G23.21-0.37} & 2.8$\pm$0.5\,(19)& 158$\pm$8 & 5.5 & 77.0 & <2.0\,(15) & [150] & 3.3 & 76.8 & 2.0$\pm$\textit{0.4}\,(16) & [100] & 4.0 & 77.2 & 5.8$^{+0.9}_{-0.6}$\,(16) & 180$\pm$20 & 5.2 & 77.8 \\
          & & & & & & & & & 8.0$\pm$\textit{1.6}\,(15) & [100] & 4.0 & 81.3 & & & & \\
         \hline
         \multirow{2}{*}{G34.30+0.20\,$^b$} & 1.2$\pm$0.3\,(19) & 130$\pm$5 & 3.0 & 56.0 & <1.8\,(15) & [150] & 3.5 & 56.5 & 1.3$\pm$\textit{0.3}\,(16) & [100] & 3.8 & 56.2 & 2.9$^{+0.6}_{-0.3}$\,(16) & 155$^{+25}_{-20}$ & 3.2 & 56.2 \\
          & 5.5$^{+1.4}_{-1.2}$\,(18) & 127$\pm$5 & 2.8 & 59.0 & & & & & & & & & & & & \\
         \hline
         \multirow{2}{*}{G34.41+0.24} & 1.7$^{+0.6}_{-0.4}$\,(19) & 130$\pm$6 & 3.5 & 60.0 & & & & & & & & & & & & \\
          & 6.0$\pm$\textit{1.5}\,(18) & [130] & [3.5] & 56.8 & <1.0\,(15) & [150] & [3.0] & 58.5 & 1.6$\pm$\textit{0.3}\,(16) & [100] & 5.5 & 58.3& 3.2$\pm$0.2\,(16) & 106$\pm$10 & 5.7 & 57.5 \\
         \hline
         \multirow{3}{*}{G345.5+1.5} & 1.3$\pm$0.3\,(18) & 120$\pm$6 & 2.5 & -15.8 & <2.5\,(14) & [150] & [2.0] & -16.0 & 8.3$^{+1.7}_{-0.5}$\,(14) & 75$^{+45}_{-30}$ & 2.3 & -15.9 & 9.5$^{+1.4}_{-1.2}$\,(15) & 120$^{+30}_{-20}$ & 2.1 & -15.3 \\
          & & & & & & & & & 3.0$\pm$\textit{0.6}\,(14) & [100] & 2.0 & -13.3 & 2.8$^{+0.7}_{-0.5}$\,(15) & 55$\pm$10 & 1.5 & -12.7 \\
          & & & & & & & & & & & & & 8.0$^{+12}_{-4.0}$\,(14) & 40$^{+40}_{-20}$ & 1.4 & -10.0 \\
         \hline
         \multirow{2}{*}{G35.03+0.35} & 2.6$^{+1.4}_{-1.1}$\,(18) & 150$\pm$15 & 3.6 & 44.2 & <6.0\,(14) & [150] & 3.5 & 47.5 & 1.8$\pm$0.4\,(15) & [100] & 3.5 & 46.0 & 6.9$\pm$0.5\,(15) & 90$\pm$15 & 3.5 & 47.6 \\
          & & & & & & & & & & & & & \textit{4.0$\pm$1.0}\,(15) & [200] & 2.5 & 53.0 \\
         \hline
         \multirow{2}{*}{G35.20-0.74N} & 1.0$\pm$0.2\,(19) & 146$\pm$5 & 5.0 & 31.5 & <1.0\,(15) & [150] & 3.5 & 32.0 & 1.0$\pm$\textit{0.2}\,(16) & [100] & 4.5 & 31.8 & 4.8$\pm$\textit{1.0}\,(15) & [200] & 2.2 & 31.0 \\
          & & & & & & & & & & & & & 4.2$^{+1.5}_{-0.7}$\,(15) & 135$^{+50}_{-30}$ & 2.1 & 33.5 \\
         \hline
         \multirow{2}{*}{IRAS 16547-4247} & 2.5$\pm$0.6\,(18) & 180$\pm$10 & 3.3 & -35.2 & <1.5\,(15) & [150] & 3.3 & -35 & 9.0$\pm$\textit{2.0}\,(15) & [100] & 4.0 & -35.3 & 1.7$^{+0.2}_{-0.1}$\,(16) & 130$^{+20}_{-10}$ & 3.3 & -34.5 \\
          & & & & & & & & & & & & & 6.9$\pm$1.0\,(15) & 115$^{+35}_{-25}$ & 2.6 & -31.4 \\
         \hline
         IRAS 18151-1208 & 1.4$^{+1.8}_{-0.7}$\,(17) & 150$\pm$30 & 2.2 & 35.0 & <1.5\,(14) & [150] & 2.0 & 35.2 & 4.2$^{+1.4}_{-0.5}$\,(14) & 50$^{+50}_{-20}$ & 2.0 & 35.2 & 2.4$^{+0.4}_{-0.3}$\,(15) & 60$^{+20}_{-10}$ & 2.5 & 34.7 \\
         \hline
         NGC 6334-38 & 6.8$\pm$2.3\,(18) & 133$\pm$10 & 3.8 & -5.0 & <6.0\,(14) & [150] & 3.0 & -6.2 & 3.5$\pm$\textit{0.7}\,(15) & 120$\pm$20 & 4.0 & -4.8 & 1.6$^{+1.5}_{-1.0}$\,(16) & 80$^{+15}_{-10}$ & 3.0 & -3.5 \\         
         \bottomrule
    \end{tabular}
     \begin{minipage}{\textwidth}
     Values in italics were estimated by visual inspection.
     \end{minipage}
    \label{tab:fit_results_other_sp}
\end{sidewaystable*}

\clearpage
\section{Transitions covered in the CoCCoA dataset}\label{appendix:transitions}
\setlength{\tabcolsep}{0.12cm}
\begin{longtable}[h]{lcccclccc}    
    \caption{Transitions of acetone, propanal, ketene, propyne, and methanol covered in the CoCCoA dataset.\label{tab:key_trans}}\\
    \toprule 
    Frequency$^{(a)}$ & Transition & $E_\mathrm{up}$ & $A_\mathrm{ij}^{(b)}$ & & Frequency & Transition & $E_\mathrm{up}$ & $A_\mathrm{ij}$ \\
    (MHz) & J K L M -- J K L M & (K) & (s$^{-1}$) & & (MHz) & J K L M -- J K L M & (K) & (s$^{-1}$) \\
    \hline\addlinespace[2pt]
    \endfirsthead
    \caption{continued.}\\
    \toprule
    Frequency & Transition & $E_\mathrm{up}$ & $A_\mathrm{ij}$ & & Frequency & Transition & $E_\mathrm{up}$ & $A_\mathrm{ij}$ \\
    (MHz) & J K L M -- J K L M & (K) & (s$^{-1}$) & & (MHz) & J K L M -- J K L M & (K) & (s$^{-1}$) \\
    \hline\addlinespace[2pt]
    \endhead
    \bottomrule 
    \endfoot
    \bottomrule 
    \endlastfoot
        \multicolumn{2}{c}{\textbf{\ce{CH3COCH3} (JPL)}} & \textbf{<600} & \textbf{>5.0(-5)} & & \multicolumn{4}{l}{\citep{Groner2002}} \\
        \hline\addlinespace[2pt] 
        238042.834$^{\star\star}$ & 32 5 27 0  --  32 5 28 1 & 317.01 & 1.05(-4) & & 241324.624 & 17 8 9 1  --  16 9 8 1 & 118.62 & 2.73(-4)\\
         & 32 6 27 0  --  32 4 28 1 & 317.01 & 1.05(-4) & & 241331.223 & 17 8 9 1  --  16 9 8 2 & 118.62 & 2.73(-4)\\
        238044.288$^{\star\star}$ & 31 4 27 1  --  31 4 28 2 & 288.66 & 9.81(-5) & & 241420.927$^{\star}$ & 17 8 9 0  --  16 9 8 1 & 118.58 & 2.73(-4)\\
         & 31 5 27 1  --  31 3 28 2 & 288.66 & 9.81(-5) & & 241513.653 & 17 8 9 0  --  16 9 8 0 & 118.55 & 2.72(-4)\\
        238044.373$^{\star\star}$ & 31 5 27 1  --  31 4 28 1 & 288.66 & 9.81(-5) & & 241559.887 & 13 10 4 0  --  12 9 3 0 & 77.71 & 3.30(-4)\\
         & 31 4 27 1  --  31 3 28 1 & 288.66 & 9.81(-5) & & 258070.737 & 17 10 8 1  --  16 9 7 2 & 121.93 & 2.63(-4)\\
        238271.013 & 30 3 27 1  --  30 2 28 2 & 261.18 & 7.71(-5) & & 258113.895 & 17 10 8 1  --  16 9 7 1 & 121.93 & 2.63(-4)\\
         & 30 4 27 1  --  30 3 28 2 & 261.18 & 7.71(-5) & & 258338.038 & 17 10 8 0  --  16 9 7 1 & 121.89 & 2.63(-4)\\
        238271.066 & 30 3 27 1  --  30 2 28 1 & 261.18 & 7.71(-5) & & 258461.738 & 14 10 5 1  --  13 9 5 2 & 87.78 & 5.70(-5)\\
         & 30 4 27 1  --  30 3 28 1 & 261.18 & 7.71(-5) & & 258472.032$^{\star\dagger}$ & 25 1 24 1  --  24 2 23 1 & 168.79 & 7.85(-4)\\
        238293.360 & 32 5 27 0  --  32 4 28 0 & 317.00 & 1.18(-4) & &  & 25 2 24 1  --  24 1 23 1 & 168.79 & 7.85(-4)\\
         & 32 6 27 0  --  32 5 28 0 & 317.00 & 1.18(-4) & & 258472.058$^{\star\dagger}$ & 25 1 24 1  --  24 2 23 2 & 168.79 & 7.85(-4)\\
        238338.954$^{\star}$ & 31 5 27 0  --  31 4 28 1 & 288.63 & 9.80(-5) & &  & 25 2 24 1  --  24 1 23 2 & 168.79 & 7.85(-4)\\
         & 31 4 27 0  --  31 3 28 1 & 288.63 & 9.80(-5) & & 258493.675$^{\star\dagger}$ & 25 1 24 0  --  24 1 23 1 & 168.71 & 1.61(-4)\\
        238354.973 & 17 10 7 0  --  16 11 6 0 & 124.02 & 9.37(-5) & &  & 25 1 24 0  --  24 2 23 1 & 168.71 & 6.24(-4)\\
        238473.550 & 29 2 27 1  --  29 2 28 2 & 234.56 & 5.40(-5) & &  & 25 2 24 0  --  24 2 23 1 & 168.71 & 1.61(-4)\\
         & 29 3 27 1  --  29 1 28 2 & 234.56 & 5.40(-5) & &  & 25 2 24 0  --  24 1 23 1 & 168.71 & 6.24(-4)\\
        238473.562 & 29 2 27 1  --  29 1 28 1 & 234.56 & 5.40(-5) & & 258515.262$^{\star\dagger}$ & 25 2 24 0  --  24 2 23 0 & 168.63 & 7.85(-4)\\
         & 29 3 27 1  --  29 2 28 1 & 234.56 & 5.39(-5) & &  & 25 1 24 0  --  24 1 23 0 & 168.63 & 7.85(-4)\\
        238611.445 & 17 10 7 0  --  16 11 6 1 & 124.07 & 9.40(-5) & & 258582.793 & 17 10 8 0  --  16 9 7 0 & 121.86 & 2.63(-4)\\
        238633.098 & 31 5 27 0  --  31 4 28 0 & 288.61 & 9.90(-5) & & 259392.224 & 18 9 9 1  --  17 10 8 1 & 134.38 & 3.02(-4)\\
         & 31 4 27 0  --  31 3 28 0 & 288.61 & 9.90(-5) & & 259406.977 & 18 9 9 1  --  17 10 8 2 & 134.38 & 3.02(-4)\\
        238695.516 & 13 10 4 1  --  12 9 3 2 & 77.73 & 6.34(-5) & & 259480.221 & 18 9 9 0  --  17 10 8 1 & 134.34 & 3.01(-4)\\
        238708.138 & 17 10 7 1  --  16 11 6 1 & 124.11 & 9.57(-5) & & 259560.545 & 18 9 9 0  --  17 10 8 0 & 134.31 & 3.01(-4)\\
        238847.035 & 23 1 22 1  --  22 1 21 1 & 144.45 & 6.14(-4) & & 259612.192$^{\star\dagger}$ & 26 0 26 1  --  25 1 25 1 & 171.05 & 8.38(-4)\\
         & 23 2 22 1  --  22 2 21 1 & 144.45 & 6.14(-4) & &  & 26 1 26 1  --  25 0 25 1 & 171.05 & 8.38(-4)\\
        238847.067$^{\star\dagger}$ & 23 1 22 1  --  22 2 21 2 & 144.45 & 6.14(-4) & & 259612.241$^{\star\dagger}$ & 26 0 26 1  --  25 0 25 2 & 171.05 & 8.38(-4)\\
         & 23 2 22 1  --  22 1 21 2 & 144.45 & 6.14(-4) & &  & 26 1 26 1  --  25 1 25 2 & 171.05 & 8.38(-4)\\
        238863.500 & 29 2 27 0  --  29 2 28 1 & 234.51 & 5.19(-5) & & 259618.411$^{\star\dagger}$ & 26 1 26 0  --  25 1 25 1 & 170.95 & 8.34(-4)\\
         & 29 3 27 0  --  29 1 28 1 & 234.51 & 5.19(-5) & &  & 26 0 26 0  --  25 0 25 1 & 170.95 & 8.34(-4)\\
        238868.953$^\ddagger$ & 23 2 22 0  --  22 2 21 1 & 144.37 & 5.90(-4) & & 259624.520$^{\star\dagger}$ & 26 0 26 0  --  25 1 25 0 & 170.85 & 8.38(-4)\\
         & 23 1 22 0  --  22 1 21 1 & 144.37 & 5.81(-4) & &  & 26 1 26 0  --  25 0 25 0 & 170.85 & 8.38(-4)\\
        238890.805$^{\star\dagger}$ & 23 2 22 0  --  22 2 21 0 & 144.29 & 6.14(-4) & & 260640.610 & 18 10 8 0  --  17 11 7 0 & 137.29 & 1.97(-4)\\
         & 23 1 22 0  --  22 1 21 0 & 144.29 & 6.14(-4) & & 260702.036 & 18 10 8 0  --  17 11 7 1 & 137.33 & 1.99(-4)\\
        238952.610 & 30 3 27 0  --  30 3 28 0 & 261.09 & 7.79(-5) & & 260715.290 & 18 10 8 1  --  17 11 7 1 & 137.36 & 2.00(-4)\\
         & 30 4 27 0  --  30 2 28 0 & 261.09 & 7.79(-5) & & 260811.801 & 18 10 8 1  --  17 11 7 2 & 137.36 & 2.00(-4)\\
        239022.408 & 17 10 7 1  --  16 11 6 2 & 124.11 & 9.33(-5) & & 261054.804 & 13 13 1 1  --  12 12 1 2 & 86.98 & 7.95(-4)\\
        239253.257 & 29 2 27 0  --  29 2 28 0 & 234.45 & 5.46(-5) & & 261078.200 & 19 8 11 1  --  18 9 10 1 & 143.24 & 4.10(-4)\\
         & 29 3 27 0  --  29 1 28 0 & 234.45 & 5.46(-5) & & 261078.367 & 19 8 11 1  --  18 9 10 2 & 143.24 & 4.10(-4)\\
        239984.730 & 24 0 24 1  --  23 0 23 1 & 146.60 & 6.59(-4) & & 261082.377 & 13 13 1 1  --  12 12 0 1 & 87.10 & 7.96(-4)\\
         & 24 1 24 1  --  23 1 23 1 & 146.60 & 6.59(-4) & & 261082.552 & 13 13 0 1  --  12 12 1 1 & 87.10 & 7.96(-4)\\
        239984.779$^{\star\star}$ & 24 1 24 1  --  23 1 23 2 & 146.60 & 6.59(-4) & & 261110.448$^{\star\dagger}$ & 13 13 0 1  --  12 12 0 2 & 87.22 & 7.96(-4)\\
         & 24 0 24 1  --  23 0 23 2 & 146.60 & 6.59(-4) & & 261169.099 & 19 9 11 1  --  18 8 10 2 & 143.24 & 4.10(-4)\\
        239991.110$^\ddagger$ & 24 1 24 0  --  23 1 23 1 & 146.50 & 6.23(-4) & & 261169.839 & 19 9 11 1  --  18 8 10 1 & 143.24 & 4.10(-4)\\
         & 24 0 24 0  --  23 0 23 1 & 146.50 & 6.23(-4) & & 261175.065$^{\star\star}$ & 19 8 11 0  --  18 9 10 1 & 143.21 & 4.10(-4)\\
        239997.383$^{\star\star}$ & 24 0 24 0  --  23 1 23 0 & 146.40 & 6.59(-4) & & 261268.570 & 19 9 11 0  --  18 8 10 1 & 143.21 & 4.11(-4)\\
         & 24 1 24 0  --  23 0 23 0 & 146.40 & 6.59(-4) & & 261271.656 & 19 8 11 0  --  18 9 10 0 & 143.18 & 4.10(-4)\\
        240475.005 & 13 10 4 0  --  12 9 3 1 & 77.73 & 1.52(-4) & & 261328.382$^{\star\dagger}$ & 13 13 1 0  --  12 12 1 1 & 87.03 & 7.98(-4)\\
        240711.378$^{\star\dagger}$ & 12 12 1 1  --  11 11 1 2 & 74.45 & 6.17(-4) & & 261356.943$^{\star\star}$ & 13 13 0 0  --  12 12 0 1 & 87.16 & 7.98(-4)\\
        240774.176$^{\star}$ & 12 12 1 1  --  11 11 0 1 & 74.57 & 6.18(-4) & & 261367.487 & 19 9 11 0  --  18 8 10 0 & 143.18 & 4.11(-4)\\
        240774.719$^{\star}$ & 12 12 0 1  --  11 11 1 1 & 74.57 & 6.18(-4) & & 261397.989 & 14 11 4 0  --  13 10 3 1 & 90.39 & 6.99(-5)\\
        240822.548 & 11 6 5 1  --  10 5 6 1 & 52.99 & 5.05(-5) & & 261602.431 & 13 13 1 0  --  12 12 0 0 & 87.09 & 8.01(-4)\\
        240838.029$^{\star\dagger}$ & 12 12 0 1  --  11 11 0 2 & 74.69 & 6.19(-4) & & 261602.595$^{\star}$ & 13 13 0 0  --  12 12 1 0 & 87.09 & 8.01(-4)\\
        240861.231 & 11 6 5 1  --  10 5 6 2 & 52.99 & 5.05(-5) & & 261755.985 & 41 11 30 1  --  41 11 31 2 & 576.34 & 2.58(-4)\\
        240941.741 & 11 6 5 0  --  10 5 6 1 & 52.91 & 5.07(-5) & &  & 41 12 30 1  --  41 10 31 2 & 576.34 & 2.58(-4)\\
        240998.755$^{\star\dagger}$ & 12 12 1 0  --  11 11 1 1 & 74.49 & 6.20(-4) & & 261756.121 & 41 12 30 1  --  41 11 31 1 & 576.34 & 2.58(-4)\\
        241042.018 & 11 6 5 0  --  10 5 6 0 & 52.83 & 5.08(-5) & &  & 41 11 30 1  --  41 10 31 1 & 576.34 & 2.58(-4)\\
        241042.122 & 13 10 4 1  --  12 9 3 1 & 77.80 & 3.26(-4) & & 261795.495 & 41 11 30 0  --  41 11 31 1 & 576.38 & 1.96(-4)\\
        241062.732$^{\star\star}$ & 12 12 0 0  --  11 11 0 1 & 74.61 & 6.20(-4) & &  & 41 12 30 0  --  41 11 31 1 & 576.38 & 1.47(-4)\\
        241286.250$^{\star}$ & 12 12 1 0  --  11 11 0 0 & 74.53 & 6.22(-4) & &  & 41 11 30 0  --  41 10 31 1 & 576.38 & 1.47(-4)\\
        241286.761$^{\star}$ & 12 12 0 0  --  11 11 1 0 & 74.53 & 6.22(-4) & &  & 41 12 30 0  --  41 10 31 1 & 576.38 & 1.96(-4)\\
        \hline\addlinespace[2pt]
        \multicolumn{2}{c}{\textbf{$s$-\ce{C2H5CHO}, v=0 (CDMS)}} & \textbf{<500} & \textbf{>1.0(-4)} & & \multicolumn{4}{l}{\citep{Zingsheim2017}} \\
        \hline\addlinespace[2pt]
        238104.321 & 20 11 10 1  --  20 10 11 1 & 172.48 & 1.09(-4) & & 259183.141 & 40 5 36 0  --  40 4 37 0 & 423.50 & 1.04(-4)\\
        238104.799 & 20 11 9 1  --  20 10 10 1 & 172.48 & 1.09(-4) & & 259367.912 & 26 12 15 1  --  26 11 16 1 & 256.58 & 1.58(-4)\\
        238105.199 & 20 11 9 0  --  20 10 10 0 & 172.48 & 1.09(-4) & &  & 26 12 14 1  --  26 11 15 1 & 256.58 & 1.58(-4)\\
        238105.214 & 20 11 10 0  --  20 10 11 0 & 172.48 & 1.09(-4) & &  & 26 12 14 0  --  26 11 15 0 & 256.58 & 1.58(-4)\\
        238350.142 & 19 11 9 1  --  19 10 10 1 & 162.33 & 1.05(-4) & &  & 26 12 15 0  --  26 11 16 0 & 256.58 & 1.58(-4)\\
        238350.636 & 19 11 8 1  --  19 10 9 1 & 162.34 & 1.05(-4) & & 259787.502$^\ast$ & 25 12 14 1  --  25 11 15 1 & 243.37 & 1.55(-4)\\
        238351.081 & 19 11 8 0  --  19 10 9 0 & 162.33 & 1.05(-4) & &  & 25 12 13 1  --  25 11 14 1 & 243.37 & 1.55(-4)\\
        238351.086 & 19 11 9 0  --  19 10 10 0 & 162.33 & 1.05(-4) & &  & 25 12 13 0  --  25 11 14 0 & 243.37 & 1.55(-4)\\
        238738.609 & 22 6 16 1  --  21 6 15 1 & 149.17 & 2.10(-4) & &  & 25 12 14 0  --  25 11 15 0 & 243.37 & 1.55(-4)\\
         & 22 6 16 0  --  21 6 15 0 & 149.17 & 2.10(-4) & & 260020.036 & 27 1 26 1  --  26 2 25 1 & 181.60 & 2.84(-4)\\
        238908.897 & 23 4 20 1  --  22 4 19 1 & 148.51 & 2.18(-4) & &  & 27 1 26 0  --  26 2 25 0 & 181.60 & 2.84(-4)\\
         & 23 4 20 0  --  22 4 19 0 & 148.51 & 2.18(-4) & & 260021.633 & 27 2 26 1  --  26 2 25 1 & 181.60 & 2.89(-4)\\
        239035.632 & 22 4 19 0  --  21 3 18 0 & 137.05 & 1.20(-4) & &  & 27 2 26 0  --  26 2 25 0 & 181.60 & 2.89(-4)\\
         & 22 4 19 1  --  21 3 18 1 & 137.05 & 1.20(-4) & & 260022.915 & 27 1 26 1  --  26 1 25 1 & 181.60 & 2.89(-4)\\
        240270.504 & 24 2 22 1  --  23 3 21 1 & 153.34 & 1.79(-4) & &  & 27 1 26 0  --  26 1 25 0 & 181.60 & 2.89(-4)\\
         & 24 2 22 0  --  23 3 21 0 & 153.34 & 1.79(-4) & & 260024.456 & 27 2 26 1  --  26 1 25 1 & 181.60 & 2.84(-4)\\
        240489.453 & 24 3 22 1  --  23 3 21 1 & 153.35 & 2.25(-4) & &  & 27 2 26 0  --  26 1 25 0 & 181.60 & 2.84(-4)\\
         & 24 3 22 0  --  23 3 21 0 & 153.35 & 2.25(-4) & & 260156.935 & 24 12 13 1  --  24 11 14 1 & 230.67 & 1.52(-4)\\
        240630.703$^{\ast}$ & 24 2 22 1  --  23 2 21 1 & 153.34 & 2.25(-4) & &  & 24 12 12 1  --  24 11 13 1 & 230.67 & 1.52(-4)\\
         & 24 2 22 0  --  23 2 21 0 & 153.34 & 2.25(-4) & &  & 24 12 12 0  --  24 11 13 0 & 230.67 & 1.52(-4)\\
        240676.214 & 40 5 35 1  --  40 4 36 1 & 435.05 & 1.04(-4) & &  & 24 12 13 0  --  24 11 14 0 & 230.67 & 1.52(-4)\\
        240677.547 & 40 5 35 0  --  40 4 36 0 & 435.05 & 1.04(-4) & & 260480.718$^\ast$ & 23 12 12 1  --  23 11 13 1 & 218.49 & 1.48(-4)\\
        240849.674 & 24 3 22 1  --  23 2 21 1 & 153.35 & 1.81(-4) & &  & 23 12 11 1  --  23 11 12 1 & 218.49 & 1.48(-4)\\
         & 24 3 22 0  --  23 2 21 0 & 153.35 & 1.81(-4) & &  & 23 12 11 0  --  23 11 12 0 & 218.49 & 1.48(-4)\\
        240953.122 & 40 6 35 1  --  40 5 36 1 & 435.06 & 1.05(-4) & &  & 23 12 12 0  --  23 11 13 0 & 218.49 & 1.48(-4)\\
        240954.428 & 40 6 35 0  --  40 5 36 0 & 435.06 & 1.05(-4) & & 260563.102 & 25 4 22 1  --  24 3 21 1 & 172.79 & 1.83(-4)\\
        241051.548 & 11 6 6 0  --  10 5 5 0 & 53.11 & 1.39(-4) & &  & 25 4 22 0  --  24 3 21 0 & 172.79 & 1.83(-4)\\
        241062.630 & 23 3 20 1  --  22 3 19 1 & 148.30 & 2.24(-4) & & 260763.017 & 22 12 11 1  --  22 11 12 1 & 206.82 & 1.43(-4)\\
         & 23 3 20 0  --  22 3 19 0 & 148.30 & 2.24(-4) & &  & 22 12 10 1  --  22 11 11 1 & 206.82 & 1.43(-4)\\
        241083.814 & 11 6 5 0  --  10 5 6 0 & 53.11 & 1.39(-4) & &  & 22 12 10 0  --  22 11 11 0 & 206.82 & 1.43(-4)\\
        241667.998$^\ast$ & 25 1 24 1  --  24 2 23 1 & 157.08 & 2.23(-4) & &  & 22 12 11 0  --  22 11 12 0 & 206.82 & 1.43(-4)\\
         & 25 1 24 0  --  24 2 23 0 & 157.08 & 2.23(-4) & & 261007.063 & 21 12 10 1  --  21 11 11 1 & 195.66 & 1.38(-4)\\
        241672.910$^\ast$ & 25 2 24 1  --  24 2 23 1 & 157.08 & 2.31(-4) & & 261007.453 & 21 12 9 1  --  21 11 10 1 & 195.66 & 1.38(-4)\\
         & 25 2 24 0  --  24 2 23 0 & 157.08 & 2.31(-4) & & 261007.926 & 21 12 9 0  --  21 11 10 0 & 195.66 & 1.38(-4)\\
        241676.670 & 25 1 24 1  --  24 1 23 1 & 157.08 & 2.31(-4) & & 261007.927 & 21 12 10 0  --  21 11 11 0 & 195.66 & 1.38(-4)\\
         & 25 1 24 0  --  24 1 23 0 & 157.08 & 2.31(-4) & & 261069.305 & 24 4 20 1  --  23 4 19 1 & 165.80 & 2.85(-4)\\
        241681.621 & 25 2 24 1  --  24 1 23 1 & 157.08 & 2.23(-4) & &  & 24 4 20 0  --  23 4 19 0 & 165.80 & 2.85(-4)\\
         & 25 2 24 0  --  24 1 23 0 & 157.08 & 2.23(-4) & & 261218.275 & 20 12 9 1  --  20 11 10 1 & 185.01 & 1.33(-4)\\
        258249.922 & 24 7 17 1  --  23 7 16 1 & 179.87 & 2.63(-4) & &  & 20 12 8 1  --  20 11 9 1 & 185.01 & 1.33(-4)\\
         & 24 7 17 0  --  23 7 16 0 & 179.87 & 2.63(-4) & &  & 20 12 8 0  --  20 11 9 0 & 185.01 & 1.33(-4)\\
        258358.429 & 28 12 17 1  --  28 11 18 1 & 284.55 & 1.61(-4) & &  & 20 12 9 0  --  20 11 10 0 & 185.01 & 1.33(-4)\\
        258358.434 & 28 12 16 0  --  28 11 17 0 & 284.55 & 1.63(-4) & & 261290.259$^{\ast\ast}$ & 28 0 28 1  --  27 1 27 1 & 184.39 & 3.29(-4)\\
        258358.633 & 28 12 16 1  --  28 11 17 1 & 284.55 & 1.61(-4) & & 261290.267$^{\ast\ast}$ & 28 0 28 0  --  27 1 27 0 & 184.39 & 3.29(-4)\\
        258359.161 & 28 12 17 0  --  28 11 18 0 & 284.55 & 1.63(-4) & & 261290.271$^{\ast\ast}$ & 28 1 28 1  --  27 1 27 1 & 184.39 & 2.96(-4)\\
        258629.931$^\ast$ & 25 3 22 1  --  24 3 21 1 & 172.70 & 2.78(-4) & & 261290.280$^{\ast\ast}$ & 28 1 28 0  --  27 1 27 0 & 184.39 & 2.96(-4)\\
         & 25 3 22 0  --  24 3 21 0 & 172.70 & 2.78(-4) & & 261290.282$^{\ast\ast}$ & 28 0 28 1  --  27 0 27 1 & 184.39 & 2.96(-4)\\
        258761.193 & 26 2 24 1  --  25 3 23 1 & 177.76 & 2.34(-4) & & 261290.291$^{\ast\ast}$ & 28 0 28 0  --  27 0 27 0 & 184.39 & 2.96(-4)\\
         & 26 2 24 0  --  25 3 23 0 & 177.76 & 2.34(-4) & & 261290.294$^{\ast\ast}$ & 28 1 28 1  --  27 0 27 1 & 184.39 & 3.29(-4)\\
        258840.070 & 26 3 24 1  --  25 3 23 1 & 177.76 & 2.82(-4) & & 261290.303$^{\ast\ast}$ & 28 1 28 0  --  27 0 27 0 & 184.39 & 3.29(-4)\\
         & 26 3 24 0  --  25 3 23 0 & 177.76 & 2.82(-4) & & 261397.487 & 19 12 8 1  --  19 11 9 1 & 174.88 & 1.26(-4)\\
        258865.207 & 16 5 12 1  --  15 4 11 1 & 82.91 & 1.02(-4) & & 261397.901 & 19 12 7 1  --  19 11 8 1 & 174.88 & 1.26(-4)\\
         & 16 5 12 0  --  15 4 11 0 & 82.91 & 1.02(-4) & & 261398.461 & 19 12 7 0  --  19 11 8 0 & 174.88 & 1.26(-4)\\
        258893.298$^{\ast}$ & 26 2 24 1  --  25 2 23 1 & 177.76 & 2.82(-4) & &  & 19 12 8 0  --  19 11 9 0 & 174.88 & 1.26(-4)\\
         & 27 12 16 1  --  27 11 17 1 & 270.31 & 1.60(-4) & & 261549.823 & 18 12 7 1  --  18 11 8 1 & 165.25 & 1.18(-4)\\
         & 26 2 24 0  --  25 2 23 0 & 177.76 & 2.82(-4) & & 261550.248 & 18 12 6 1  --  18 11 7 1 & 165.25 & 1.18(-4)\\
         & 27 12 15 1  --  27 11 16 1 & 270.31 & 1.60(-4) & & 261550.849 & 18 12 6 0  --  18 11 7 0 & 165.25 & 1.18(-4)\\
         & 27 12 15 0  --  27 11 16 0 & 270.31 & 1.61(-4) & &  & 18 12 7 0  --  18 11 8 0 & 165.25 & 1.18(-4)\\
         & 27 12 16 0  --  27 11 17 0 & 270.31 & 1.61(-4) & & 261677.566$^\ast$ & 17 12 6 1  --  17 11 7 1 & 156.14 & 1.09(-4)\\
        258971.997 & 26 3 24 1  --  25 2 23 1 & 177.76 & 2.34(-4) & & 261678.002$^\ast$ & 17 12 5 1  --  17 11 6 1 & 156.14 & 1.09(-4)\\
         & 26 3 24 0  --  25 2 23 0 & 177.76 & 2.34(-4) & & 261678.447$^\ast$ & 13 6 8 1  --  12 5 7 1 & 65.82 & 1.50(-4)\\
        259161.713 & 40 4 36 1  --  40 3 37 1 & 423.50 & 1.04(-4) & & 261678.642$^\ast$ & 17 12 5 0  --  17 11 6 0 & 156.14 & 1.09(-4)\\
        259163.316 & 40 4 36 0  --  40 3 37 0 & 423.50 & 1.04(-4) & &  & 17 12 6 0  --  17 11 7 0 & 156.14 & 1.09(-4)\\
        259181.673 & 40 5 36 1  --  40 4 37 1 & 423.50 & 1.04(-4) & & 261679.645$^\ast$ & 13 6 8 0  --  12 5 7 0 & 65.82 & 1.54(-4)\\
        \hline\addlinespace[2pt]
        \multicolumn{2}{c}{\textbf{\ce{CH2CO} (CDMS)}} & \textbf{<800} & & & \multicolumn{4}{l}{\small \citep{Johnson1952, Fabricant1977, Brown1990}} \\
        \hline\addlinespace[2pt]
        240185.794$^{\star\dagger}$ & 12 1 12 -- 11 1 11 & 87.99 & 1.55(-4) & & 260191.982$^{\star\dagger}$ & 13 1 13 -- 12 1 12 & 100.47 & 1.98(-4) \\
        \hline\addlinespace[2pt]
        \multicolumn{2}{c}{\textbf{\ce{CH3CCH} (CDMS)}} & \textbf{<800} & & & \multicolumn{4}{l}{\citep{Cazzoli2008}} \\
        \hline\addlinespace[2pt]
        238883.441 & 14 9  --  13 9 & 670.50 & 2.76(-5) & & 239179.281$^{\star\star}$ & 14 4  --  13 4 & 201.70 & 4.34(-5)\\
        238960.695 & 14 8  --  13 8 & 548.00 & 3.17(-5) & & 239211.215$^{\star\star}$ & 14 3  --  13 3 & 151.14 & 4.51(-5)\\
        239028.931 & 14 7  --  13 7 & 439.85 & 3.54(-5) & & 239234.034$^{\star\dagger}$ & 14 2  --  13 2 & 115.02 & 4.63(-5)\\
        239088.122 & 14 6  --  13 6 & 346.07 & 3.85(-5) & & 239247.728$^{\star\star}$ & 14 1  --  13 1 & 93.35 & 4.71(-5)\\
        239138.245 & 14 5  --  13 5 & 266.68 & 4.12(-5) & & 239252.294$^{\star\star}$ & 14 0  --  13 0 & 86.12 & 4.73(-5)\\
        \hline\addlinespace[2pt]
        \multicolumn{2}{c}{\textbf{\ce{CH3OH}, vt=0--2 (CDMS)}} & \textbf{<2000} & \textbf{>1.0(-7)} & & \multicolumn{4}{l}{\citep{Xu2008}} \\
        \hline\addlinespace[2pt]
        238440.805$^{\star\star}$ & 25 5 21 3  --  24 6 19 3 & 1171.71 & 4.20(-5) & & 240960.557$^\ddagger$ & 5 1 5 3  --  4 1 4 3 & 359.95 & 5.76(-5)\\
        238440.974$^{\star\star}$ & 25 5 20 3  --  24 6 18 3 & 1171.71 & 4.20(-5) & & 241042.589$^{\star\dagger}$ & 22 6 16 2  --  23 5 19 2 & 775.57 & 2.31(-5)\\
        238665.976$^\star$ & 30 3 27 0  --  30 2 28 0 & 1128.27 & 8.47(-5) & & 241159.199$^\ddagger$ & 5 4 2 4  --  4 4 1 4 & 398.11 & 2.15(-5)\\
        238723.283$^\star$ & 7 7 1 5  --  8 6 3 5 & 588.21 & 1.33(-6) & & 241166.580$^\ddagger$ & 5 3 2 4  --  4 3 1 4 & 452.14 & 3.86(-5)\\
        238729.425$^\star$ & 31 3 28 0  --  31 2 29 0 & 1200.13 & 8.61(-5) & & 241178.445$^\ddagger$ & 5 4 1 3  --  4 4 0 3 & 515.80 & 2.18(-5)\\
        238890.424$^{\star\star}$ & 29 3 26 0  --  29 2 27 0 & 1058.72 & 8.37(-5) & &  & 5 4 2 3  --  4 4 1 3 & 515.80 & 2.18(-5)\\
        239142.749$^{\star\dagger}$ & 32 3 29 0  --  32 2 30 0 & 1274.32 & 8.80(-5) & & 241179.886$^\ddagger$ & 5 3 3 5  --  4 3 2 5 & 357.36 & 3.83(-5)\\
        239219.963 & 24 15 10 0  --  25 12 13 3 & 1806.78 & 7.08(-6) & & 241184.189$^\ddagger$ & 5 4 1 5  --  4 4 0 5 & 440.13 & 2.16(-5)\\
        239219.963 & 24 15 9 0  --  25 12 14 3 & 1806.78 & 7.08(-6) & & 241187.428$^\ddagger$ & 5 2 4 5  --  4 2 3 5 & 399.34 & 5.07(-5)\\
        239345.054$^\star$ & 28 3 25 0  --  28 2 26 0 & 991.49 & 8.29(-5) & & 241192.856$^\ddagger$ & 5 2 4 3  --  4 2 3 3 & 333.40 & 5.03(-5)\\
        239397.997$^{\star\star}$ & 16 3 13 2  --  17 0 17 1 & 378.28 & 6.77(-7) & & 241196.430$^\ddagger$ & 5 2 3 3  --  4 2 2 3 & 333.40 & 5.03(-5)\\
        239659.467 & 33 2 31 2  --  33 2 32 1 & 1332.63 & 7.22(-6) & & 241198.285$^\ddagger$ & 5 3 3 3  --  4 3 2 3 & 430.84 & 3.83(-5)\\
        239746.219$^\ddagger$ & 5 1 5 0  --  4 1 4 0 & 49.06 & 5.66(-5) & & 241198.291$^\ddagger$ & 5 3 2 3  --  4 3 1 3 & 430.84 & 3.83(-5)\\
        239971.367$^\star$ & 33 3 30 0  --  33 2 31 0 & 1350.82 & 9.04(-5) & & 241203.706$^\ddagger$ & 5 1 5 4  --  4 1 4 4 & 326.20 & 5.75(-5)\\
        239977.050$\star$ & 27 3 24 0  --  27 2 25 0 & 926.58 & 8.24(-5) & & 241206.035$^\ddagger$ & 5 0 5 4  --  4 0 4 4 & 335.31 & 6.00(-5)\\
        240241.490$^\ddagger$ & 5 3 3 1  --  6 2 5 1 & 82.53 & 1.44(-5) & & 241210.764$^\ddagger$ & 5 2 3 4  --  4 2 2 4 & 434.63 & 5.04(-5)\\
        240321.199$^{\star\dagger}$ & 27 8 19 0  --  28 7 22 0 & 1196.59 & 2.17(-5) & & 241238.144$^\ddagger$ & 5 1 4 5  --  4 1 3 5 & 448.11 & 5.75(-5)\\
        240321.205$^{\star\dagger}$ & 27 8 20 0  --  28 7 21 0 & 1196.59 & 2.17(-5) & & 241267.862$^\ddagger$ & 5 0 5 3  --  4 0 4 3 & 458.39 & 6.00(-5)\\
        240454.848 & 5 1 5 6  --  4 1 4 6 & 717.41 & 5.69(-5) & & 241283.133 & 34 3 31 0  --  34 2 32 0 & 1429.63 & 9.33(-5)\\
        240738.926$^\ddagger$ & 26 3 23 0  --  26 2 24 0 & 863.98 & 8.21(-5) & & 241364.143$^\star$ & 5 1 4 6  --  4 1 3 6 & 717.54 & 5.75(-5)\\
        240752.863$^\star$ & 5 3 2 8  --  4 3 1 8 & 949.67 & 3.79(-5) & & 241441.270$^\ddagger$ & 5 1 4 3  --  4 1 3 3 & 360.02 & 5.79(-5)\\
        240757.889 & 5 2 4 6  --  4 2 3 6 & 911.34 & 5.05(-5) & & 241588.758$^\ddagger$ & 25 3 22 0  --  25 2 23 0 & 803.70 & 8.19(-5)\\
        240757.920$^\star$ & 5 2 3 6  --  4 2 2 6 & 911.34 & 5.05(-5) & & 241615.470 & 30 0 30 1  --  29 3 26 2 & 1082.44 & 2.04(-6)\\
        240784.498 & 5 4 1 7  --  4 4 0 7 & 920.95 & 2.13(-5) & & 241700.159$^\ddagger$ & 5 0 5 1  --  4 0 4 1 & 47.94 & 6.03(-5)\\
        240817.972$^\star$ & 5 1 4 7  --  4 1 3 7 & 833.64 & 5.75(-5) & & 258142.379 & 16 5 12 4  --  16 3 13 4 & 744.14 & 1.16(-7)\\
        240861.406$^{\star\star}$ & 5 4 2 8  --  4 4 1 8 & 779.22 & 2.18(-5) & & 258567.037 & 17 9 8 3  --  18 7 11 3 & 1037.76 & 7.43(-7)\\
        240869.551$^\star$ & 5 0 5 7  --  4 0 4 7 & 768.92 & 5.97(-5) & &  & 17 9 9 3  --  18 7 12 3 & 1037.76 & 7.43(-7)\\
        240916.172$^\ddagger$ & 5 3 3 6  --  4 3 2 6 & 692.78 & 3.87(-5) & & 258780.248$^\ddagger$ & 19 3 17 0  --  19 2 18 0 & 490.59 & 9.01(-5)\\
        240916.173$^\ddagger$ & 5 3 2 6  --  4 3 1 6 & 692.78 & 3.87(-5) & & 259273.686$^\ddagger$ & 17 2 15 3  --  16 1 15 3 & 652.67 & 5.58(-5)\\
        240932.051$^\star$ & 5 4 1 6  --  4 4 0 6 & 649.19 & 2.14(-5) & & 259581.398$^\ddagger$ & 24 1 23 1  --  24 0 24 1 & 717.01 & 4.91(-5)\\
         & 5 4 2 6  --  4 4 1 6 & 649.19 & 2.14(-5) & & 260064.318$^{\star\star}$ & 20 8 13 2  --  21 7 14 2 & 808.20 & 2.13(-5)\\
        240936.742$^\ddagger$ & 5 2 3 8  --  4 2 2 8 & 679.63 & 5.00(-5) & & 260381.463$^\ddagger$ & 20 3 18 0  --  20 2 19 0 & 536.95 & 9.14(-5)\\
        240938.974$^\ddagger$ & 5 0 5 6  --  4 0 4 6 & 542.89 & 5.98(-5) & & 260947.124 & 18 14 5 5  --  19 13 7 5 & 1645.96 & 1.95(-5)\\
        240948.343$^\ddagger$ & 5 3 3 7  --  4 3 2 7 & 656.13 & 3.81(-5) & & 261061.320$^\ddagger$ & 21 4 18 2  --  20 5 16 2 & 623.60 & 3.02(-5)\\
        240952.056$^\ddagger$ & 5 2 4 7  --  4 2 3 7 & 620.93 & 5.05(-5) & & 261704.409$^\ddagger$ & 12 6 7 1  --  13 5 8 1 & 359.77 & 1.78(-5)\\
        240958.922$^\ddagger$ & 5 1 5 8  --  4 1 4 8 & 566.72 & 5.75(-5) & & & & & \\
        
\end{longtable}
\begin{minipage}{0.98\textwidth}
\textbf{Notes}:
(a) Denotations after frequency: 
For the robustly detected species, 
$\star\star$ = key transitions (i.e., unblended and optically thin) that can be used to constrain the column densities and excitation temperatures in most sources (might be blended in a few very bright sources); 
$\star$ = key transitions that can be used in sources without too much line blending; 
$\star\dagger$ = strong but blended transitions (not considered);
$\ddagger$ = optically thick transitions.
For the tentatively detected species, 
$\ast\ast$ = key transitions that are used to determine the upper limits of column densities;
$\ast$ = unblended but weaker transitions that are checked for anti-correlation in determining the upper limits.
(b) The format of $A_\mathrm{ij}$: $m(n)=m\times10^n$.
\end{minipage}

\section{Figures for spectral fitting results}\label{appendix:full_fit}
\begin{figure*}[!h]
    \centering
    \includegraphics[width=0.98\textwidth]{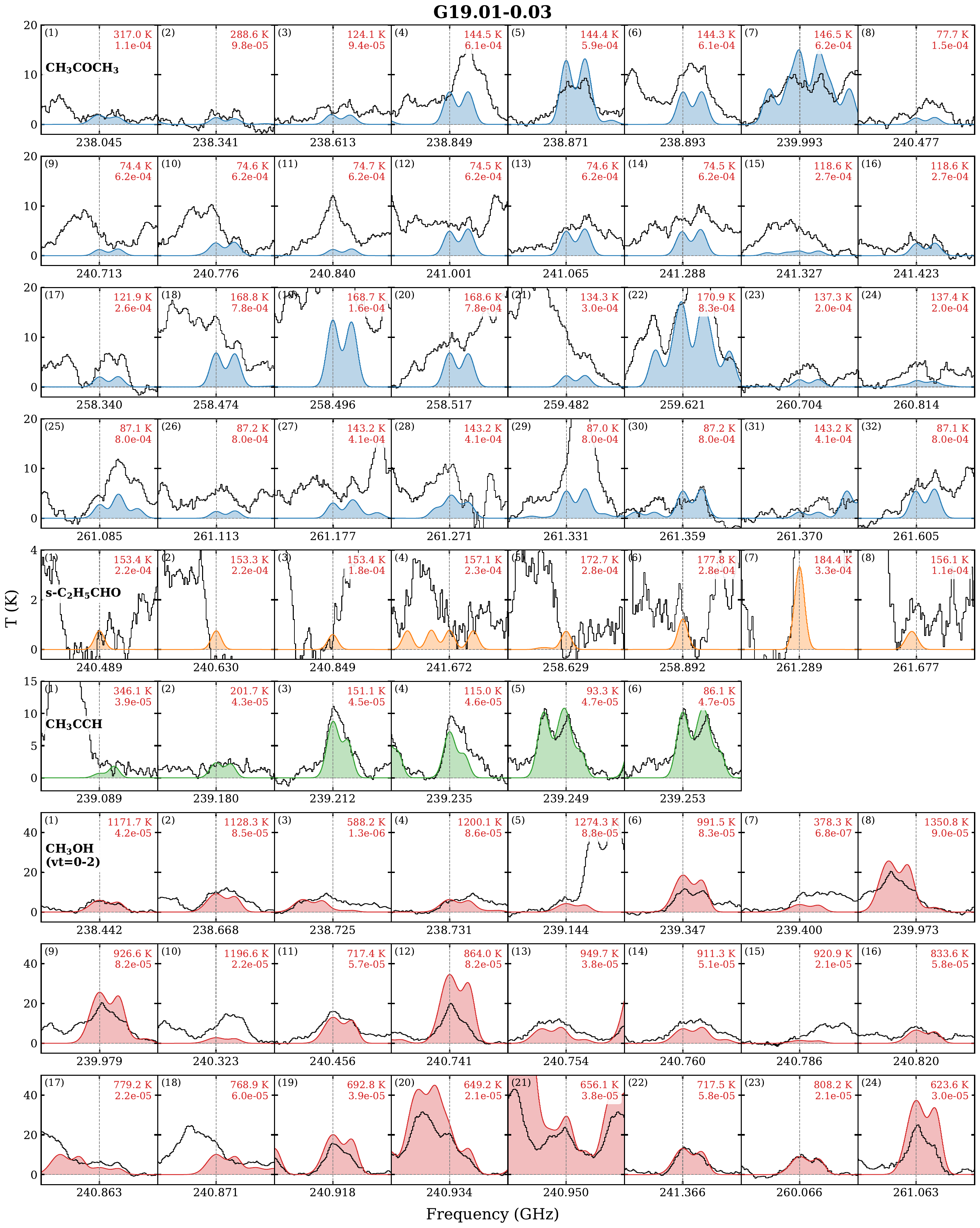}
    \caption{Best-fit LTE models of acetone (red), propanal (orange), propyne (green), and methanol (red) overlaid on the ALMA spectrum of G19.01-0.03 (black). The species name is indicated in bold black in the upper left in the first panel. The upper energy level and Einstein A coefficient are annotated in red in the upper right. We only labeled the central frequency of each line on the x axis, but the span of each panel is equal to [--15, +15] km~s$^{-1}$. Not all transitions listed in Table~\ref{tab:key_trans} are displayed here; the selected transitions are either strong or not very blended. For methanol, the very optically thick transitions were now shown here, but there are still several lines (e.g., no.~9, 12, 19, 24) that are overestimated due to optical depth effects, especially in brighter sources.}
    \label{fig:fit_all_G19.01}
\end{figure*}

\begin{figure*}[!h]
    \centering
    \includegraphics[width=\textwidth]{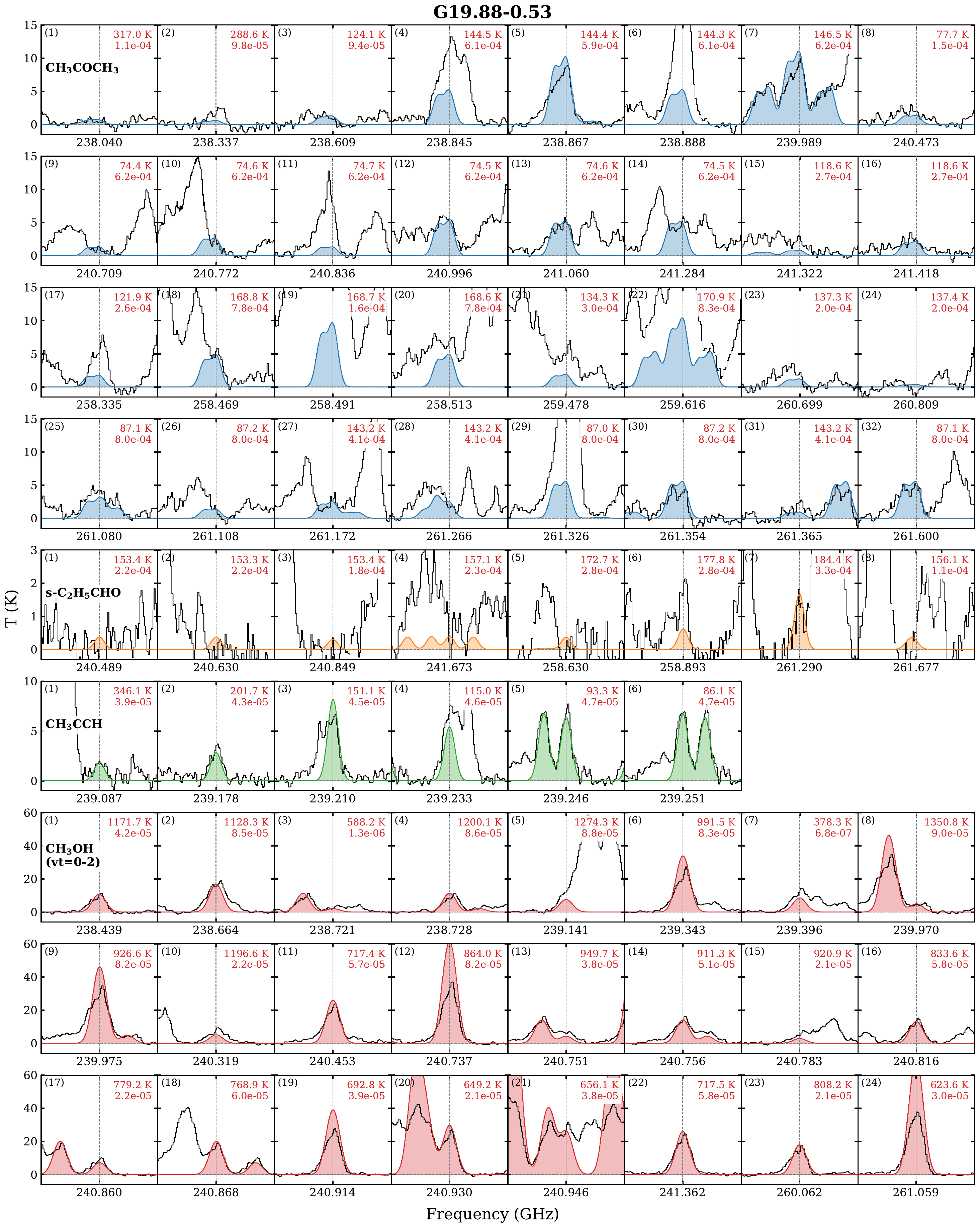}
    \caption{Same as Fig.~\ref{fig:fit_all_G19.01} but for G19.88-0.53.}
    \label{fig:fit_all_G19.88}
\end{figure*}

\begin{figure*}[!h]
    \centering
    \includegraphics[width=\textwidth]{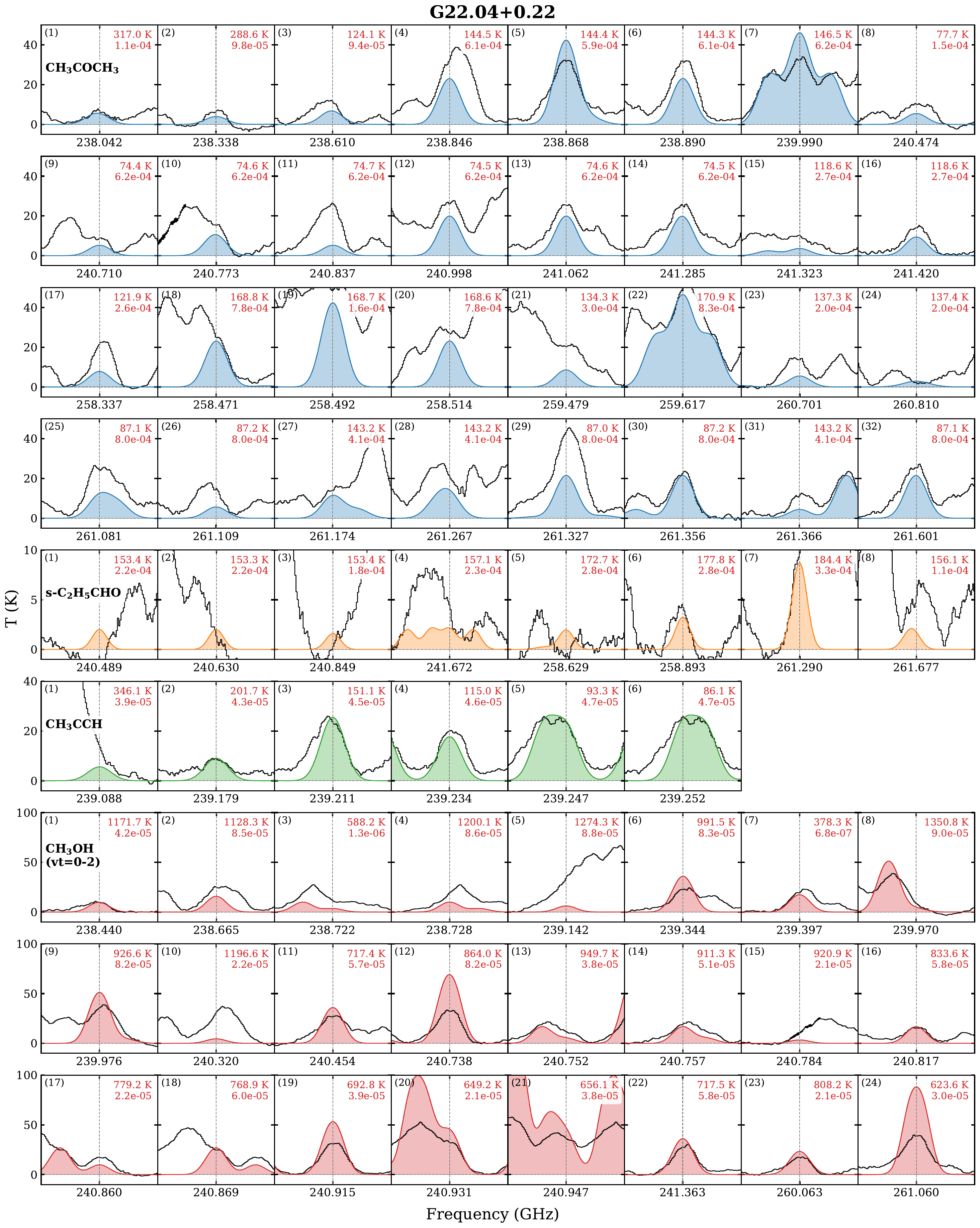}
    \caption{Same as Fig.~\ref{fig:fit_all_G19.01} but for G22.04+0.22.}
    \label{fig:fit_all_G22.04}
\end{figure*}

\begin{figure*}[!h]
    \centering
    \includegraphics[width=\textwidth]{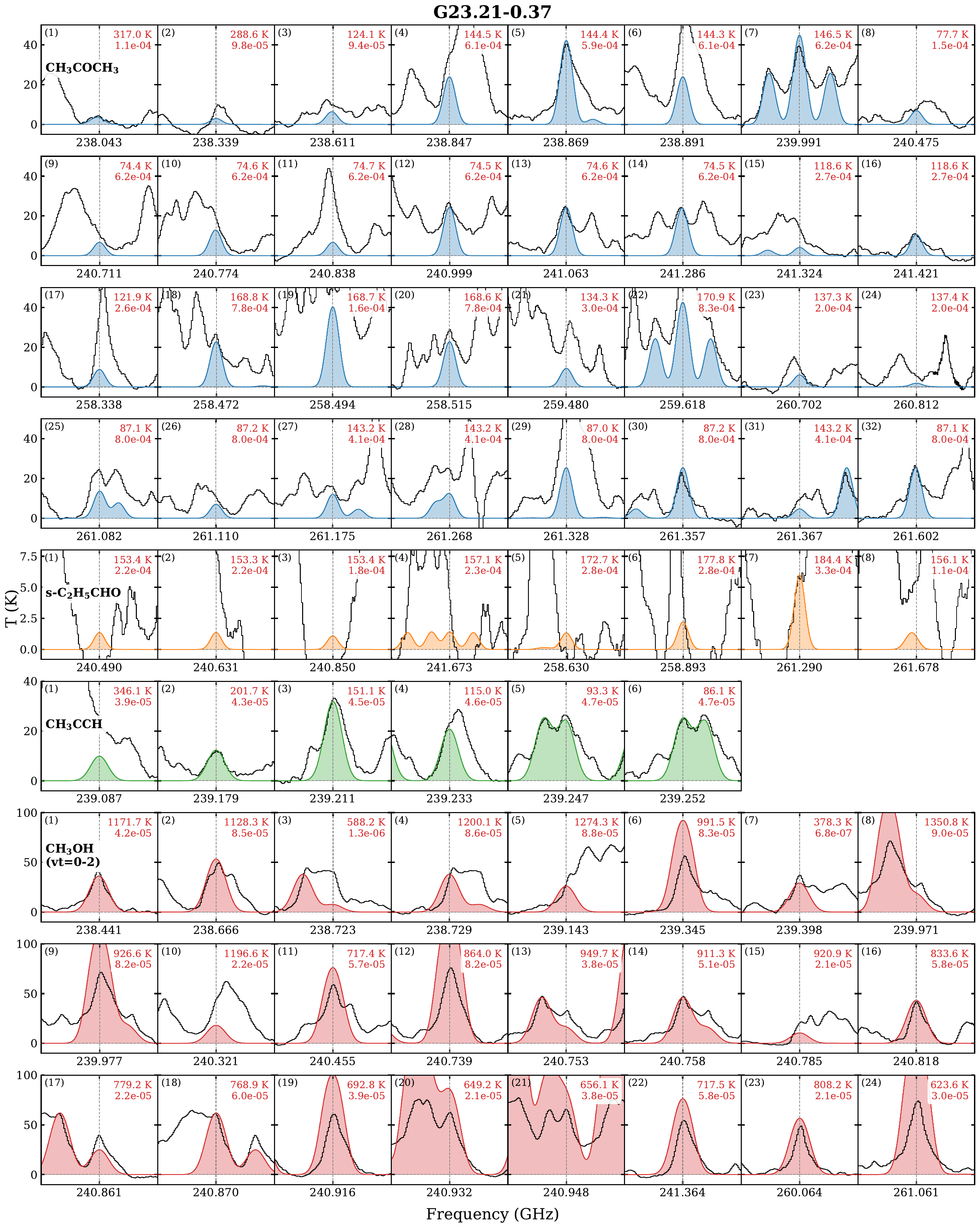}
    \caption{Same as Fig.~\ref{fig:fit_all_G19.01} but for G23.21-0.37.}
    \label{fig:fit_all_G23.21}
\end{figure*}

\begin{figure*}[!h]
    \centering
    \includegraphics[width=\textwidth]{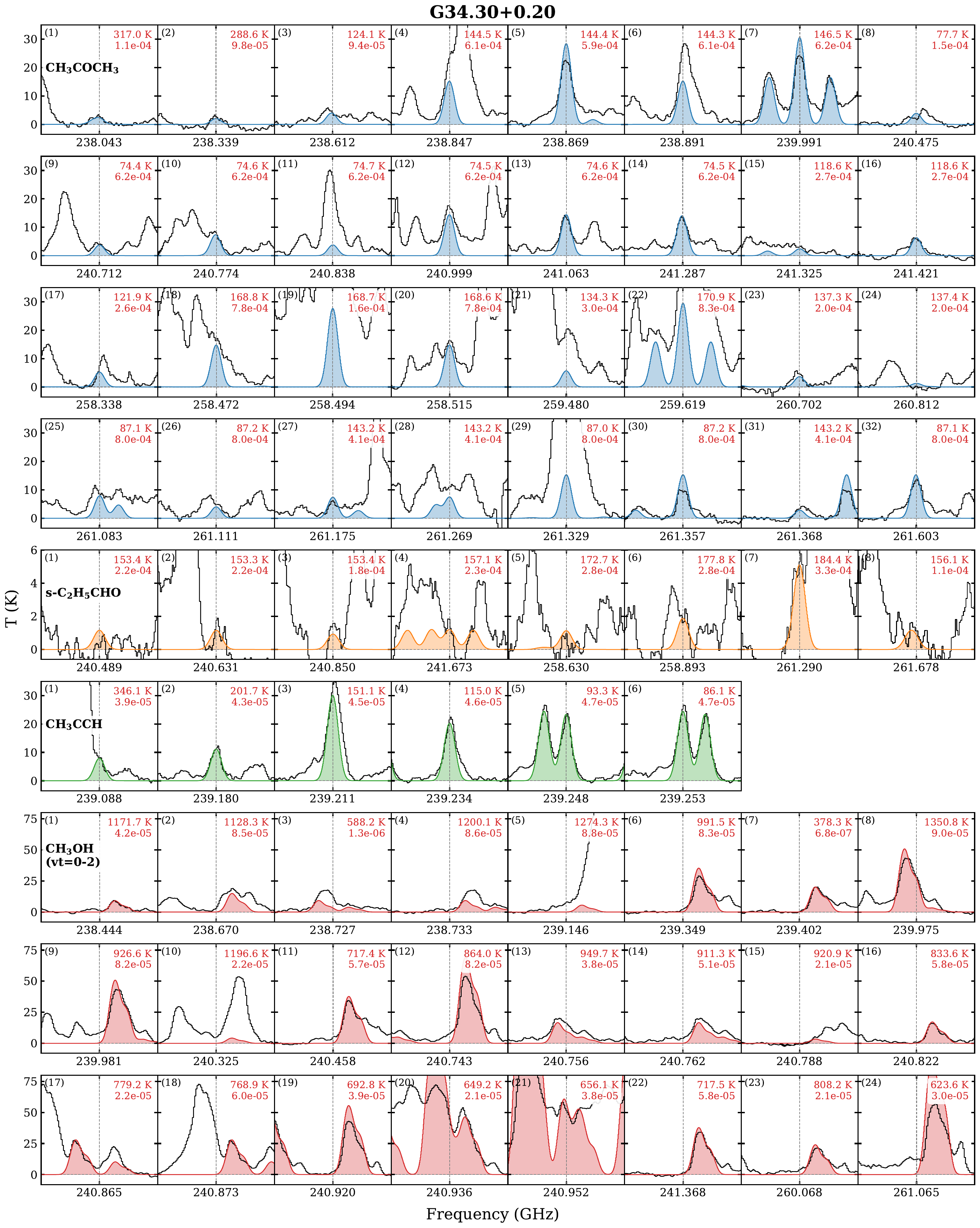}
    \caption{Same as Fig.~\ref{fig:fit_all_G19.01} but for G34.30+0.20.}
    \label{fig:fit_all_G34.30}
\end{figure*}

\begin{figure*}[!h]
    \centering
    \includegraphics[width=\textwidth]{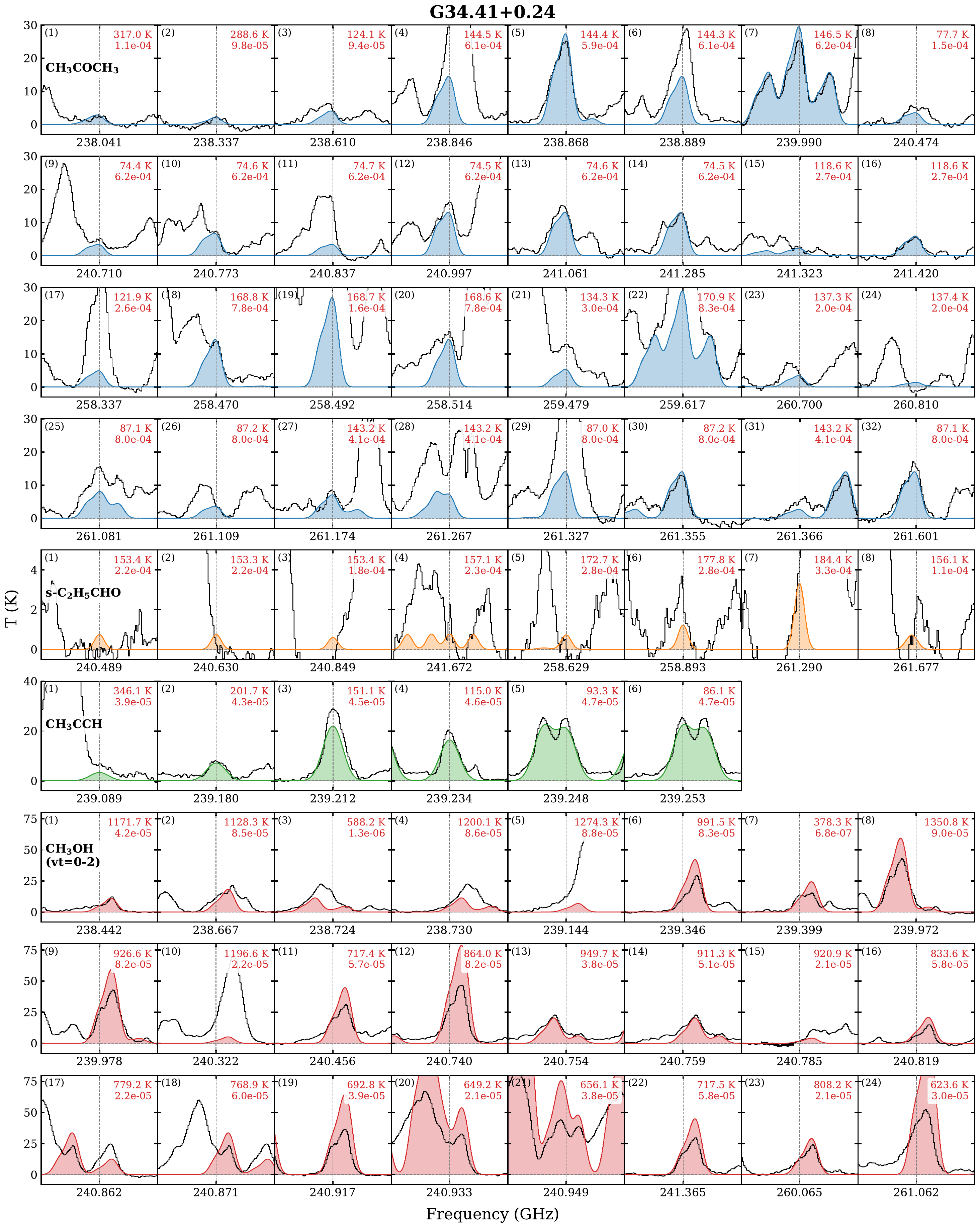}
    \caption{Same as Fig.~\ref{fig:fit_all_G19.01} but for G34.41+0.24.}
    \label{fig:fit_all_G34.41}
\end{figure*}

\begin{figure*}[!h]
    \centering
    \includegraphics[width=\textwidth]{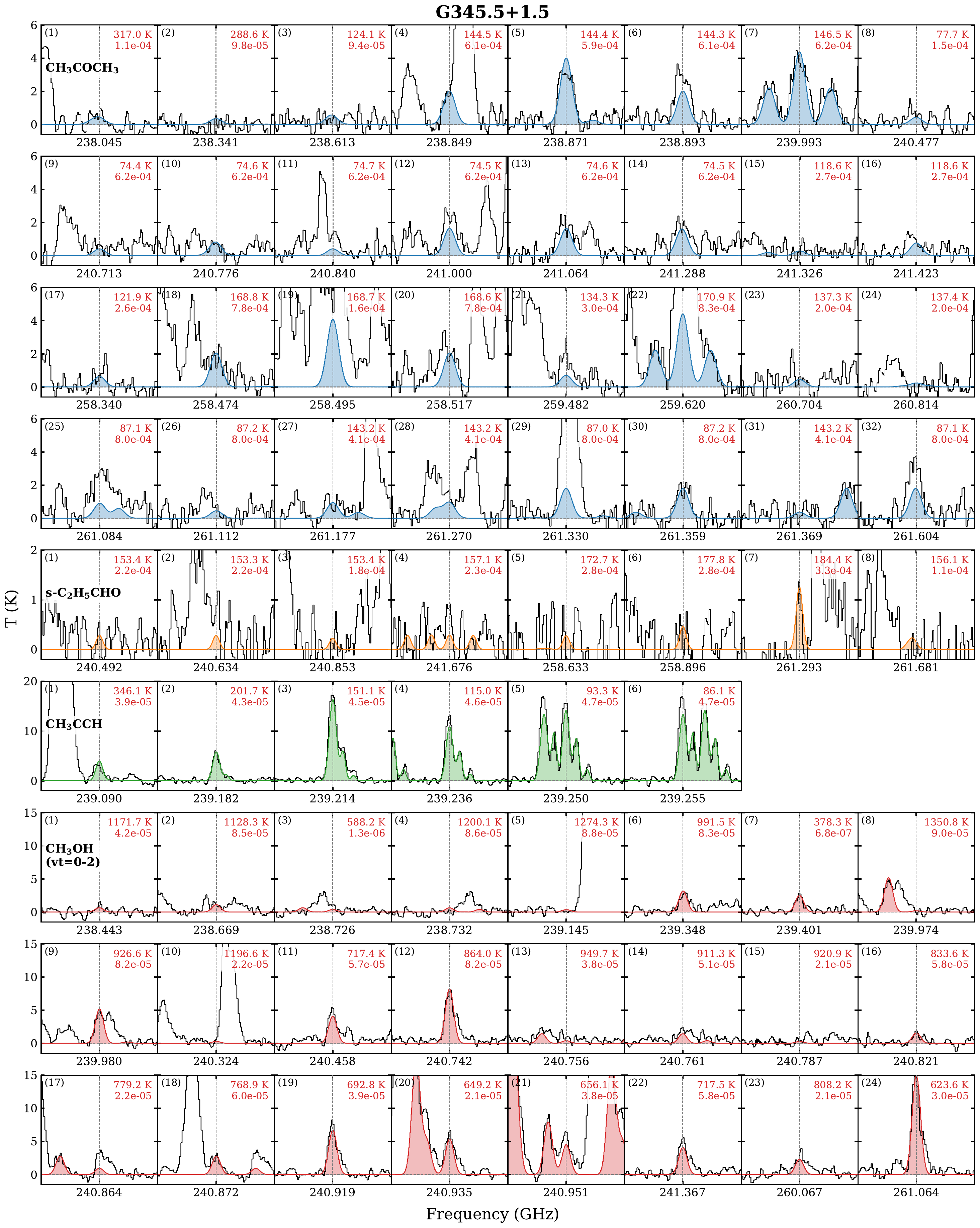}
    \caption{Same as Fig.~\ref{fig:fit_all_G19.01} but for G345.5+1.5.}
    \label{fig:fit_all_G345.5}
\end{figure*}

\begin{figure*}[!h]
    \centering
    \includegraphics[width=\textwidth]{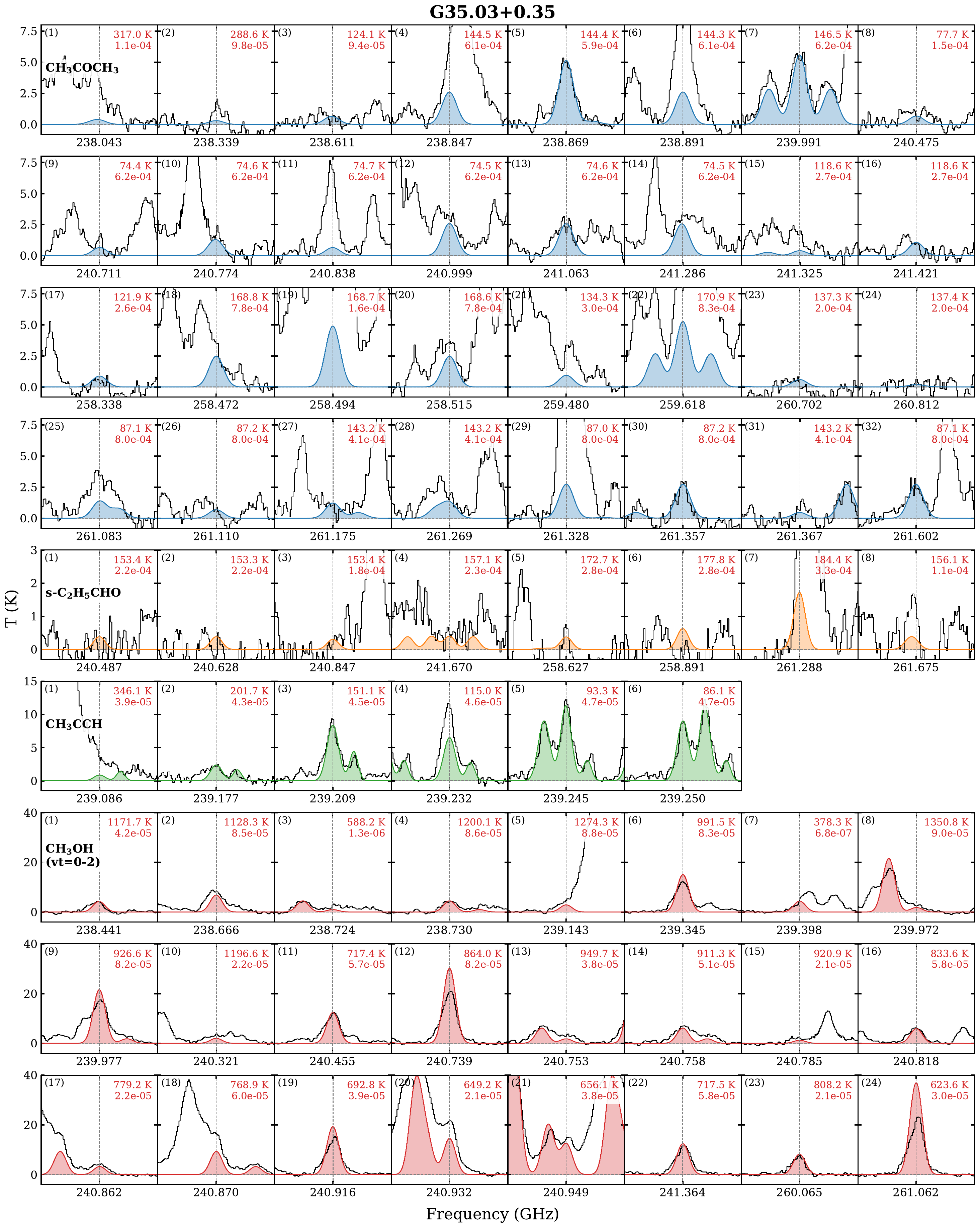}
    \caption{Same as Fig.~\ref{fig:fit_all_G19.01} but for G35.03+0.35.}
    \label{fig:fit_all_G35.03}
\end{figure*}

\begin{figure*}[!h]
    \centering
    \includegraphics[width=\textwidth]{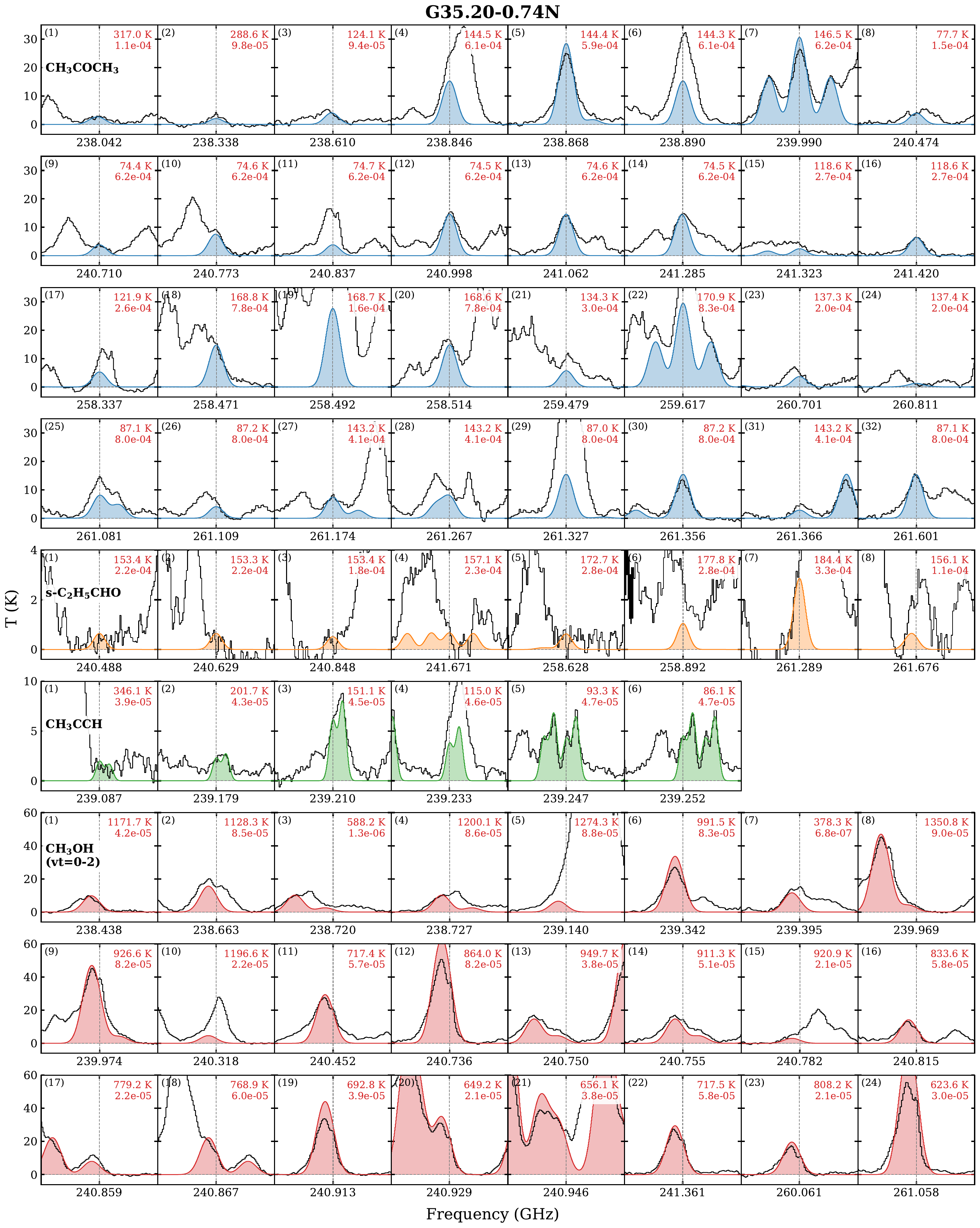}
    \caption{Same as Fig.~\ref{fig:fit_all_G19.01} but for G35.20-0.74N.}
    \label{fig:fit_all_G35.20}
\end{figure*}

\begin{figure*}[!h]
    \centering
    \includegraphics[width=\textwidth]{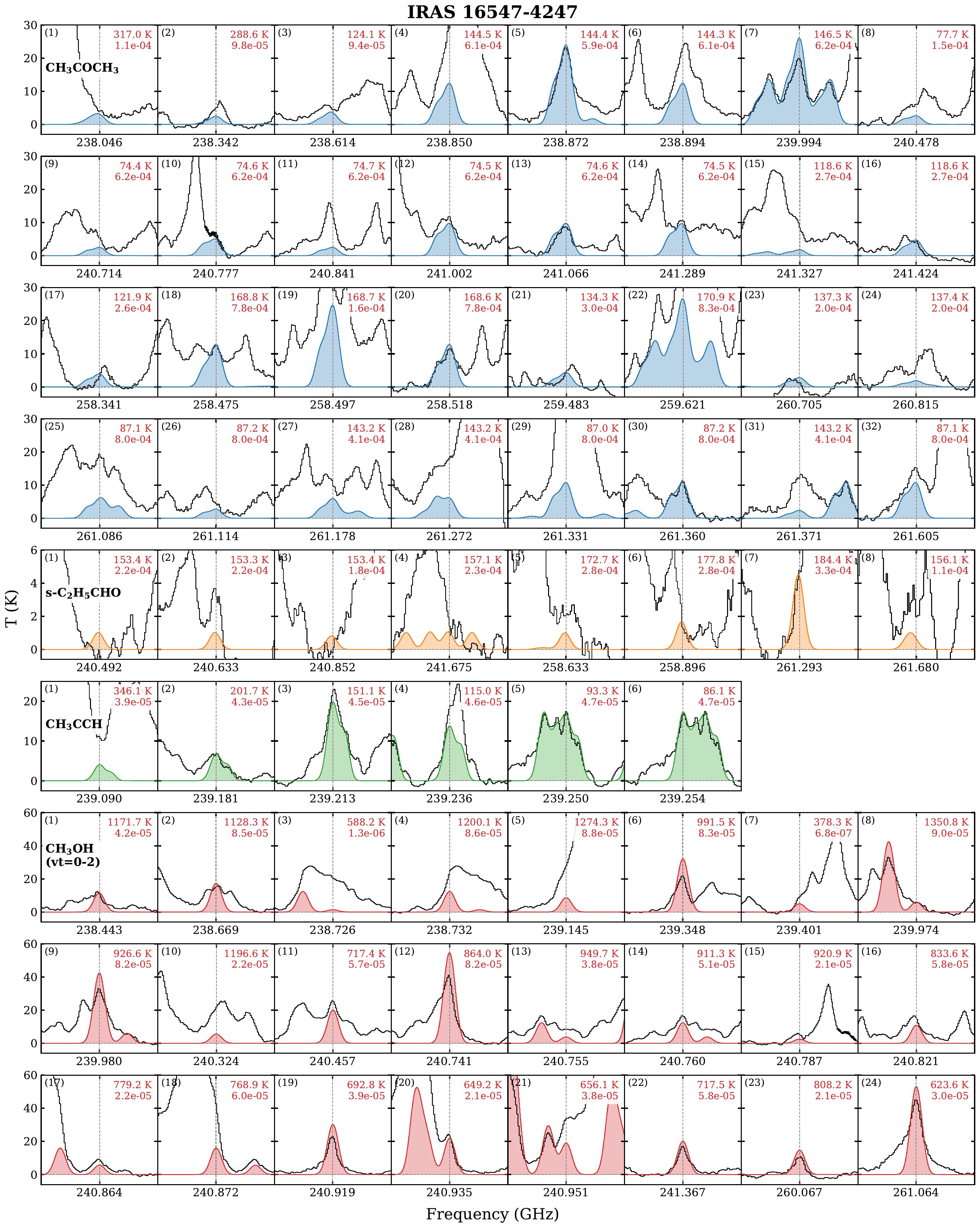}
    \caption{Same as Fig.~\ref{fig:fit_all_G19.01} but for IRAS~16547-4247.}
    \label{fig:fit_all_IRAS16547}
\end{figure*}

\begin{figure*}[!h]
    \centering
    \includegraphics[width=\textwidth]{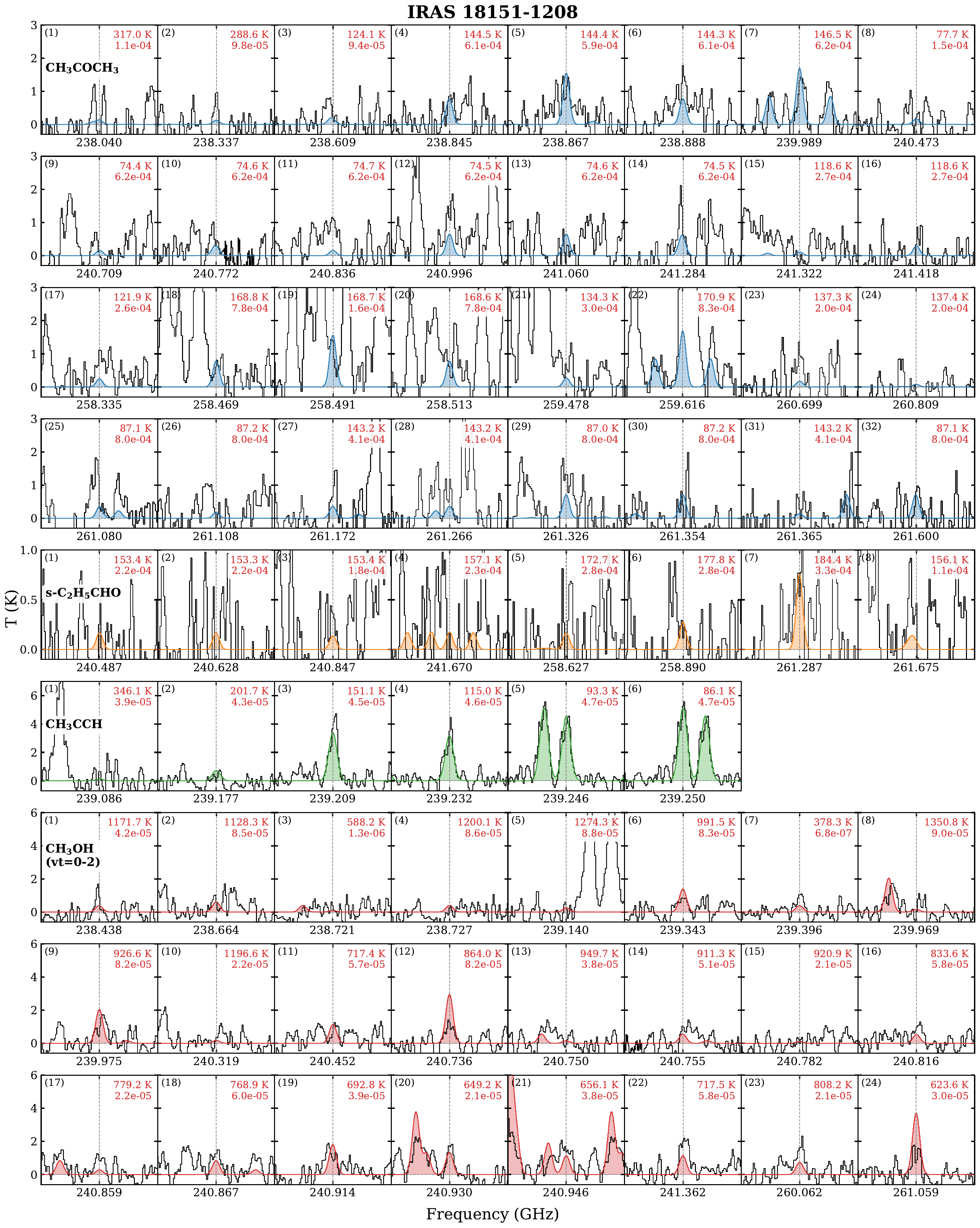}
    \caption{Same as Fig.~\ref{fig:fit_all_G19.01} but for IRAS~18151-1208.}
    \label{fig:fit_all_IRAS18151}
\end{figure*}

\begin{figure*}[!h]
    \centering
    \includegraphics[width=\textwidth]{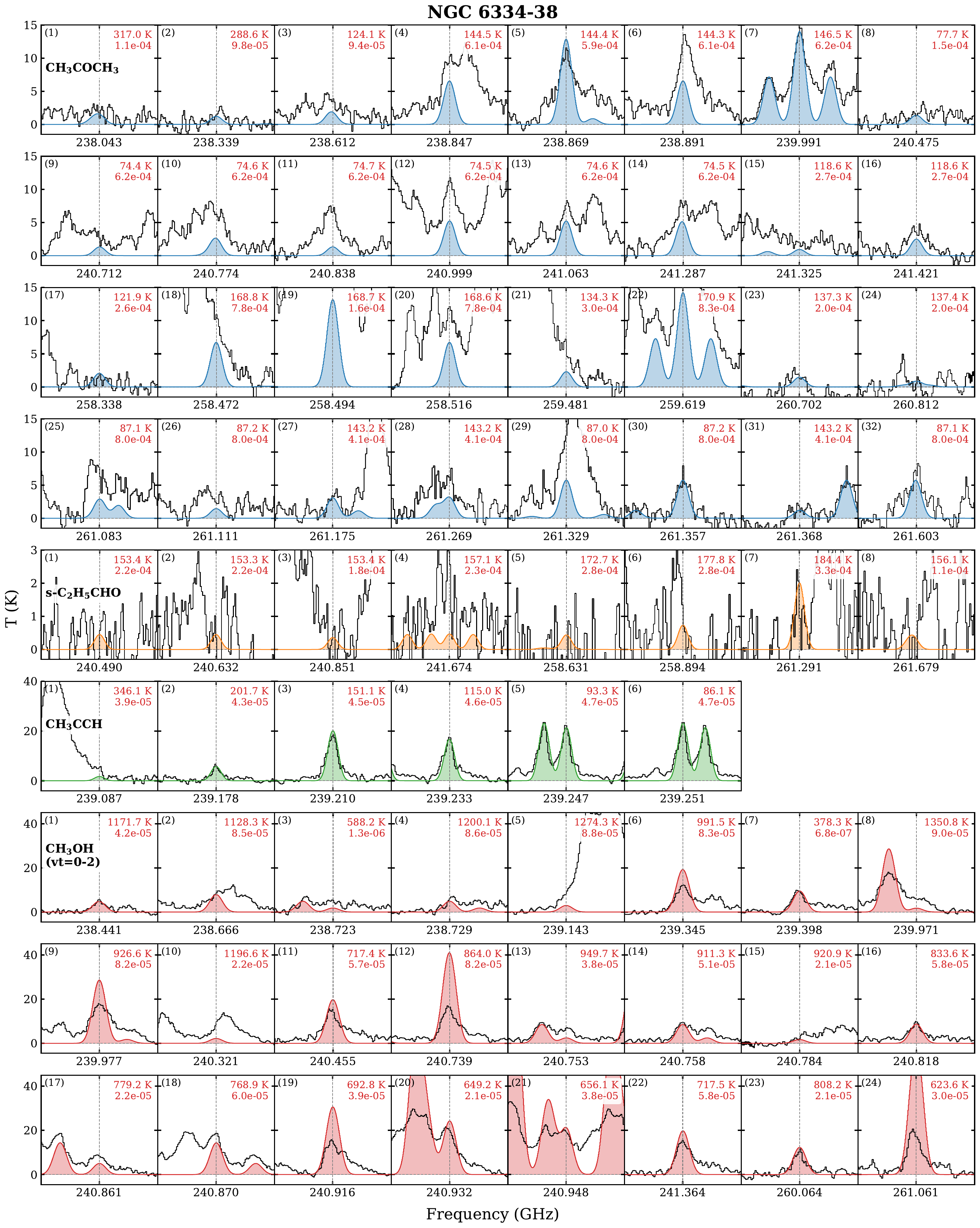}
    \caption{Same as Fig.~\ref{fig:fit_all_G19.01} but for NGC~6334-38.}
    \label{fig:fit_all_NGC6334-38}
\end{figure*}

Figures~\ref{fig:fit_all_G19.01}--\ref{fig:fit_all_NGC6334-38} present the best-fit LTE-modeled spectra of acetone, propanal, methanol, and propyne, overlaid on the ALMA spectra. For each species, a number of strong transitions are selected to be displayed, including both the unblended lines and the optically thick ones that are overfit by the models.

\end{appendix}

\end{document}